\documentclass[11pt, a4paper]{article}
\usepackage[labelfont=bf,textfont=normal]{caption}
\usepackage[font=small, format=hang, margin={1cm, 1cm}]{caption}
\usepackage{jheppub}
\usepackage{amsfonts} 
\usepackage{amsmath} 
\usepackage{mathrsfs}  %
\usepackage{amssymb}
\usepackage{empheq}
\usepackage{mathtools}
\usepackage{slashed}
\usepackage{relsize}
\usepackage{setspace}   
\usepackage{color}  
\usepackage{wasysym}
\usepackage{booktabs}
\usepackage[utf8,applemac]{inputenc} 
\usepackage{tensor}
\usepackage{tikz}
\usetikzlibrary{calc, arrows.meta}
\usepackage{graphicx}
\usepackage{subfig}
\usepackage[normalem]{ulem}
\usepackage{hyperref}
\usepackage{epstopdf}
\usepackage{comment}
\usepackage{mathtools}

\usepackage[hang,flushmargin]{footmisc}

\numberwithin{equation}{section}
\usepackage{dcolumn}
\usepackage{bm}
\usepackage[toc,page]{appendix}

\usepackage{vmargin}
\setmargrb{2cm}{1cm}{2cm}{2cm}

\usepackage{multirow}

\usepackage{enumitem}

\setlist{
  topsep = 0em,
  }

\newcommand{\dub}{\,\,} 
\renewcommand{\SS}{\mathcal{S}}

\newcommand{\dd}{\text{d}}

\newcommand{\eps}{\epsilon}

\newcommand\nts{\negthickspace}
\newcommand\bns{\nts \nts \nts}

\newcommand\defeq{\mathrel{\mathop:}=}

\newcommand\VV{\mathcal{V}}
\newcommand\UU{\mathcal{U}}
\newcommand\RR{\mathcal{R}}

\newcommand\EE{\mathcal{E}}

\newcommand\DD{\mathcal{D}}
\newcommand\OO{\mathcal{O}}
\newcommand\CC{\mathcal{C}}
\newcommand\NN{\mathcal{N}}
\newcommand\MM{\mathcal{M}}

\renewcommand\AA{\mathcal{A}}
\newcommand\KK{\mathcal{K}}
\newcommand\TT{\mathcal{T}}
\newcommand\PP{\mathcal{P}}

\newcommand\JJ{\mathcal{J}}
\newcommand\WW{\mathcal{W}}

\newcommand\Dv{\text{D}_{v}}

\definecolor{bluegreen}{RGB}{0,102,102}

\usepackage{pict2e}
\makeatletter
\newcommand{\loplus}{\mathbin{\mathpalette\dog@lsemi{+}}}
\newcommand{\dog@lsemi}[2]{\dog@semi{#1}{#2}{270,90}}
\newcommand{\dog@semi}[3]{%
  \begingroup
  \sbox\z@{$\m@th#1#2$}%
  \setlength{\unitlength}{\dimexpr\ht\z@+\dp\z@\relax}%
  \makebox[\wd\z@]{\raisebox{-\dp\z@}{%
    \begin{picture}(1,1)
    \linethickness{\variable@rule{#1}}
    \roundcap
    \put(0.5,0.5){\makebox(0,0){\raisebox{\dp\z@}{$\m@th#1#2$}}}
    \put(0.5,0.5){\arc[#3]{0.5}}
    \end{picture}%
  }}%
  \endgroup
}
\newcommand{\variable@rule}[1]{%
  \fontdimen8  
  \ifx#1\displaystyle\textfont3\else
    \ifx#1\textstyle\textfont3\else
      \ifx#1\scriptstyle\scriptfont3\else
        \scriptscriptfont3\relax
  \fi\fi\fi
}
\makeatother

\hypersetup{
    colorlinks,%
    citecolor=brick,%
    filecolor=blue,%
    linkcolor=cornflower,%
    urlcolor=cornflower,
    linktoc=page
}

\definecolor{scarlet}{rgb}{0.8, 0, 0}
\definecolor{brick}{rgb}{0.64314, 0, 0}
\definecolor{cornflower}{rgb}{0.12549, 0.29020, 0.52941}

\def\mathcolor#1#{\@mathcolor{#1}}
\def\@mathcolor#1#2#3{%
  \protect\leavevmode
  \begingroup
    \color#1{#2}#3%
  \endgroup
}

\begin{document}

\title{(Anti)-de Sitter with leaky boundaries and corners}
\author{ Robert McNees${}^1$ and C\'eline Zwikel${}^{2,3}$}
\affiliation{$^1$Department of Physics, Loyola University ChicagoChicago, IL, USA\\$^2$Perimeter Institute for Theoretical Physics, 31 Caroline Street North, Waterloo, Ontario, Canada N2L 2Y5\\$^3$Coll\`ege de France, 11 place Marcelin Berthelot, 75005 Paris, France}
\emailAdd{rmcnees@luc.edu}\emailAdd{celine.zwikel@college-de-france.fr}

\abstract{
We construct charges for four-dimensional spacetimes with a non-vanishing cosmological constant, including charges that are not conserved because of a leaky boundary and charges associated with corner terms in the symplectic current. The construction leads to manifestly finite charges for any choice of boundary conditions, and reveals new charges in partial Bondi gauge for an enlarged class of field configurations which are not fully on-shell, including a Weyl charge for conformal boundary conditions. In addition, we generalize the variational principle for AdS$_4$ gravity with Dirichlet boundary conditions to a wedge of spacetime where the conformal boundary includes a corner. The boundary data is completely general, with no conditions restricting time dependence or the determinant of the metric. The Ward identities associated with boundary diffeomorphisms are shown to reproduce the evolution equations for the mass and angular momentum of fully on-shell field configurations.
}

\maketitle

\section{Introduction}

In gauge theories such as gravity, charges are defined on a codimension-two surface or ``cut'' of the boundary. One may be interested in the evolution of a charge between two cuts. If the values differ then the quantity is not conserved, and in this sense the boundary is ``leaky.'' For example, radiation carried away can reduce the energy of the system, as encoded in the Trautman--Bondi mass-loss formula for asymptotically flat spacetimes \cite{Trautman:1958zdi, Bondi:1962px}. It has been argued that for asymptotically de Sitter (dS) spacetimes the boundary must be leaky \cite{Ashtekar:2014zfa}. This follows from the fact that its asymptotic boundary is a spacelike surface where initial data can be prescribed, so forbidding leaks represents an \textit{ad hoc} removal of physical data. For anti-de Sitter (AdS) spacetimes, the situation is different: the asymptotic boundary is timelike, and consistent non-leaky boundary conditions have been extensively discussed. However, it is also interesting to consider leaky AdS, where energy transported along null geodesics reaches the asymptotic boundary and can then escape into an external bath rather than reflect back into the interior. This is precisely the scenario considered in recent discussions of the information-loss paradox in two-dimensional AdS with conformal matter \cite{Almheiri:2019yqk, Almheiri:2019qdq, Almheiri:2020cfm}; see also the covariant phase space analysis of \cite{Grumiller:2023ahv}.

An important aspect of gravity with a non-vanishing cosmological constant is that leakiness requires a fluctuating boundary metric \cite{Anninos:2010zf, Ashtekar:2015lla, Chrusciel:2020rlz, Compere:2019bua, Compere:2020lrt, Fernandez-Alvarez:2021yog, Compere:2023ktn, Bonga:2023eml, Ciambelli:2024kre, McNees:2024iyu, Compere:2024ekl, Arenas-Henriquez:2025rpt}. This is in contrast with asymptotically flat spacetimes where leakiness occurs even with a fixed boundary  metric. Indeed, when $\Lambda \neq 0$ the Einstein equations couple the boundary data to quantities associated with gravitational radiation. Another aspect worth exploring is the role of finite portions or corners of the asymptotic boundary. In AdS the spacetime is not globally hyperbolic, and therefore initial data on a spacelike slice must be supplemented with boundary conditions. The compatibility between boundary conditions and initial conditions is subtle \cite{An_2021}, and the corner -- the intersection of the spacelike slice with the timelike boundary -- makes this explicit. For instance, in \cite{Anninos:2024xhc} the authors analyze finite tubes in AdS$_4$ and show the emergence of a new corner mode that arises using what they refer to as conformal boundary conditions. Likewise, there are subtleties associated with the formulation in terms of the initial data on a null hypersurface in Bondi coordinates \cite{Poole:2018koa}.  

Motivated by these considerations, we study leaky boundaries with corners in four-dimensional asymptotically locally (A)dS spacetimes. We analyze the consequences of considering finite portions of leaky boundaries for the the symplectic structure and the construction of variational principles, as well as the residual symmetries and their generators. While most of our results apply for both $\Lambda < 0$ and $\Lambda >0$, there are of course significant global differences between these two cases. For that reason we sometimes focus on $\Lambda<0$, especially when there are connections to holography.

Our setup approaches the asymptotic boundary using Bondi coordinates based on a family of outgoing null hypersurfaces. These coordinates are well-suited to discussing radiation in asymptotically flat spacetimes \cite{Bondi:1962px,Sachs:1962zza,Newman:1962cia} and they can also be used for a non-vanishing cosmological constant \cite{Poole:2018koa, Compere:2019bua, Fiorucci:2020xto, Compere:2020lrt, Compere:2023ktn, Geiller:2022vto,McNees:2024iyu}. However, we cannot work in the Bondi--Sachs (BS) \cite{Bondi:1962px,Sachs:1962zza} or Newman--Unti (NU) \cite{Newman:1962cia} gauges since either condition would restrict the allowed field configurations to those with fixed boundary metric. Fortunately it is possible to relax some of the conditions associated with these gauges. An immediate consequence of relaxing the gauge is the enlargement of this asymptotic symmetry algebra.\,\footnote{In certain cases, these enlargements have a clear physical explanation by being related to memory effects or soft theorems through infrared triangles, see the review \cite{Strominger:2017zoo}.} These generalizations may be obtained by allowing the induced metric on the boundary to fluctuate \cite{Barnich:2009se,Barnich:2016lyg,Barnich:2011mi,Barnich:2010eb,Campiglia:2015yka,Campiglia:2020qvc,Campiglia:2014yka,Flanagan:2015pxa,Compere:2018ylh,Freidel:2021fxf}, restoring the covariance of boundary Carrollian structure \cite{Ciambelli:2018wre,Ciambelli:2018xat,Campoleoni:2023fug,Mittal:2022ywl, Geiller:2025dqe}, or relaxing the conditions on the radial coordinate \cite{Geiller:2022vto,Geiller:2024amx}. The latter option, known as the partial Bondi gauge (PBG), will be adopted here. Unlike BS gauge, which fixes the radial coordinate to be the areal distance, or NU gauge, which identifies it with an affine parameter, the PBG only requires that the radial coordinate run along the null geodesics without further specification.  One benefit of working in PBG is that it is easier to reach from other gauges. For example, what we refer to as PBG aligns with the translation of Kerr-dS$_4$ in Boyer-Lindquist coordinates to the ``generalized Bondi coordinates'' of \cite{Hoque:2021nti}. 

This work generalizes previous analyses of asymptotic boundaries of (A)dS in Bondi gauge in two ways. First, the induced boundary metric in our setup is completely general. There are no restrictions on any of its components, which allows for arbitrary dependence on the retarded Bondi time, rotating frames, conformal class, etc. This is motivated by applications to holography, where the derivation of the Ward identities for the dual theory require varying the action with respect to arbitrary boundary data (or ``sources'' from the point of view of dual theory). And second, we work in PBG and do not impose the BS gauge condition. In \cite{Geiller:2024amx} it was shown that additional charges are accessible via this gauge relaxation in asymptotically flat spacetimes. A natural question is whether such charges also occur for $\Lambda \neq 0$; we show that they do.

Another important motivation for our work concerns the general framework of the covariant phase space for leaky boundaries and in the presence of corners. The present analysis is concerned with three aspects: finiteness and integrability of the charges, the emergence of new charges associated with corner terms in the symplectic current, and relaxing select on-shell conditions on the fields. 
This work is the continuation of \cite{McNees:2023tus,McNees:2024iyu}.

First, computations involving asymptotic boundaries typically yield divergent quantities that require regularization and renormalization in order to extract meaningful results. Many prescriptions exist in the literature, but we will focus on \cite{McNees:2023tus,McNees:2024iyu}, for related earlier work see \cite{Andrade:2006pg}. This approach has the advantage of not being tied to a specific choice of boundary conditions and is therefore well suited to leaky boundaries. It ensures that the symplectic potential is finite up to a total variation in the field space, which implies a finite symplectic current and hence finite charges. Once boundary conditions are chosen, an action and variational principle can be constructed by including appropriate surface terms in the action to address the $\delta$-exact divergences in the symplectic potential. For $\Lambda\neq0$ and Dirichlet boundary conditions, working in PBG, we show that the surface terms in the action are those usually required by holographic renormalization \cite{deHaro:2000xn,Compere:2008us,Hoque:2025iar,Compere:2020lrt} along with new corner terms. These corner terms take a different form than the joint contributions obtained in previous analyses of non-smooth boundaries \cite{Hayward:1993my, Hawking:1996ww, Lehner:2016vdi, Jubb:2016qzt}.

The second aspect concerns progress in understanding the nature of corner charges. Corner charges differ fundamentally from the more familiar Brown--York type charges. The key distinction is that they are not constrained by the equations of motion, so any conditions on their evolution must come from imposing additional boundary conditions or boundary dynamics. As emphasized in \cite{McNees:2024iyu}, the relevant effects localize at corners at the level of the symplectic potential, which becomes manifest after a suitable rearranging of terms. In this work, we extend this analysis to the charges themselves and demonstrate how their status differs from that of Brown--York charges. 

Third, we relax the on-shell assumption by allowing what we refer to as ``partially on-shell'' configurations. These conditions, which impose only components of the Einstein equations conjugate to the non-zero components of the metric in PBG, are sufficient to determine the large-$r$ behavior of the fields. But they do not enforce the evolution equations for the mass and angular momentum of the spacetime. This implies that the Brown-York quasilocal stress tensor is traceless but not conserved. Although the fields are not fully on-shell, the construction described in appendix A of \cite{McNees:2024iyu}, and reviewed here in appendix \ref{app:Conserved}, nevertheless gives a conserved current built from the presymplectic and weakly vanishing Noether currents. The associated codimension-2 form defines charges which may be integrable but not conserved, as expected for leaky systems. Since the leakiness in this case is tied directly to relaxing some of the on-shell conditions, one might refer to the system as ``off shell leaky.'' In lower dimensional examples, relaxing the on-shell conditions in this manner reveals how an extended off-shell symmetry is broken on-shell in the holographic dual. For example, how the conformal symmetry of SYK-type models reduces to a simple $U(1)$ on-shell \cite{Grumiller:2017qao}.

\paragraph{Organization of the paper:} 
We begin in section \ref{sec:PBG} by reviewing the partial Bondi gauge \cite{Geiller:2022vto} and introducing the geometric framework needed to describe the asymptotic boundary. Our starting point in section \ref{sec:BeforeAndAfterConstraints} is the finite symplectic potential obtained in \cite{McNees:2024iyu}. The latter was compatible with a flat limit, as the condition relating the boundary geometry and the normal shear was not imposed. Here, since we focus on the case $\Lambda \neq 0$, we impose this relation and analyze the structure of the symplectic potential: a boundary contribution that we identify with the usual Brown--York tensor, a corner contribution that explicitly carries the new fields arising from the relaxation of the gauge, and $\delta$-exact terms.
In section \ref{sec:symmcharges}, we review the residual symmetries of the PBG, derive the transformation laws of the fields, and compute the associated charges. We find that new corner charges may appear in PBG, similar to the case of asymptotically flat spacetimes \cite{Geiller:2024amx}. In section \ref{sec:FiniteAction} we construct the variational principle for Dirichlet conditions at a boundary with an explicit corner and show that $\delta$-exact divergences in the action can be removed by appropriate surface and corner Lagrangians. We then examine the Ward identities and find that they correspond to the evolution equations for the mass and angular momentum aspects. We conclude with a brief summary of our results. Several appendices elaborate on technical details, present quantities associated with projections of the bulk Weyl tensor, and develop the Kerr-AdS$_4$ spacetime in PBG.

\section{Review of Gauge, On-Shell Conditions, and Geometry}\label{sec:PBG}

In this section, we briefly review the partial Bondi gauge (PBG) \cite{Geiller:2022vto,Geiller:2024amx} and introduce the notations used to describe the asymptotic geometry. Additional details appear in appendices \ref{app:Geometry} and \ref{app:AsymptoticExpansions}.

\subsection{Partial Bondi Gauge and Partially On-Shell Fields}

We use coordinates $(u,r,x^A)$, where $u$ is the retarded time, $r$ a radial coordinate, and $x^A$ a pair of angular coordinates. The partial Bondi gauge (PBG) is defined by setting 
\begin{equation}
    g_{rr}=g_{rA}=0 ~,
\end{equation}
which fixes three gauge conditions. These imply that the retarded time $u$ labels the null geodesics and that the angular coordinates are constant along these null rays. This partial gauge fixing is a convenient starting point since it encompasses both Bondi-Sachs (BS) gauge \cite{Bondi:1962px,Sachs:1962zza}, which can be imposed by using the remaining gauge freedom to set $\sqrt{\gamma}= r^2\sqrt{q_0}$ where $q_0$ is the determinant of the 2-sphere, and the Newman-Unti (NU) gauge \cite{Newman:1962cia}, by setting $g_{ur} =-1$. 
These two gauges are widely used to describe null infinity and have also been used to study Al(A)dS spacetimes \cite{Poole:2018koa, Compere:2019bua, Fiorucci:2020xto, Compere:2020lrt,Compere:2023ktn}. 
The line element in PBG has the form 
\begin{gather}\label{eq:PartialBondiMetric}
	\dd s^{2} = e^{2\,\beta}\,\VV\,\dd u^{2} - 2\,e^{2\beta}\,\dd u\,\dd r + \gamma_{AB}\,\Big(\dd x^{A} - \UU^{A} \dd u\Big)\Big(\dd x^{B} - \UU^{B} \dd u\Big) ~,
\end{gather}
where $\VV$, $\UU^{A}$, and $\beta$ are arbitrary function of the coordinates $x^{\mu} = (u,r,x^A)$. 

Rather than placing the fields fully on-shell, we enforce only those components of the Einstein equations conjugate to the non-zero terms in \eqref{eq:PartialBondiMetric}. Specifically, if $G^{\mu\nu}$ is the Einstein tensor then the equations of motion involve 
\begin{gather}\label{eq:EinsteinEquations}
    \mathcal{G}^{\mu\nu} = - \frac{1}{2\kappa^{2}}\,\Big( G^{\mu\nu} + \Lambda\,g^{\mu\nu} \Big) ~,
\end{gather}
where $\kappa^{2}=8\pi G$ will usually be set to 1. Accounting for field variations that are fixed by PBG, we enforce the minimum set of on-shell conditions that ensure $\mathcal{G}^{\mu\nu}\,\delta g_{\mu\nu} = 0$. This is just
\begin{gather}\label{eq:PartiallyOnShellConditions}
    \mathcal{G}^{uu} = \mathcal{G}^{ur} = \mathcal{G}^{uA} = \mathcal{G}^{AB} = 0 ~.
\end{gather}
Since $\delta g_{rr} = \delta g_{rA} = 0$ in PBG, the fields need not satisfy the $\mathcal{G}^{rr}$ and $\mathcal{G}^{rA}$ components of the equations of motion. As we will show later on, those components encode the evolution of the mass aspect and angular momentum of the spacetime.

The partially on-shell conditions \eqref{eq:PartiallyOnShellConditions} fix the four dimensional curvature scalar $R = 4\,\Lambda$ (but not the full Ricci tensor) and constrain the dependence of the fields on $r$. At large $r$ we take
\begin{gather}
    \label{eq:gammaAB}
    \gamma_{AB} = r^{2}\,\left(\gamma^{0}_{AB} + \frac{1}{r}\,\gamma^{1}_{AB} + \frac{1}{r^2}\,\gamma^{2}_{AB} + o(r^{-2}) \right) ~,
\end{gather}
which is compatible with conformal compactification.\footnote{The notation $o(r^{-2}) $ indicates terms that fall off faster than $r^{-2}$. The expansion becomes polyhomogeneous in $1/r$ and $\ln r$ after the $1/r^2$ term. For $\Lambda\neq0$, the equations of motion eliminate any traceless terms of the form $\AA_{\langle AB \rangle} \log r^m/r^n$. However, trace terms of the form $\gamma^{0}_{AB} \tilde{\gamma}^{{(m,n)}} \log r^m / r^{n}$, with $n \geq 3$ and $m\leq n-2$, are allowed.} Then the $r$-dependence of leading and subleading terms in the large-$r$ expansion of the fields $\VV$, $\UU^{A}$, and $\beta$ is determined by \eqref{eq:PartiallyOnShellConditions} -- the explicit form of these fields is given in appendix \ref{app:FieldsAtLarger}. Throughout the rest of the paper, indices on two dimensional quantities appearing at a specific order in the large-$r$ expansion, such as $\gamma^{n}_{AB}$ or $\UU_{n}^{A}$, are raised and lowered with $\gamma^{0}_{AB}$. This includes traces, which we denote by 
\begin{gather}
    \gamma_{n} = \gamma_{0}^{AB}\,\gamma^{n}_{AB}  ~.
\end{gather}
The trace-free part of a two dimensional tensor is indicated with a hat, as
\begin{gather}
    \hat{\gamma}^{n}_{AB} = \gamma^{n}_{AB} - \frac{1}{2}\,\gamma^{0}_{AB}\,\gamma_{n} ~.
\end{gather}
An important example is the (normal) shear $\hat{\gamma}^{1}_{AB}$, which is the traceless part of the first subleading term in \eqref{eq:gammaAB}. Brackets $\langle\,\,\rangle$ on a pair of indices refers to the symmetric, trace-free part, defined with an overall factor of $1/2$. 

Unlike the BS and NU gauges, the partial gauge fixing used here does not constrain the traces $\gamma_n$. Recall that in those gauges they satisfy
\begin{gather}
    \text{Bondi-Sachs:} \quad \gamma_1=0\,,\, \gamma_2=\frac12\gamma_1^{AB}\gamma^1_{AB}\,,\\
    \text{Newman-Unti:} \quad \gamma_1=0\,,\, \gamma_2=\frac14\hat\gamma_1^{AB}\hat\gamma^1_{AB} ~.
\end{gather}
In section \ref{sec:charges} we show that, similar to the flat case \cite{Geiller:2024amx}, the traces $\gamma_1$ and $\gamma_2$ may be associated to large gauge symmetry generators and for some boundary conditions cannot be gauged away in presence of a boundary with corners.

\subsection{Asymptotic geometry}\label{sec:asympgeom}
In most of this work we are interested in an asymptotic boundary that includes a timelike component $B$, located at $r \to \infty$, with a corner $\partial B$. The corner is taken to be the intersection of $B$ with a hypersurface defined on a neighborhood of $B$ by the condition $u=$ constant.

The coordinates on $B$ are $x^i=(u,x^A)$. At large $r$ the induced metric on $B$ behaves as $h_{ij} = r^{2}\,h^{0}_{ij} + \ldots$ with leading part 
\begin{equation}\label{eq:meth0}
h_{ij}^0 \, \dd x^{i} \dd x^{j} = \frac{\Lambda}3\,e^{4\beta_0} \, \dd u^2+\gamma^0_{AB}\Big(\dd x^A- \UU_0^A \dd u\Big)\Big(\dd x^B-\UU_0^B \dd u\Big) ~.
\end{equation}
We usually refer to the collection of two dimensional fields $\gamma^{0}_{AB}$, $\UU_{0}^{\,A}$, and $\beta_0$ as ``boundary data,'' though this term should be reserved for quantities on $B$ that are fixed for a particular choice of boundary conditions. This is simply for convenience, and at this point does not imply any particular choice of boundary conditions.

The asymptotic boundary $B$ is foliated by surfaces $\CC$ of constant $u$. Such surfaces are defined by a timelike and future-pointing unit vector $v^{i}$ with components 
\begin{equation}\label{def:vasymp}
    v_i=- e^{2\beta_0} \sqrt{-\frac{\Lambda}{3}} \, \delta_{i}{}^{u}\,, \quad v^i=e^{-2\beta_0} \sqrt{-\frac{3}{\Lambda}} \,\Big( \delta^{i}{}_u+\delta^{i}{}_A \,\UU^A_0\Big)    
\end{equation}
Three dimensional tensors are projected normal or tangent to a surface $\CC$ by contracting with $v^{i}$ or the projector $\perp^{i}{\!}_{j} = \delta^{i}{\!}_{j} + v^{i} \, v_{j}$, with components
\begin{gather}\label{def:perpasymp}
    \perp^{u}{\!}_{i} = 0 \qquad \perp^{A}{\!}_{B} = \delta^{A}{}_{B} \qquad \perp^{A}{\!}_{u} = - \,\UU_0^{A} ~.
\end{gather}
The induced metric on $\CC$ is $\gamma_{AB} = r^{2} \, \gamma^{0}_{AB} + \ldots\,$, and the covariant derivative compatible with $\gamma^{0}_{AB}$ is denoted $\DD_{A}$. 

Evolution equations and other results in this paper involve projections of three dimensional derivatives tangent or normal to the constant-$u$ surfaces $\CC$. Derivatives tangent to $\CC$ are written in terms of $\partial_{A}$ or $\DD_{A}$, while derivatives normal to $\CC$ act on scalars and spatial (orthogonal to $v^{i}$) covariant tensors as 
\begin{gather}\label{def:Dv}
 \Dv \defeq e^{-2\beta_0}\big(\partial_{u} + \pounds_{\UU_{0}} \big) ~.    
\end{gather}
Here $\pounds_{\UU_0}$ is the Lie derivative along $\UU^{A}_0$ and $e^{2 \beta_0}$ is the lapse (up to a $\Lambda$ factor). 
Then the Lie derivative of a covariant spatial tensor $S_{i j \ldots}$ along an arbitrary  three-dimensional vector $X^{k}$ can be written
\begin{gather}
    \pounds_{X} S_{i j \ldots} = e^{2\beta_0} X^{u} \,\Dv  S_{i j \ldots} + \pounds_{\perp X}  S_{i j \ldots} ~.
\end{gather}
Many of our results include the normal derivative $\Dv$ acting on $\ln\sqrt{\gamma_0}$. This quantity, denoted $\theta$, is given by
\begin{gather}
    \theta = \Dv \ln \sqrt{\gamma_0} = e^{-2\beta_0} \Big( \, \partial_{u} \ln\sqrt{\gamma_0} + \DD_{A} \UU_{0}^{\,A} \, \Big) ~.
\end{gather}
Additional discussion of the geometry of $B$ and its foliation by constant-$u$ surfaces $\CC$ can be found in appendix \ref{app:Geometry}.

Using the definitions above, an important kinematic constraint on the fields (obtained from the equations of motion) can be written 
\begin{gather} \label{eq:Constraintshear} 
    \Dv \gamma^{0}_{AB} = \frac{\Lambda}{3} \, \hat{\gamma}^{1}_{AB} + \gamma^{0}_{AB} \, \theta ~. 
\end{gather}
This condition relates the shear along $B$ to the normal shear $\hat{\gamma}^{1}_{AB}$. It encodes a striking difference between asymptotically flat spacetimes and asymptotically locally (A)dS spacetimes: the (normal) shear is unconstrained for the former, but tied to the boundary geometry for the latter. For $\Lambda \neq 0$ this is related to the fact that gravitational radiation reaching the boundary would result in a boundary metric with non-trivial dependence on $u$.

\section{Presymplectic potential}
\label{sec:BeforeAndAfterConstraints}

In this section we discuss the symplectic potential in PBG with arbitrary boundary data. The starting point is the finite part of the shifted presymplectic potential described in equation (4.31) of \cite{McNees:2024iyu}. For completeness, that result is generalized here to include terms proportional to $\UU_{0}^{A}$. It is valid for arbitrary $\Lambda$, including the flat limit $\Lambda \to 0$. The fields are assumed to satisfy the partially on-shell conditions described in the previous section, except for the kinematic constraint \eqref{eq:Constraintshear} which is specific to $\Lambda \neq 0$.

Focusing on the  $\Lambda \neq 0$ case, we then impose \eqref{eq:Constraintshear} and express the result in a form suited to the variational problem with Dirichlet boundary conditions. A careful treatment of total derivatives allows the result to be written as a boundary part related to the Brown-York stress tensor, and a corner part containing new contributions apparent in the relaxed gauge. The finite $\delta$-exact part, which is necessary for the analysis of the variational principle in section \ref{sec:FiniteAction}, is also given.

\subsection{Before imposing the constraint}\label{sec:beforeconstrain}
The starting point is the shifted presymplectic potential $\widetilde{\Theta}$ obtained in \cite{McNees:2024iyu}, which is valid with or without a cosmological constant. It is finite as $r \to \infty$ for $\Lambda = 0$, while for $\Lambda \neq 0$ the only divergences in this limit are $\delta$-exact terms. In either case, the finite part of $\widetilde{\Theta}$ on $B$ takes the form
\begin{align}\label{eq:PresymplecticPotentialBefore}
    \widetilde{\Theta}^{r}_{0} = & \,\, \sqrt{\gamma_0}\,e^{2 \beta_0} \left[ \,\frac{1}{4}\,N^{AB} \, \delta \Big( e^{-2\beta_0}\,\hat{\gamma}^{1}_{AB} \Big) + \frac{1}{2}\,\hat{\mathscr{T}}^{AB} \, \delta\gamma^{0}_{AB} + \mathscr{T}_{A} \, \delta \UU_{0}^{A} + \frac{1}{2} \, \mathscr{T} \, \delta \Big(\ln\sqrt{\gamma_0} - 4\,\beta_0 \Big)\right] \\ \nonumber
     & \,\, + \partial_{u} \left[ \, \frac{1}{2} \, \sqrt{\gamma_0} \, \hat{t}^{AB} \, \delta \gamma^{0}_{AB} + \frac{1}{2} \, \sqrt{\gamma_0} \, t \, \delta  \Big(\ln\sqrt{\gamma_0} - 4\,\beta_0 \Big)  \right] + \delta \Big( \mathscr{A}_{B} + \partial_u \mathscr{A}_{\partial B} \Big) ~.
\end{align}
The metric $\gamma^{0}_{AB}$, vector $\UU_{0}^{A}$, and scalar $\beta_0$ may be arbitrary (sufficiently smooth) functions of $u$ and $x^{A}$. When $\Lambda = 0$, the shear $\hat{\gamma}^{1}_{AB}$ is not related to those fields and contributes an independent variation. 

Coefficients of the field variations in \eqref{eq:PresymplecticPotentialBefore} are proportional to the news $N_{AB}$ and the components normal and tangent to $\CC$ of a three-dimensional stress tensor. The news is given by
\begin{align} \label{eq:NewsN}
    N_{AB} = & \,\,e^{2\,\beta_0}  \left( \Dv - \frac{1}{2}\,\theta \right) \hat{\gamma}^{1}_{AB}  - \frac{\Lambda}{6}\,e^{2\beta_0}\,\gamma^{0}_{AB}\,\hat{\gamma}_{1}^{CD} \, \hat{\gamma}^{1}_{CD} ~.
\end{align}
and we can check that it is tracefree. 
Imposing the partially on-shell conditions on subleading terms in the large-$r$ expansions of the fields,\,\footnote{Strictly speaking, we impose the $\Lambda \neq 0$ partially on-shell conditions. The $\Lambda \to 0$ limit does not fully capture the log branch of the asymptotically flat spacetimes with $\Lambda = 0$. For example, $\hat{\gamma}^{2}_{AB}$ is proportional to $\gamma_1 \, \hat{\gamma}^{1}_{AB}$ when $\Lambda \neq 0$, but contains an additional degree of freedom when $\Lambda = 0$. See \cite{Geiller:2022vto, McNees:2024iyu} for details.} the stress tensor components are
\begin{subequations}
\begin{align}
    \mathscr{T} = & \,\, 2\,\MM - \frac{1}{4} \, \Big( \Dv + \theta \Big) \Big( \hat{\gamma}_{1}^{AB} \hat{\gamma}^{1}_{AB} \Big) + \frac{1}{2} \, \DD_{A} \DD_{B} \hat{\gamma}_{1}^{AB} - \frac{1}{2} \, \hat{\gamma}_{1}^{AB} e^{-2\beta_0} \DD_{A}\DD_{B} e^{2\beta_0} \\    
    \mathscr{T}_{A} = & \,\, e^{-2\beta_0} \, \left( \PP_{A} - \frac{1}{8} \, \hat{\gamma}_{1}^{BC} \DD_{A} \hat{\gamma}^{1}_{BC} - \frac{1}{4} \, \hat{\gamma}^{1}_{AB} \DD_{C} \hat{\gamma}_{1}^{BC} \right) \\
    \hat{\mathscr{T}}_{AB} = & \,\, - \frac{\Lambda}{6}\,\EE_{AB} + \frac{1}{2}\,\RR \, \hat{\gamma}^{1}_{AB} - \frac{1}{4} \, \DD_{C} \DD^{C} \hat{\gamma}^{1}_{AB} + \frac{\Lambda}{24} \, \hat{\gamma}^{1}_{AB} \hat{\gamma}_{1}^{CD} \hat{\gamma}^{1}_{CD} + \frac{3}{4} \, \hat{\gamma}^{1}_{AB} \, e^{-2\beta_0} \,\DD_{C} \DD^{C}  e^{2\beta_0} \\ \nonumber
    & \,\, + 4\,\DD^{C}\beta_0 \, \DD_{\langle A} \hat{\gamma}^{1}_{B \rangle C} - 3 \, \DD^{C}\beta_0 \,\DD_{C} \hat{\gamma}^{1}_{AB} ~.
\end{align}
\end{subequations}
Definitions of $\MM$, $\PP_{A}$, and other quantities appearing in these expressions (which are related to the Newman-Penrose scalars) are given in equations \eqref{NPscalars1} and \eqref{NPscalars2} of appendix \ref{app:AsymptoticExpansions}. The only partially on-shell condition that is not enforced at this point, which we discuss below, is the constraint \eqref{eq:Constraintshear} relating $\hat{\gamma}^{1}_{AB}$ to derivatives of the boundary data. 

In addition to the terms on $B$, there are corner terms in $\widetilde{\Theta}_{0}^{r}$. They consist of a traceless tensor $\hat{t}_{AB}$ coupled to the metric variation and a scalar $t$ coupled to the variation of the Weyl invariant quantity $\ln\big(e^{-4\beta_0}\sqrt{\gamma_0}\big)$,
\begin{subequations}
\begin{align}    
    \hat{t}_{AB} = & \,\, \frac{1}{8} \, \gamma_{1} \hat{\gamma}^{1}_{AB} \\
    t = & \,\, \frac{1}{4} \, \hat{\gamma}_{1}^{AB} \hat{\gamma}^{1}_{AB} - \frac{1}{2} \, \gamma_{2} ~.
\end{align}
\end{subequations}
There are also $\delta$-exact terms on $B$ and $\partial B$, given by 
\begin{subequations}
\begin{align}\label{eq:FirstAB}
    \mathscr{A}_{B} = & \,\, \sqrt{\gamma_0} \, e^{2\beta_0} \, \left[ \, \MM + \frac{\Lambda}{4} \, \gamma_{3} + \frac{\Lambda}{16} \, \gamma_{1} \left( \gamma_2 - \frac{3}{4} \, \hat{\gamma}_{1}^{AB} \hat{\gamma}^{1}_{AB} - \frac{1}{24} \, (\gamma_1)^{2} \right) \right] \\ \label{eq:FirstAdB}
    \mathscr{A}_{\partial B} = & \,\, \sqrt{\gamma_0} \, \left[ \,\frac{1}{2} \, \gamma_2 - \frac{3}{8} \, \hat{\gamma}_{1}^{AB} \hat{\gamma}^{1}_{AB}\right] ~.
\end{align}
\end{subequations}
While these $\delta$-exact terms do not contribute to the symplectic current, they are relevant for the variational formulation of the theory once boundary conditions are chosen.

\subsection{After imposing the constraint}
For $\Lambda \neq 0$, the constraint \eqref{eq:Constraintshear} means that $\delta \hat{\gamma}^{1}_{AB}$ must be expressed in terms of variations of the boundary data $\gamma^{0}_{AB}$, $\UU_{0}^{A}$, and $\beta_0$. Imposing this condition, the finite part \eqref{eq:PresymplecticPotentialBefore} is now
\begin{subequations}\label{eq:PresymplecticPotentialAfter}  
\begin{align}\label{eq:boundarysymplpot}
    \widetilde{\Theta}_{0}^{r} = &\,\,\sqrt{\gamma_0}\,e^{2\beta_0}  \left[ \, \frac{1}{2}\,\hat{\TT}^{AB}\,\delta\gamma^{0}_{AB} + \TT_{A}\,\delta\UU_{0}^{A} + \frac{1}{2}\,\TT  \,\delta\Big(\ln\sqrt{\gamma_0}-4\,\beta_0\Big)\right] \\ \label{eq:cornersymplpot}
    &\,\, + \partial_{u} \left[ \, \frac{1}{2}\,\sqrt{\gamma_0}\,\hat{\tau}^{AB}\,\delta\gamma^{0}_{AB} + \frac{1}{2}\,\sqrt{\gamma_0}\,\tau\,\delta\ln\sqrt{\gamma_0} + \sqrt{\gamma_0} \, \chi \, \delta\Big(\ln\sqrt{\gamma_0} - 4\,\beta_0\Big) \right] \\ 
    &\,\, + \delta \Big(A_{B} + \partial_{u}A_{\partial B} \Big) ~.
\end{align}
\end{subequations}
Here, $\widetilde{\Theta}_{0}^{r}$ has been organized into parts which are linear in the field variations but not their derivatives, and parts which are $\delta$-exact. Derivatives of field variations have been rewritten using integration-by-parts, discarding total derivatives tangent to $\CC$ and retaining derivatives normal to $\CC$ (corner terms) with support on $\partial B$.

Schematically, \eqref{eq:PresymplecticPotentialAfter} has the form
\begin{equation}\label{eq:ThetaSplit}
  \widetilde{\Theta}_{0}^{r} = \widetilde{\Theta}^r_B+\partial_u(\widetilde{\Theta}^{r}_{\partial B})+\delta \big(A_{B} + \partial_{u}A_{\partial B} \big) ~.
\end{equation}
The $\delta$-exact part is not unique; one may always move terms around via $f \, \delta g = \delta( f\,g ) - g \, \delta f.$ The split described here is made with Dirichlet boundary conditions in mind, and is relevant for the formulation of the variational problem in section \ref{sec:FiniteAction}. For other boundary conditions \cite{Compere:2008us,Odak:2021axr}, one would instead arrange \eqref{eq:ThetaSplit} so that the non $\delta$-exact part is linear in variations of appropriate boundary data. Let us now consider the terms in \eqref{eq:ThetaSplit} one at a time and describe the quantities appearing in \eqref{eq:PresymplecticPotentialAfter}.

\paragraph{Boundary term $\widetilde{\Theta}^r_B$ in the symplectic potential}
The coefficients of the field variations on $B$ are a symmetric tensor $\TT^{AB}$ with traceless part $\hat{\TT}^{AB}$ and trace $\TT = \TT^{C}{}_{C}$, and a vector $\TT_{A}$. They are given by
\begin{subequations}\label{eq:boundarysources}
\begin{align}
    \TT = &\,\, 2\,\MM \\ 
    \TT_{A} = &\,\, - \frac{3}{\Lambda}\,e^{-2\beta_0}\,\left(\mathcal{J}_{A} - \frac{\Lambda}{6}\,\PP_{A} \right) \\
    \widehat{\TT}^{AB} = &\,\, -\frac{3}{\Lambda}\,{\NN}^{AB} - \frac{\Lambda}{12}\,\EE^{AB} ~.
\end{align}
\end{subequations}
The scalar $\MM$, vectors $\JJ_{A}$ and $\PP_{A}$, and traceless tensors $\NN_{AB}$ and $\EE_{AB}$, are related to the Newman-Penrose scalars. Expressions for these quantities are given in appendix \ref{app:AsymptoticExpansions}. The combinations appearing in $\TT_{A}$ and $\TT^{AB}$ are precisely the components of the electric part of the four-dimensional Weyl tensor projected normal and tangent to $\CC$, as shown in appendix \ref{sec:WeylParts}. Note that the traces $\gamma_1$ and $\gamma_2$, which appear when relaxing Bondi gauge to PBG, are not independent variables in this part of the symplectic potential, in the sense that their contributions could be absorbed in a redefinition of the fields at third subleading order.

Together, $\TT_{A}$ and $\TT^{AB}$ describe the components of a traceless three-dimensional stress tensor $T^{ij}$. To see this, we first write 
\begin{gather}\label{eq:FullThetarOnB}
    \widetilde{\Theta}^{r}_{B} = \frac{1}{2}\,\sqrt{-h_{0}} \, T^{ij}\,\delta h^{0}_{ij} ~.
\end{gather}
Using the leading part \eqref{eq:meth0} of the metric on $B$, $T^{ij} \, \delta h^{0}_{ij}$ is given by 
\begin{align}
    T^{ij} \, \delta h^{0}_{ij} = & \,\,  - \frac{\Lambda}{3} \, e^{4\beta_0} \, T^{uu}  \, \delta \ln\Big(e^{-4\beta_0} \, \sqrt{\gamma_0} \,\Big) + T^{k}{}_{k} \, \delta\ln\sqrt{\gamma_0} + 2\,\Big(T^{uu} \, \gamma^{0}_{AB} \, \UU_{0}{\!}^{B} - T^{uB} \gamma^{0}_{AB} \Big) \, \delta \UU_{0}{\!}^{A} \\ \nonumber
    & + \Big( T^{uu} \, \UU_{0}{\!}^{\langle A} \, \UU_{0}{\!}^{B \rangle} - 2 \, T^{u\langle A} \, \UU_{0}{\!}^{B\rangle} + T^{\langle AB \rangle} \Big) \, \delta \gamma^{0}_{AB} ~,
\end{align}
where $T^{k}{}_{k}$ is the trace $h^{0}_{ij} \, T^{ij}$. Comparison with \eqref{eq:boundarysymplpot} shows that $T^{k}{}_{k}$ is zero, and the components of the traceless three dimensional stress tensor are related to $\TT_{A}$ and $\TT^{AB}$ by
\begin{subequations}
\begin{align}
    \TT & = \frac{1}{\ell}\,e^{2\beta_0}\big(-\VV_0\big) \, T^{uu} = \frac{1}{\ell^{3}}\,e^{4\beta_0} \, T^{uu} \\
    \TT_{A} & = \frac{1}{\ell}\,\Big( T^{uu} \, \gamma^{0}_{AB} \, \UU_{0}^{B} - \gamma^{0}_{AB} T^{uB} \Big) \\
    \TT^{AB} & = \frac{1}{\ell} \, \Big( T^{AB} - T^{uB} \, \UU_0^{A} - T^{Au} \, \UU_0^{B} + \UU_0^{A} \UU_0^{B} \, T^{uu} \Big) ~.
\end{align}
\end{subequations}
These are simply the components of $T^{ij}$ projected normal and tangent to $\CC$
\begin{gather}\label{3dstresstensor}
    T^{ij} = v^{i}v^{j} \,v_{k}v_{l} \, T^{kl} - v^{i} \perp^{j}{\!}_{l}\Big(T^{kl} v_{k} \Big) - v^{j} \perp^{i}{\!}_{k}\Big(T^{kl} v_{l} \Big) + \perp^{i}{\!}_{k} \perp^{j}{\!}_{l} \,T^{kl} ~,
\end{gather}
where $v^{k}$ is the normal vector \eqref{def:vasymp} and $\perp^{i}{\!}_{j}$ is the projector \eqref{def:perpasymp}. Additional details are given in appendix \ref{app:Geometry}. 

The partially on-shell conditions \eqref{eq:PartiallyOnShellConditions} ensure that the stress tensor $T^{ij}$ is traceless. But it is only conserved, ${}^{3}\nabla_{j} T^{ij} = 0$, once the remaining equations of motion are enforced and the theory is fully on-shell. This will be carefully accounted for when constructing charges in section \ref{sec:symmcharges}.

The expression \eqref{eq:FullThetarOnB} is equivalent to the usual result in Fefferman--Graham gauge \cite{Anninos:2010zf,Fiorucci:2020xto}. The change of coordinates between the Fefferman--Graham and Bondi gauges was constructed in \cite{Compere:2019bua,Poole:2018koa}, and it was shown that the Fefferman--Graham stress tensor can be expressed in terms of the Bondi variables as in \eqref{3dstresstensor}. Those results assumed additional restrictions on the boundary data that are not imposed here, but our results agree once those restrictions are applied.

\paragraph{Corner term in the symplectic potential $\widetilde{\Theta}^{r}_{\partial B}$}
The response functions on the corner $\partial B$ involve a two-dimensional tensor $\tau^{AB}$, with trace $\tau$ and trace-free part $\hat{\tau}^{AB}$, and a scalar $\chi$ conjugate to the variation of the Weyl invariant quantity $\ln(\sqrt{\gamma_0}) - 4\beta_0$.\,\footnote{Under a Weyl rescaling $h^{0}_{ij} \to e^{2\sigma}\,h^{0}_{ij}$ the two dimensional fields transform as $\gamma^{0}_{AB} \to e^{2\sigma} \gamma^{0}_{AB}$ and $\beta_0 \to \beta_0 + \frac{1}{2}\,\sigma$. } They are
\begin{subequations}\label{eqcornersources}
\begin{align}
    \hat{\tau}^{AB} = & \,\, \frac{3}{2\, \Lambda}\,\lambda^{AB} + \frac{1}{8}\,\gamma_{1}\,\hat{\gamma}_{1}^{AB} \\
    \tau = &\,\, \frac{3}{2\,\Lambda}\,\RR \\
    \chi = &\,\, - \frac{1}{4} \, \gamma_2 + \frac{1}{8}\,\hat{\gamma}_{1}^{AB} \, \hat{\gamma}^{1}_{AB}
\end{align}
\end{subequations}
where
\begin{equation}\label{eq:lamAB}
\lambda_{AB}=\left(\Dv - \frac{1}{2}\,\theta \right)\hat{\gamma}^{1}_{AB}  - \frac{\Lambda}{6}\,\gamma^{0}_{AB}\,\hat{\gamma}_{1}^{CD} \, \hat{\gamma}^{1}_{CD} + 2\,e^{-2\beta_0}\,\DD_{\langle A}\DD_{B\rangle} e^{2\beta_0} ~.
\end{equation}
While terms in $\widetilde{\Theta}^{r}_{B}$ are invariant under Weyl rescalings of $h^{0}_{ij}$, this is not the case for $\tilde{\Theta}_{\partial B}^{r}$. The non-zero trace $\tau$ couples to $\delta\sqrt{\gamma_0}$ and has the form expected for a conformal anomaly in two dimensions. The scalar $\chi$ is, up to an overall factor, the second subleading term in the large-$r$ expansion of the two dimensional volume element $\sqrt{\gamma}$. 

The functions $\gamma_1$ and $\gamma_2$ associated with relaxing BS gauge to PBG give independent contributions here, since the corner symplectic pairs do not vanish under a redefinition of the fields that would absorb their contributions in \eqref{eq:boundarysources}. Hence, they will play a role in obtaining new charges due to the presence of the corner.

\paragraph{$\delta$-exact terms}
The $\delta$-exact parts of $\widetilde{\Theta}_{0}^{r}$ are also shifted from the previous expressions \eqref{eq:FirstAB} and \eqref{eq:FirstAdB}. They are
\begin{align}\label{eq:ABexact}
    A_{B} = \sqrt{\gamma_0}\,e^{2\beta_0} & \left( \, \MM + \frac{\Lambda}{4}\,\gamma_{3} + \frac{\Lambda}{16}\,\gamma_{1} \left( \gamma_{2}-\frac{1}{24}\,\gamma_{1}^{\,2} - \frac{3}{4}\,\hat{\gamma}_{1}^{AB} \, \hat{\gamma}^{1}_{AB} \right)  \right. \\ \nonumber
    & \,\, \left. \dub + \, \DD_{A}\beta_0\,\DD_{B}\hat{\gamma}_{1}^{AB} 
    + \frac{3}{4\Lambda}\,\Dv \RR \,  \right) \\     \label{eq:AdBexact}
    A_{\partial B} =  \sqrt{\gamma_0} & \left(\frac{1}{2}\,\gamma_2 - \frac{3}{8}\, \hat{\gamma}_{1}^{AB} \, \hat{\gamma}^{1}_{AB} - \frac{3}{4\,\Lambda}\,\RR \right) ~.
\end{align}
There are two new terms in $A_{B}$ compared to \eqref{eq:FirstAB}. One of these is the normal derivative $\Dv$ of the two-dimensional Ricci tensor on a constant-$u$ surface $\CC$, while the other involves derivatives tangent to $\CC$ of $\beta_0$ and the shear. There is also a new term in $A_{\partial B}$ that is proportional to $\sqrt{\gamma_0} \,\RR$ and hence topological on the two-dimensional corner.

In section \ref{sec:FiniteAction} we show that \eqref{eq:ABexact} and \eqref{eq:AdBexact} are canceled in the action by contributions from appropriate surface terms, after revising an assumption made in \cite{McNees:2024iyu} about the finite part of the corner contribution to $\widetilde{\Theta}$. But the symmetry analysis and charges, which we discuss in the next session, are insensitive to these $\delta$-exact terms.

\section{Symmetries and Charges}
\label{sec:symmcharges}

In this section we discuss residual symmetries preserving PBG and the transformation laws of various quantities. General expressions for the charges and flux are obtained from the codimension-2 form, accounting for novel corner contributions in addition to the usual stress-tensor charges. The charges are then analyzed for different sets of boundary conditions. The term ``boundary data'' is again used throughout this section to refer to leading terms in the large-$r$ expansion of the metric, but is not meant to imply a specific choice of boundary conditions. A few additional results, which may be helpful when verifying specific expressions, can be found in appendix \ref{app:Diffeos}.

\subsection{Residual symmetries and algebra}
The residual symmetry $\xi$ preserving $g_{rr}=g_{rA}=0$ and \eqref{eq:gammaAB} is 
\begin{subequations}\label{eq:AKV}
\begin{align} \label{xiu}
 \xi^u&=f(u,x^A)\,, \\ \label{xir}
\xi^r&=h(u,x^A)\,r+\xi_1^r(u,x^A)+\xi^r_2(u,x^A)\frac1r+o(r^{-1}) \\ \label{xiA}
 \xi^A&=Y^A(u,x^A)-\int^\infty_r \bns \dd r' \,e^{2\beta(r')} \gamma^{AB}(r') \, \partial_Bf 
\end{align}
\end{subequations}
The functions $(f,Y^A)$ generate a boundary diffeomorphism $\xi^i$
\begin{equation}
 \xi^i= f \, \partial_u + Y^A \, \partial_A ~.
\end{equation}
It will be convenient to use the projection of $\xi^{i}$ onto components normal and tangent to $\CC$, as in section \ref{sec:asympgeom}
\begin{gather}
    \xi^{i} = - v^{i} v_{j} \xi^{j} + \perp^{i}{\!}_{j} \xi^{j} = \Big( \, \delta^{i}{}_{u} + \delta^{i}{}_{A} \, \UU_{0}^{\,A} \,\Big)  f + \delta^{i}{}_{A}\Big( \, Y^{A} - \UU_{0}^{\,A} f \, \Big)  ~.
\end{gather}
The vector $h\,r \, \partial_r$ is referred to as a Penrose-Brown-Henneaux (PBH) diffeomorphism \cite{Brown:1986nw,Imbimbo:1999bj, deHaro:2000xn, Skenderis:2000in}.\,\footnote{Unlike Fefferman-Graham gauge, where a local rescaling of the AdS radial coordinate induces subleading (at large $r$) terms in the components of the diffeomorphism tangent to the conformal boundary, a PBH diffeomorphism in partial Bondi gauge is contained in the $\xi^{r}$ component.} Such transformations induce Weyl rescalings of the fields on $B$. The subleading terms $\xi^r_1$ and $\xi^r_2$ in \eqref{xir} are parameterized in terms of two functions $k_1$ and $k_2$ as
\begin{align}
\xi^r_1  & = k_1(u,x^A) +\frac12 \, e^{-2\beta_0} \DD_AX^A\\\label{def:k2}
\xi^r_2  & = k_2(u,x^A)-\frac{k_1\gamma_1}{4}-\frac12 \, e^{-2\beta_0}\left(\DD^B\hat\gamma^1_{AB} -\frac{1}{2} \, \DD_A\gamma_1\right)X^A ~,
\end{align}
where 
\begin{equation}
 X^{A} \defeq  \,\, e^{4\beta_{0}} \DD^{A} f ~.
\end{equation}
For each subleading order in $\xi^r$ a new arbitrary function $k_{n}$ appears. However the functions $k_{n}$ with $n>2$ will turn out to be pure gauge.\,\footnote{The expansion becomes polyhomogeneous in $1/r$ and $\ln r$ beginning at $\ln r / r^3$. The coefficients of these terms, $(\ln r)^m / r^n$ with $n>2$ and $1\leq m < n-2$, are also pure gauge.}  This is similar to the flat case studied in \cite{Geiller:2024amx}. The residual symmetries $\xi$ are therefore parametrized by the six functions $f,Y^A,h,k_1,k_2$. If we were to further restrict the gauge with the Bondi-Sachs gauge condition, then $h$, $k_1$, and $k_2$ would not be part of the asymptotic symmetry algebra -- see the discussion at the end of section \ref{sec:examplesofBC}.

One motivation for introducing $k_1,k_2$ is that the residual symmetry algebra is particularly simple in terms of these functions. The modified Lie bracket of two residual symmetries is
\begin{align}
\big[\xi_1,\xi_2\big]_*
&:=\big[\xi_1,\xi_2\big]-\big(\delta_{\xi_1}\xi_2-\delta_{\xi_2}\xi_1\big)=\xi(f_{12},Y_{12},h_{12},(k_1)_{12},(k_2)_{12}),
\end{align}
where $\xi_a=\xi(f_a,Y_a,h_a,(k_1)_a,(k_2)_a)$, $a=1,2$ and 
\begin{equation}
\begin{split}
f_{12}&=f_1\partial_uf_2+Y^A_1\partial_Af_2-\delta_{\xi_1}f_2-(1\leftrightarrow2),\\
Y_{12}^A&=f_1\partial_uY^A_2+Y^B_1\partial_BY^A_2-\delta_{\xi_1}Y^A_2-(1\leftrightarrow2)\\
h_{12}&=f_1\partial_uh_2+Y^A_1\partial_Ah_2-\delta_{\xi_1}f_2-(1\leftrightarrow2),\\
(k_1)_{12}&=f_1\partial_u(k_1)_2+Y^A_1\partial_A(k_1)_2-h_1(k_1)_2-\delta_{\xi_1}(k_1)_2-(1\leftrightarrow2),\\
(k_2)_{12}&=f_1\partial_u(k_2)_2+Y^A_1\partial_A(k_2)_2-2h_1(k_2)_2-\delta_{\xi_1}(k_2)_2-(1\leftrightarrow2)
\end{split}
\end{equation}
When $\delta f=\delta Y^A=\delta h=\delta k_1=\delta k_2=0$, the algebra is \begin{equation}
(\text{Diff}(B)_{\xi^i} \loplus \mathbb R_{h} )  \loplus(\mathbb R_{k_1}\oplus\mathbb R_{k_2})
\end{equation}
It is a semi-direct sum of the diffeomorphisms of $B$ generated by $\xi^i$ with the PHB transformations $h$, followed by a semi-direct sum with two abelian sectors generated by $k_1$ and $k_2$. All generators are arbitrary functions of the retarded time and angular coordinates. As we will show in section \ref{sec:charges}, all these functions may be associated with non-vanishing infinitesimal surface charges. Further comments on additional boundary condition constraints will be provided in section \ref{sec:examplesofBC}. In particular we will comment on Dirichlet, conformal and $\Lambda-$BMS boundary conditions \cite{Compere:2019bua,Compere:2020lrt}.

\subsection{Transformation laws}\label{sec:transfos}
The fields transform under a diffeomorphism as
\begin{gather}
    \delta_{\xi} F = \Big( \,\xi^{v} \, \Dv  + \pounds_{\perp \xi}  + w h \,\Big) \,F + \tilde{\delta}_{\xi} F ~,   
\end{gather}
where $\xi^{v} = e^{2\beta_0} f$ and $\perp\!\xi^A = Y^{A} - \UU_{0}^{\,A} f $ are the components of $\xi^{i}$ normal and tangent to $\CC$, $\Dv$ is the derivative normal to $\CC$ defined in \eqref{def:Dv}, the constant $w$ is the conformal weight of the field, and $\tilde{\delta}_{\xi} F$ represents any additional terms. 

For the boundary data $\gamma^{0}_{AB}$, $\UU_{0}^{\,A}$, and $\beta_0$ the transformations are
\begin{align}\label{eq:GammaTransformation}
    \delta_{\xi} \gamma^{0}_{AB} = &\,\,  \Big( \,e^{2\beta_0} f \,\Dv+ \pounds_{\perp \xi} + 2\,h\,\Big) \,\gamma^{0}_{AB} \\     
    \label{eq:UTransformation}
    \delta_{\xi} \UU_{0}^{\,A} = & \,\,  \Big( \,e^{2\beta_0} f\,\Dv  + \pounds_{\perp \xi}\,\Big) \, \UU_{0}^{\,A} - \partial_{u} Y^{A} + \UU_{0}^{\,A} \partial_{u}f - \frac{\Lambda}{3}\,X^{A} \\    
    \label{eq:BetaTransformation}
    \delta_{\xi} \beta_0 = & \,\,  \Big( \,e^{2\beta_0} f\,\Dv+ \pounds_{\perp\xi}\,\Big)\,\beta_{0} + \frac{1}{2}\,h  + \frac{1}{2} \, e^{2\beta_0} \Dv f ~.
\end{align}
The metric has conformal weight $2$ and $\UU_{0}^{A}$ has conformal weight 0. The field $\beta_0$, on the other hand, shifts by $h/2$, so that $e^{n \beta_0}$ has conformal weight $n/2$. From the transformations of the metric and $\beta_0$, the volume elements $\sqrt{\gamma_0}$ on $\CC$ and $\sqrt{-h_0} = e^{2\beta_0}\sqrt{\gamma_0}$ on $B$ transform as
\begin{align}
    \delta_{\xi} \sqrt{\gamma_0} = & \,\, \Big( e^{2\beta_0} f\,\Dv + \pounds_{\perp \xi} + 2\,h \Big) \sqrt{\gamma_0} \\   \label{eq:3DVolumeTransform} 
    \delta_{\xi} \sqrt{-h_0} = & \,\, \Big( e^{2\beta_0}f \,\Dv + e^{2\beta_0} \Dv f + \pounds_{\perp\xi} + 3\,h \Big) \sqrt{-h_0} \\ \nonumber
    = & \,\, \partial_{i}\Big(\sqrt{-h_0} \,\xi^{i} \Big) + 3\,h\sqrt{-h_0} ~.    
\end{align}
This last result is the expected behavior for the volume element on $B$ under a Weyl rescaling and a three dimensional diffeomorphism $x^i \to x^i + \xi^i$. Taken together, the transformations above reflect the fact that the leading part of the metric on $B$ \eqref{eq:meth0} transforms as
\begin{equation}\label{transfoh0}
\delta_\xi h_{ij}^{0}= \big( \pounds _{\xi^i} + 2 h \big) h_{ij}^{0} ~.
\end{equation}
It is also apparent that the combination $\sqrt{\gamma_0}\,e^{-4\beta_0}$ is Weyl-invariant, as mentioned in the previous section. It transforms as
\begin{align}
    \delta_{\xi}\Big(\sqrt{\gamma_0} \, e^{-4\,\beta_0} \Big) = \Big( e^{2\beta_0} f \, \Dv + \pounds_{\perp\xi} - 2\,e^{2\beta_0} \Dv f \Big) \sqrt{\gamma_0} \, e^{-4\,\beta_0}  ~.
\end{align}
The factors of $h$ cancel, so it does not respond to a PBH diffeomorphism. 

Rather than recording the transformations of all of the various fields, we record only those quantities that are relevant for the calculations of charges. 
The transformation of the shear is
\begin{align}
    \delta_{\xi} \hat{\gamma}^{1}_{AB} = & \,\, \Big(e^{2\beta_0} f \, \Dv + \pounds_{\perp\xi} + h \Big)\,\hat{\gamma}^{1}_{AB} -2 \, e^{-2\beta_0} \DD_{\langle A} X_{B\rangle} ~,
\end{align}
while the traces $\gamma_1$ and $\gamma_2$ transform as
\begin{subequations}
\begin{align}
\delta_\xi \gamma_1&= \Big( e^{2\beta_0} f \, \Dv  + \pounds_{\perp \xi}  - \,h \Big)\gamma_1+4\,k_1\\
 \delta_\xi \gamma_2&= \Big( e^{2\beta_0} f \, \Dv  + \pounds_{\perp \xi}  -2  \,h \Big)\gamma_2+4\,k_2-e^{-2\beta_0}\hat\gamma^{AB}_1\DD_AX_B ~.
\end{align}
\end{subequations}
It follows that the news tensors $N_{AB}$ \eqref{eq:NewsN} and $\lambda_{AB}$ \eqref{eq:lamAB}, constructed from the shear and its derivatives, transform as
\begin{subequations}
\begin{align}
    \delta_{\xi}N_{AB} = & \,\, \Big( e^{2\beta_0}f \, \Dv + \pounds_{\perp\xi} + h + e^{2\beta_0} \Dv f \Big) N_{AB} - 2 \,  e^{2\beta_0} \,\Dv\Big(e^{-2\beta_0} \DD_{\langle A} X_{B\rangle} \Big) + \theta \,\DD_{\langle A} X_{B\rangle} \\ \nonumber
    & \,\, - \frac{\Lambda}{3} \, \left(\pounds_{X}\hat{\gamma}^{1}_{AB} - \frac{1}{2}\,\hat{\gamma}^{1}_{AB}\pounds_{X}\ln\sqrt{\gamma_0} \right)  + \frac{2 \Lambda}{3}\,\gamma^{0}_{AB} \hat{\gamma}_{1}^{CD}\, \DD_{\langle C} X_{D\rangle} \\
    \delta_{\xi}\lambda_{AB} = &\,\, \Big( e^{2\beta_0} f \, \Dv + \pounds_{\perp\xi}\Big) \lambda_{AB} +2\,\DD_{\langle A} \DD_{B \rangle} h + e^{-2\beta_0} \theta \, \DD_{\langle A} X_{B\rangle} + 2\,e^{-2\beta_0}X_{\langle A} \DD_{ B\rangle}\theta  \\ \nonumber
    &\,\, + \frac{\Lambda}{6} \, e^{-2\beta_0} \hat{\gamma}^{1}_{AB} \, \DD_{C} X^{C} - \frac{2\Lambda}{3}\,e^{-2\beta_0} X_{\langle A} \DD^{C} \hat{\gamma}^{1}_{B \rangle C} ~.
\end{align}
\end{subequations}
Compared to those expressions, which are fairly complicated, the transformations of quantities related to the Newman-Penrose scalars are very simple:
\begingroup
\allowdisplaybreaks
\begin{subequations}\label{eq:NPDiffeoTransformations}
\begin{align}
    \delta_{\xi} \MM = & \,\, \Big( e^{2\beta_0} f\,\Dv + \pounds_{\perp \xi} - 3\,h \Big)\, \MM  + \Big(\JJ_{A} - \frac{\Lambda}{6}\,\PP_{A}\Big) \, e^{-2\beta_0} X^{A} \\ 
    \delta_{\xi} \widetilde{\MM} = & \,\, \Big( e^{2\beta_0} f \,\Dv + \pounds_{\perp \xi} - 3\,h \Big)\, \widetilde{\MM}  + \eps^{AB}\Big(\JJ_{A} + \frac{\Lambda}{6}\,\PP_{A}\Big) 
    \, e^{-2\beta_0} X_{B} \\ 
    \delta_{\xi} \PP_{A} =  & \,\, \Big( e^{2\beta_0}  f \,\Dv + \pounds_{\perp\xi} -2\,h \Big)\,\PP_{A} + \left( 3\,\MM \, \gamma^{0}_{AB} - 3\,\widetilde{\MM} \, \eps_{AB} - \frac{\Lambda}{6} \EE_{AB}\right) e^{-2\beta_0} X^{B} \\  
    \delta_{\xi} \JJ_{A} = & \,\, \Big( e^{2\beta_0}  f \, \Dv + \pounds_{\perp\xi} -2\,h\Big)\, \JJ_{A} + \left( -\frac{\Lambda}{2}\,\MM\,\gamma^{0}_{AB} -\frac{\Lambda}{2}\, \widetilde{\MM} \, \eps_{AB} + \NN_{AB} \right) e^{-2\beta_0} X^{B} \\  
    \delta_{\xi}\NN_{AB} = & \,\, \Big( e^{2\beta_0} f \,\Dv + \pounds_{\perp\xi} - h \Big)\, \NN_{AB}  - \frac{2\Lambda}{3} \,e^{-2\beta_0} \JJ_{\langle A} X_{B\rangle} \\ 
    \delta_{\xi}\EE_{AB} = & \,\, \Big( e^{2\beta_0}  f\,\Dv + \pounds_{\perp\xi} - h \Big)\, \EE_{AB} + 4\,e^{-2\beta_0} \PP_{\langle A} X_{B\rangle} ~.
\end{align}
\end{subequations}
\endgroup
An important difference with asymptotically flat spacetimes concerns the transformation laws of $\NN_{AB}$, $\JJ_A$, and $\widetilde{\MM}$. In asymptotically flat spacetimes, setting these three quantities to zero corresponds to Penrose's condition of non-radiation, which is preserved under residual symmetries. Here, preserving the analogous conditions $\NN_{AB}=0$, $\JJ_A=0$, and $\widetilde{\MM}=0$, without also restricting the residual symmetries,\,\footnote{Specifically, setting $X^{A} \sim \DD^{A}f = 0$, which eliminates the angle-dependent supertranslations and leaves only the zero mode of $f$ on $\CC$.} forces us to set $\MM$ and $\PP$ to zero and leads to a solution space that contains only vacuum solutions.

The results above make it straightforward to write out the transformations for the components of the stress tensor on $B$ \eqref{eq:boundarysources}
\begin{subequations}
\begin{align}\label{eq:TTtransform}
    \delta_{\xi}\TT = & \,\, \Big( e^{2\beta_0} f \, \Dv + \pounds_{\perp\xi} -3\,h \Big) \, \TT - \frac{2\Lambda}{3}\,\TT_{A}\,X^{A} \\ \label{eq:TTAtransform}
    \delta_{\xi}\TT_{A} = & \,\, \Big( e^{2\beta_0} f \, \Dv + \pounds_{\perp\xi} -3\,h \Big) \, \TT_{A} - \TT_{A} \, e^{2\beta_0} \Dv f + e^{-4\beta_0} \left( \widehat{\TT}_{AB} + \frac{3}{2}\,\gamma^{0}_{AB} \, \TT \right) X^{B} \\ \label{eq:TTABtransform}
    \delta_{\xi}\widehat{\TT}_{AB} = & \,\, \Big( e^{2\beta_0} f \, \Dv + \pounds_{\perp\xi} - h \Big) \, \widehat{\TT}_{AB} - \frac{2\Lambda}{3} \, \TT_{\langle A} X_{B \rangle}~.
\end{align}
\end{subequations}
Similarly, we could propose a non-radiation solution by setting $\widehat{\TT}_{AB}$ to zero. However this will force us to set $\TT_A$ and $\TT$ to zero, trivializing the solution space.

These combine non-trivially, via \eqref{3dstresstensor}, into the simple transformation law expected for the three-dimensional stress tensor $T^{ij}$ 
\begin{equation}\label{eq:transfTij}
\delta_\xi T^{ij}= \Big( \pounds _{\xi^i}-5 h \Big) \, T^{ij}  ~.
\end{equation}
Relaxing the gauge and on-shell conditions has not generated any corrections to the usual transformation law for $T^{ij}$: the components of the stress tensor \eqref{eq:boundarysources} do not transform under the subleading $k_1$ and $k_2$. The same is true for the boundary data $\gamma^0_{AB},\UU_0^A,\beta_0$. Therefore, the effect of working in PBG (as opposed to fully fixing the gauge with the BS or NU conditions) when computing the charges will manifest itself solely through corner contributions to the symplectic potential \eqref{eq:PresymplecticPotentialAfter}. This is discussed in subsection \ref{sec:charges}.

The corner terms \eqref{eqcornersources} in the symplectic potential transform as
\begin{subequations}
\begin{align}
\delta_\xi \hat\tau_{AB}= & \,\, \Big( e^{2\beta_0} f \, \Dv + \pounds_{\perp\xi}\Big) \hat\tau_{AB} +\frac{3}{\Lambda} \, \DD_{\langle A} \DD_{B \rangle} h 
    + \frac{3}{2\,\Lambda} e^{-2\beta_0}\Big( \theta \, \DD_{\langle A}X_{B\rangle}  + 2\,\DD_{\langle A}\theta\,X_{B\rangle}\Big) \\ \nonumber
    & \,\, + \frac{1}{2} \, k_1 \hat\gamma_{AB}^1 +\frac{1}{4} \, e^{-2\beta_0} \left( \hat{\gamma}^{1}_{AB} \, \DD_{C}X^C-4X_{\langle A} \DD^{C} \hat{\gamma}^{1}_{B\rangle C} -\gamma_1 \DD_{\langle A}X_{B\rangle} \right)
&\\
\delta_\xi {\tau}=&\,\, \Big( e^{2\beta_0} f \, \Dv + \pounds_{\perp\xi} -2\,h \Big) {\tau} - \frac{3}\Lambda\DD^2 h - \frac{3}{2\,\Lambda} \, e^{-2\beta_0} \Big( \theta \, \DD_A X^A + 2 \, \DD_A \theta\, X^A\Big) \\\nonumber
& + e^{-2\beta_0} \Big( \frac{1}{2} \, \hat{\gamma}_{1}^{AB} \DD_A X_B +  \DD_B \hat\gamma^{AB}_1 \, X_A \Big) \\
\delta_\xi\chi=&\,\, \Big( e^{2\beta_0} f \, \Dv + \pounds_{\perp\xi} -2\,h \Big) \chi - k_2
-\frac14 e^{-2\beta_0}\hat\gamma^{AB}_1\DD_AX_B
\end{align}
\end{subequations}
We are now fully equipped to perform the computation of the charges in PBG.

\subsection{Charges} \label{sec:charges}
The infinitesimal surface charge density $k_\xi^{r u}$ is obtained by contracting the symplectic current $\delta \widetilde \Theta^r(\delta)$ with a residual symmetry 
\begin{align}\label{eq:chargedef}
\delta \widetilde\Theta^r [\delta_\xi] =  \partial_i k^{ri}[\xi]
\end{align}
after imposing the partially on-shell conditions, see for instance \cite{McNees:2023tus} and references therein for more details. The infinitesimal surface charge is defined by integrating the form over a cut $\CC$ of $B$
\begin{equation}
\slashed \delta Q[\xi]=\int_{\CC} \nts \dd^{2} x \, k^{r u}[\xi] ~.
\end{equation}
The notation $\slashed \delta$ is a reminder that the infinitesimal charge $Q$ is in general not integrable in field space. This is to be expected for leaky boundary conditions.

The symplectic potential \eqref{eq:ThetaSplit} includes terms $\widetilde{\Theta}_{B}^{r}$ on $B$ and corner terms (total $u$-derivatives). We consider their contributions to $k$ separately, writing
\begin{equation}
    k^{ru}[\xi]=k_B^{ru}[\xi]+k_{\partial B}^{ru}[\xi] ~,
\end{equation}
where $\delta \widetilde\Theta^r_B [\delta_\xi] =  \partial_i k_B^{ri}[\xi]$ and $\delta \widetilde\Theta^{r}_{\partial B} [\delta_\xi]=k_{\partial B}^{ru}[\xi]$.

\subsubsection{Boundary contribution}
It is easier to compute the charges using three-dimensional quantities instead of their $2+1$ decomposition.
We start from \eqref{eq:FullThetarOnB} and use the transformations of the three dimensional stress tensor \eqref{eq:transfTij} and boundary metric $h_{ij}^0$ \eqref{transfoh0}. We then integrate by parts to extract $k^{ru}[\xi]$ and use the fact that the stress tensor is trace-free. The derivation follows similar steps as in Appendix B4 of \cite{Compere:2020lrt}, with two important differences. First, the three dimensional volume element is not fixed. And second, the three dimensional stress tensor need not be conserved. This construction, which involves a contribution from the weakly vanishing Noether current, is given in appendix \ref{app:Conserved}; additional details may be found in appendix A of \cite{McNees:2024iyu}. The result is
\begin{equation}\label{eq:boundarycharge}
k^{ri}_B [\xi] =\delta\left(\sqrt{-h_{0}}\,T^{i}{}_{j}\,\xi^j\right)- \sqrt{-h_{0}}\,T^{i}{}_{j}\,\delta \xi^j -\frac12 \, \sqrt{-h_{0}} \, \xi^i \, T^{jk}\delta h_{jk}^0 ~.
\end{equation}
Only the boundary diffeomorphism $\xi^i$ is present in this expression -- there is no contribution from $h$, $k_1$, or $k_2$. The absence of a Weyl charge here can be traced back to the fact that there is no conformal anomaly in odd (boundary) dimension. We recover the results for the charges obtained in \cite{Anninos:2010zf,Fiorucci:2020xto} upon enforcing the remaining on-shell conditions, and generalize the results of previous analyses \cite{Henneaux:1985tv,Ashtekar:1999jx,Aros:1999kt,Balasubramanian:2001nb,Papadimitriou:2005ii,Marolf:2012vvz} by allowing a non-conserved boundary stress tensor.

The infinitesimal charge is manifestly non-integrable as expected for leaky boundary conditions. We identify the first term of \eqref{eq:boundarycharge} as the integrable charge $ \delta q_{B}[\xi]$ and the rest as the non-integrable flux $ F_B[\xi]$
\begin{equation}\label{boundarycharges}
    q_B[\xi]=\int_{\CC} \nts d^{2}x\,\sqrt{-h_{0}} \, v_i\,T^{i}{}_j \xi^j\,, \quad F_B[\xi]= - \int_{\CC} \nts d^{2}x \sqrt{-h_{0}} \, v_i\left( T^{i}{}_{j} \delta \xi^{j} + \frac12 \, \xi^i \,  T^{jk}\delta h_{jk}^0 \right) ~.
\end{equation}
The integrable part $q_{B}[\xi]$ is the Brown-York charge constructed from the stress tensor on $B$. In terms of the $2+1$ split we have
\begin{gather}
   q_B[\xi]= \int_{\CC} \nts d^{2}x \sqrt{\gamma_0} \, \Big( \, \TT \, \xi^{v} + e^{2\beta_0} \, \TT_{A} \, (\perp\!\xi)^{A} \, \Big) ~,
\end{gather}
with $\xi^{v} = -v^{0}_{i} \xi^{i} = e^{2\beta_0} \, \xi^{u}$. Using the expressions for $\TT$ and $\TT_{A}$, and restoring factors of $\kappa^{2}=8\pi G$ that were previously set to unity, this is
\begin{gather}\label{eq:StressTensorCharges}
 q_B[\xi]= \int_{\CC} \nts d^{2}x \sqrt{\gamma_0} \, \left[ \, \frac{2\,\MM}{\kappa^{2}} \, \xi^{v} + \frac{1}{2\,\kappa^{2}}\,\left(\PP_{A} - \frac{6}{\Lambda}\,\JJ_{A} \right) \, (\perp\!\xi)^{A} \, \right] ~.
\end{gather}
In appendix \ref{app:Kerr} we apply this to the Kerr-AdS$_4$ spacetime in partial Bondi gauge, and recover the usual charges.

For field variations that either fix the boundary metric or preserve its conformal class, the compatible diffeomorphisms $\xi^{k}$ on $B$ are conformal Killing vectors. Then the flux term in \eqref{boundarycharges} vanishes, leaving just the integrable infinitesimal charge. This is the usual condition in AdS/CFT. However, unlike the usual AdS/CFT construction, the fields here are \textit{not} fully on-shell. Recall that the components of the Einstein equations conjugate to $\delta g_{rr} = \delta g_{rA} = 0$ have not been enforced. Those equations describe the evolution of the mass aspect and angular momentum and are equivalent to conservation of the three dimensional stress tensor on $B$, which is not required for \eqref{eq:boundarycharge}. Thus, the partially on-shell conditions allow field configurations for which the charges \eqref{eq:StressTensorCharges} are integrable but not conserved. The fields need not satisfy the evolution equations for the mass and angular momentum so those charges will in general depend on the choice of cut $\CC$. The connection between the divergence of the three dimensional stress tensor and the remaining components of the Einstein equations is discussed further in section \ref{sec:WardIdentities}.

\subsubsection{Corner contribution}
The contribution of the corner term to the charge is easier to extract as it is already a total derivative. We have 
\begin{equation}
    k_{\partial B}^{ru}[\xi]=\delta \widetilde\Theta_{\partial B}^r[\delta_\xi]
\end{equation}
where $\widetilde\Theta_{\partial B}^r[\delta_\xi]$ is given in \eqref{eq:cornersymplpot}. There is a choice to be made regarding the corner symplectic potential that depends on the boundary conditions we will choose on the corner. So far, we have only required finiteness of the symplectic current. In \cite{McNees:2024iyu}, we gave a prescription for the remaining finite ambiguity. In section \ref{sec:FiniteAction} we will see that this should be modified for Dirichlet boundary conditions in the presence of a corner, though only by a $\delta$-exact term that has no effect on the charges. Nevertheless, in order to track the impact on the charges of different choices for the finite corner term in $\widetilde{\Theta}^{r}$, let us briefly consider a parameterized version with arbitrary constants $a_i$ multiplying different terms which might appear. Here, we will include the various contributions appearing in \eqref{eq:cornersymplpot}, as well as two new terms related to the non $\delta$-exact part of our prescription: 
\begin{equation} \label{eq:barThetaDefintion}
\bar\Theta_{\partial B}^{r}[\delta]= \sqrt{\gamma_0}\left(   \frac{a_1}{2}\,\hat{\tau}^{AB}\,\delta\gamma^{0}_{AB} +\frac{a_2}{2}\,\tau\,\delta\ln\sqrt{\gamma_0} + a_3\, \chi \, \delta\Big(\ln\sqrt{\gamma_0} - 4\,\beta_0\Big)+a_4\chi\delta\beta_0+a_5  \delta\beta_2\right) ~.
\end{equation}
Setting $a_1=a_2=a_3=1$ and $a_4=a_5=0$ in this expression recovers $\widetilde\Theta^{r}_{\partial B}$. 
Recall that $\beta_2$ is determined algebraically by the first and second order subleading terms in the metric, as in \eqref{eq:Beta2Condition}. Then from the results in the previous subsection we have
\begin{gather}
    \delta_\xi\beta_2=\Big( e^{2\beta_0} f \, \Dv + \pounds_{\perp\xi} -2\,h \Big) {\beta_2}-\frac12 k_2+\frac18 k_1 \,\gamma_1  ~.
\end{gather} 
Taking a second, antisymmetrized variation of $\bar{\Theta}_{\partial B}$ and contracting with the $h$, $k_1$, and $k_2$ components of the diffeomorphism gives
\begin{subequations}
\begin{align}\label{totder}
\delta\bar\Theta_{\partial B}^{r}[\delta_h] = & \,\, a_2 \,\frac3{2\Lambda}\sqrt{\gamma_0}\left( \frac{h}{\sqrt{\gamma_0}} \delta\big(\sqrt{\gamma_0} \RR \big) -\DD^{\langle A}\DD^{B\rangle}h\,\delta \gamma_{AB}^0+\DD^2h \,\delta \ln \sqrt{\gamma_0} \right)   \\ \label{cornerchargeh}
&+\frac3{2\Lambda}(a_2-a_1)\sqrt{\gamma_0}\DD^{\langle A}\DD^{B\rangle}h\,\delta \gamma_{AB}^0+h\,\delta\left( \sqrt{\gamma_0}\left( \frac{a_4}2\chi-2a_5\beta_2\right)\right)\\
\label{cornerchargek1}
\delta\bar\Theta_{\partial B}^{r}[\delta_{k_1}] = & \,\,  \frac18\sqrt{\gamma_0}\,k_1\left(-2a_1\hat\gamma^{AB}_1\delta\gamma^0_{AB}+a_5\gamma_1\delta\ln\sqrt{\gamma_0}\right)\\
\label{cornerchargek2}
\delta\bar\Theta_{\partial B}^{r}[\delta_{k_2}] = & \,\, \sqrt{\gamma_0}\,k_2\left( \left(a_3-\frac{a_5}2\right)\delta \ln\sqrt{\gamma_0}+(a_4-4a_3)\delta\beta_0\right)
\end{align}
\end{subequations}
The first line \eqref{totder} vanishes identically when integrated over $\CC$, so there is no (Weyl) charge associated with $h$ for $\widetilde \Theta_{\partial B}^r[\delta_\xi]$.  The contribution associated with the diffeomorphism $\perp\!\xi$ tangent to $\CC$ is
\begin{equation}\label{cornerchargeY}
\delta\bar\Theta_{\partial B}^{r}[\delta_{\perp\xi}] =\delta \left[ \sqrt{\gamma_0} \left(
\DD_A\left(a_5\beta_2-\frac{a_2}2\tau-a_3\chi\right)
-a_1 \DD^{B }\hat{\tau}_{AB} + \left(a_4-4a_3\right)\chi \DD_A\beta_0  \right)  \right] \perp\!\xi^A 
\end{equation}
Contributions from $f$ can be separated into parts proportional to $f$, its normal derivative $\Dv f$, and its angular derivatives through $X^{A} = e^{4\beta_0}\,\DD^{A}f$.
\begingroup
\allowdisplaybreaks
\begin{subequations}
\begin{align}
\label{cornerchargef}
\delta\bar\Theta_{\partial B}^{r}[\delta_{f}] = & \,\, e^{2\beta_0} f \left[ \vphantom{\left(\frac{a}{b}\right)}\frac{a_1}{2}\, \delta\left( \sqrt{\gamma_0}\,\hat{\tau}^{AB}\right) \Dv \gamma^{0}_{AB} - \frac{a_1}{2}\, \Dv\left(\sqrt{\gamma_0} \,\hat{\tau}^{AB}\right) \delta \gamma^{0}_{AB} \right. \\ \nonumber
&+ \frac{a_2}{2}\,\delta\left( \sqrt{\gamma_0}\,\tau\right)   \Dv   \ln\sqrt{\gamma_0}- \frac{a_2}{2}\, \Dv\left(\sqrt{\gamma_0}\,\tau\right) \delta\ln\sqrt{\gamma_0} \\\nonumber
&+ a_3\,\delta\left( \sqrt{\gamma_0}\, \chi\right) \Dv\Big(\ln\sqrt{\gamma_0} - 4\,\beta_0\Big) - a_3\,\Dv\left(\sqrt{\gamma_0} \, \chi\right) \delta \Big(\ln\sqrt{\gamma_0} - 4\,\beta_0\Big)\\\nonumber
& \left. + a_4\,\delta\left( \sqrt{\gamma_0}\,\chi\right) \Dv\beta_0- a_4 \, \Dv\left(\sqrt{\gamma_0} \,\chi\right) \delta\beta_0 + a_5 \, \delta\left( \sqrt{\gamma_0}\right) \Dv\beta_2 - a_5\,\Dv\left(\sqrt{\gamma_0} \right) \delta \beta_2 \vphantom{\frac{a_1}{2}}\right]\\ 
\label{cornerchargeduf}
\delta\bar\Theta_{\partial B}^{r}[\delta_{\Dv f}]  = & \,\, \left(\frac{a_4}2-2a_3\right)  e^{2\beta_0} \,\Dv f \, \delta\big(\sqrt{\gamma_0}\,\chi \big) \\
\label{cornerfangularderv}
\delta\bar\Theta_{\partial B}^{r}[\delta_{X_A}] = & \,\, \frac{\sqrt{\gamma_0}}4e^{-2\beta_0}\left[ -\frac{3}{\Lambda}\,a_1 \Big( \theta \, \DD^{\langle A}X^{B\rangle}  + 2\,\DD^{\langle A}\theta\,X^{B\rangle}\Big) \delta \gamma_{AB}^0 \right. \\ \nonumber
&-\frac{a_1}{2}  \left( \hat{\gamma}_{1}^{AB} \, \DD_{C}X^C-4X^{\langle A} \DD_{C} \hat{\gamma}_{1}^{B\rangle C} -\gamma_1 \DD^{\langle A}X^{B\rangle} \right) \delta \gamma_{AB}^0 \\\nonumber
&+a_2\left(  \frac{3}{\Lambda} \,  \Big( \theta \, \DD_A X^A + 2 \, \DD_A \theta\, X^A\Big)  -  \hat{\gamma}_{1}^{AB} \DD_A X_B -2\DD_B \hat\gamma^{AB}_1 \, X_A  \right)\delta \ln \sqrt{\gamma_0} \\\nonumber
&+\left. \hat\gamma^{AB}_1\DD_AX_B \Big(a_3 \delta\left(\ln \sqrt{\gamma_0}-4\beta_0\right) +a_4\delta\beta_0\Big) \vphantom{\frac{3}{\Lambda}} \right]
\end{align}
\end{subequations}
\endgroup
Taking into account the boundary and corner contributions, we have shown that there can be non-vanishing infinitesimal charges associated with any of the six functions $\xi^i(f,Y^A),h,k_1,k_2$ parameterizing the residual symmetries. This is one of our main results.

Due to the complicated structure of the charges, we will not discuss the split between integrable and non-integrable parts. This would require a careful treatment of the notion of radiation \cite{Wald:1999wa}, and of the field dependence of residual symmetries via a choice of slicing. Instead, we will now consider the charges for specific choices of boundary conditions.

\subsubsection{Examples of boundary conditions}\label{sec:examplesofBC}

Here we work out the charges for three common choices of boundary conditions in AdS$_4$: Dirichlet boundary conditions that completely fix all boundary data, conformal boundary conditions that allow variations within the conformal class of the three dimensional metric $h^{0}_{ij}$, and a generalization of the $\Lambda$-BMS boundary conditions studied in \cite{Compere:2019bua,Compere:2020lrt}.

\paragraph{Case 1: Dirichlet boundary conditions} They are given by $\delta h_{ij}^0=0$, which translates to $\delta \gamma_{AB}^0=\delta\beta_0=\delta \UU^A_0=0$. We take $\UU_A^0=0=\beta_0$, and fix $\gamma_{AB}^0$ to be the round metric on the 2-sphere. This last condition implies that the shear $\hat{\gamma}^{1}_{AB}$ and other quantities involving $u$-derivatives vanish. Since $\UU_{0}^{A} = 0$, the projection of $\xi^{i}$ tangent to $\CC$ is just $Y^{A}$. Preserving the boundary conditions imposes the following conditions on the symmetry generators
\begin{equation}\label{DirichletBCAKV}
\partial_u Y^{A} = -\frac\Lambda3X^A\,, \quad h=-\partial_u f=-\frac12 D_AY^A     \,, \quad \DD_AY_B+\DD_BY_A-\gamma^{0}_{AB}\,D_{C} Y^{C}  = 0 ~.
\end{equation}
First we note that \eqref{cornerchargek1}, \eqref{cornerchargek2}, \eqref{cornerchargef} and \eqref{cornerfangularderv} are zero. The non-vanishing contributions, from \eqref{cornerchargeh}, \eqref{cornerchargeduf} and \eqref{cornerchargeY} combine into
\begin{equation}
    \oint \delta\bar\Theta_{\partial B}^{ru}[\delta_\xi] =\oint \delta\Big( -a_1 \, \sqrt{\gamma_0} \,\DD^B\hat\tau_{AB} \Big) Y^A 
\end{equation}
For Dirichlet boundary conditions, the equation of motion \eqref{eq:Constraintshear} sets the shear $\hat{\gamma}^1_{AB}$ to zero, and consequently $\hat{\tau}_{AB}$ vanishes. The presence of corner terms in the symplectic potential has no effect on the charges. There are no charges associated with $k_1$ and $k_2$, which therefore correspond to pure gauge transformations.\,\footnote{Using this gauge freedom to fix the traces $\gamma_n$ or the sub-leading $\beta_n$ allows one to reach either the BS or NU gauges, respectively.} One might try to source some of the derivative terms in \eqref{cornerchargeduf}-\eqref{cornerfangularderv} by fixing $\gamma^{0}_{AB}$ to be the round metric on the sphere times a $u$-dependent scale factor. This simple generalization, which generates non-zero $\theta$ but still sets $\hat{\gamma}^{1}_{AB} = 0$, changes some of the intermediate steps but not the final result. The corner terms do not affect the charges in that case, either. 

The total charges, accounting for both boundary and corner contributions, are given by the integrable expression \eqref{eq:StressTensorCharges} (assuming $\delta Y^A = \delta f = 0$). Working in the relaxed PBG has no effect on the observables in the Dirichlet case. In other words, the values of $\gamma_1$ and $\gamma_2$ can be chosen freely, without physical cost. The symmetry algebra is generated by $f,Y^A$ which gives $SO(3,2)$ for $\Lambda<0$ and $SO(1,4)$ for $\Lambda>0$.

\paragraph{Case 2: Conformal boundary conditions} They are given by $\delta h_{ij}^0=\frac23 \delta\ln \sqrt{-h_0}\,h_{ij}^0$ which translates into $\delta \gamma_{AB}^0= \delta\ln\sqrt{\gamma_0}\,\gamma_{AB}^0$,  $\delta \UU^A_0=0$, $\delta\beta_0=\frac14 \delta \ln\sqrt{\gamma_0}$. We also take $\UU^A_0=0$. We assume for simplicity that $\gamma_{AB}^0=\sqrt{\gamma_0}\,\bar\gamma_{AB}^0$ where $\bar\gamma_{AB}^0$ is unimodular and $\delta \bar\gamma_{AB}^0=0=\partial_u\bar\gamma_{AB}^0$. This implies
\begin{equation}\label{confBCAKV}
\partial_uY^A=-\frac\Lambda3X^A\,,  \quad \DD_AY_B+\DD_BY_A=\DD_CY^C\,\gamma_{AB}^0
\end{equation}
and 
\begin{equation}
  \label{confBCrelpf}  \partial_uf=-2Y^A\partial_A\beta_0+\frac12\DD_AY^A +\frac{1}{2} \, f \, \partial_u(\ln\sqrt{\gamma_0}-4\beta_0) ~.
\end{equation}
The equation of motion \eqref{eq:Constraintshear} and the conformal boundary conditions imply  $\hat\gamma^1_{AB}=0$ and $\hat\tau_{AB} =(3/\Lambda) \, e^{-2\beta_0} \DD_{\langle A} \DD_{B \rangle} e^{2\beta_0}$. Implementing these, we consider the contribution to the charges of $Y$ and $\partial_uf$,  using \eqref{confBCrelpf}, 
it is 
\begin{align}
\oint \delta\bar\Theta_{\partial B}^{r}[\delta_{Y,\partial_uf}] = & \oint 
 \left( \vphantom{\bigg|} \delta \left[\sqrt{\gamma_0} \, \DD_A \left(a_5\beta_2-\frac{a_2}2\tau - \frac{a_4}4\chi\right)
   - a_{1} \, \sqrt{\gamma_0}\, \DD^B\hat{\tau}_{AB} \right]  Y^A  \right. \\ \nonumber
&  \qquad \left. +f\partial_u \big(\ln\sqrt{\gamma_0}-4\beta_0 \big) \, \delta\left[\left(\frac{a_4}4-a_3 \right) \sqrt{\gamma_0} \chi \right] \vphantom{\bigg|}\right)
\end{align}
Using the conformal Killing equation, the contribution proportional to $a_1$ vanishes when evaluated on field variations preserving the boundary conditions. Adding the remaining contributions to the charges gives 
\begin{align}
\oint \delta\bar\Theta_{\partial B}^{r}[\delta_{\xi}] &=\oint
\delta\left[\sqrt{\gamma_0} 
\, \DD_A \left(a_5\beta_2-\frac{a_2}2\tau-\frac{a_4}4\chi\right)
 \right]Y^A\\\nonumber
&+\delta\left[\sqrt{\gamma_0}\left( \frac{a_4}2\chi-2a_5\beta_2\right)\right]h+\Big[\left(\frac{a_4}4-\frac{a_5}2\right)k_2+\frac{a_5}8\gamma_1\,k_1 \Big]\delta\sqrt{\gamma_0}\\\nonumber
&+f\,\delta\left[\frac{a_2}{2}\,\tau+\frac{a_4}4\chi-a_5\beta_2\right]   \partial_u \sqrt{\gamma_0}- f\,\partial_u\left[\frac{a_2}{2}\,\tau+\frac{a_4}4\chi-a_5\beta_2\right]\delta\sqrt{\gamma_0}\\\nonumber
&+\frac{3a_2}{4\Lambda}e^{-2\beta_0}\Big[ \theta \, \DD_A X^A + 2 \, \DD_A \theta\, X^A \Big]\delta \sqrt{\gamma_0}
\end{align}
Since $\tau = 3\,\RR / (2\Lambda)$, a short calculation shows that the terms proportional to $a_2$ in the last two lines cancel. This leaves
\begin{align}\label{eq:ConformalCornerCharges}
\oint \delta\bar\Theta_{\partial B}^{r}[\delta_{\xi}] &=\oint
\delta\left[\sqrt{\gamma_0} \left(
\DD_A\left(a_5\beta_2-\frac{a_2}2\tau-\frac{a_4}4\chi\right)
 \right)  \right]Y^A\\\nonumber
&+\delta\left[\sqrt{\gamma_0}\left( \frac{a_4}4\chi-a_5\beta_2\right)\right] \big( 2h + f\, \partial_u \ln\sqrt{\gamma_0} \big) \\ \nonumber
&+\Big[\left(\frac{a_4}4-\frac{a_5}2\right)k_2+\frac{a_5}8\gamma_1\,k_1 - f\,\frac{1}{\sqrt{\gamma_0}}\partial_u\left(\sqrt{\gamma_0}\left(\frac{a_4}4\chi-a_5\beta_2\right)\right)\Big]\delta\sqrt{\gamma_0}
\end{align}
Note that for our prescription, with $a_4 = a_5 = 0$ and $a_2 = 1$, most of these terms vanish and conformal boundary conditions result in a correction to the charge associated with $Y^A$ due to the anomaly of the corner. For a different choice of corner term, with $a_4$ and/or $a_5$  non-zero, there are two possibilities.
\begin{itemize}[itemsep=0.75em,topsep=0.75em]
    \item If $(a_5 - a_4/2 \neq 0, a_5 \neq 0)$ or $(a_4 \neq 0, a_5 = 0 )$, then $k_1$ is pure gauge. In the first case this is because the function $k_1$ in \eqref{eq:ConformalCornerCharges} can be absorbed into a redefinition (shift) of $k_2$. In the second case $k_1$ does not appear. Then a change of slicing that splits $h$ and $k_2$ into field-independent and field-dependent parts (the latter of which absorb accompanying terms in \eqref{eq:ConformalCornerCharges}) leads to two new charges. One of these is associated with the PBH diffeomorphism $h$, and the other with the subleading function $k_2$. In addition, the boundary charge associated with $Y^A$ receives a correction.  
    \item If $a_5- a_4/2=0$, with $a_4$ and $a_5$ both non-zero, then $k_2$ is pure gauge. However, the combination $\chi-2\,\beta_2$ that now appears in the charges does not transform under $k_2$, so this gauge freedom can not be used to set it to zero. Now a change of slicing involving $h$ and $k_1$ gives two new charges: a Weyl charge associated with $h$ and a charge associated with $k_1$. There is also a corner correction to the $Y^{A}$ charge. 
\end{itemize}
In both cases, the result is a Weyl charge associated with $h$ and a charge related to the area of $\CC$ associated with $k_1$ or $k_2$. The choice of slicing that renders the charges integrable in field space redefines those parameters, but does not affect the $f, Y^{A}$ terms in \eqref{eq:boundarycharge}.

The above result shows that one can define a symplectic form for conformal boundary conditions that allows a Weyl charge at the asymptotic boundary. This is a different construction than the one used in \cite{Anninos:2024xhc}, where a Weyl charge appears for a boundary at finite $r$ but vanishes in the $r \to \infty$ limit. We leave a detailed comparison of the two approaches for future work.

\paragraph{Case 3: $\Lambda-$BMS boundary conditions}
The $\Lambda$--BMS boundary conditions are an example of boundary conditions compatible with a notion of radiation for a non-vanishing cosmological constant \cite{Compere:2019bua,Compere:2020lrt,Compere:2023ktn,Compere:2024ekl}. They are
\begin{equation}\label{LambdaBMSBC}
 \sqrt{\gamma}=r^2 \sqrt{q_0} \,,\quad \beta_0=0\,,\quad \UU_0^A=0 ~,
\end{equation}
where $q_0$ is the determinant of the round metric on the 2-sphere. This implies $\delta \sqrt{\gamma_0}=0$, as well as $\gamma_1 = \chi=0$ for the subleading terms in $\sqrt{\gamma}$. In particular, these boundary conditions do not restrict the symmetric trace-free part of $\delta\gamma^{0}_{AB}$, compatible with non-trivial contributions from $\hat\gamma_{AB}^1$. They are preserved by diffeomorphisms satisfying
\begin{equation}\label{eq:AKVLambdaBMS}
 \partial_uY^A= - \frac{\Lambda}{3}\,X^{A}  \,, \quad h=-\partial_uf=-\frac12\DD_AY^A\,,  \quad k_1=0\,, \quad k_2=-\frac14 e^{-2\beta_0} \hat\gamma^{AB}_1 \DD_A X_B ~.
\end{equation}
The natural generalization to PBG is to require only $\sqrt{\gamma_0} = \sqrt{q_0}$, with $\delta\sqrt{\gamma_0} = 0$, rather than the full condition $\sqrt{\gamma} = r^{2} \,\sqrt{q_0}$. The conditions for the residual symmetries are then \eqref{eq:AKVLambdaBMS} without the conditions fixing $k_1$ or $k_2$. However, when considering the resulting corner charges, we find
\begin{align}
\oint \delta\bar\Theta_{\partial B}^{r}[\delta_{\xi}] =\oint &
\delta\left[\sqrt{\gamma_0}\left(-\frac{a_2}2\,\DD_A\tau -a_1 \DD^B\hat\tau_{AB}  \right)\right]Y^A\\\nonumber
&+\sqrt{\gamma_0}\left(-\frac{a_1}4\,k_1\hat\gamma_{1}^{AB} + \frac3{2\Lambda}(a_2-a_1)\DD^{\langle A}\DD^{B\rangle}h\right)\delta \gamma_{AB}^0\\\nonumber
&+ \frac{a_1}{2}\, \Big( f  \delta\left( \sqrt{\gamma_0}\hat{\tau}^{AB}\right) \partial_u \gamma^{0}_{AB} - f\, \partial_u\left(\sqrt{\gamma_0} \hat{\tau}^{AB}\right) \delta \gamma^{0}_{AB} \Big) \\\nonumber
&- \frac{a_1}{8} \,\sqrt{\gamma_0} \left( \hat{\gamma}_{1}^{AB} \, \DD_{C}X^C-4X^{\langle A} \DD_{C} \hat{\gamma}_{1}^{B\rangle C} -\gamma_1 \DD^{\langle A}X^{B\rangle} \right) \delta \gamma_{AB}^0 ~,
\end{align}
which shows that there are no charges associated with $k_2$, making it pure gauge. However $k_1$ still appears. We leave the analysis of this term for future work.

\section{The Action for Dirichlet Boundary Conditions with Corner}
\label{sec:FiniteAction}

In this section we return to $\widetilde{\Theta}^{r}$ and consider its application to the variational formulation of the theory. The focus here is Dirichlet boundary conditions -- the split \eqref{eq:ThetaSplit} was made with this example in mind. 

The variational principle for gravity with $\Lambda < 0$ and Dirichlet boundary conditions is well understood, but is worth revisiting here for two reasons. First, both the gauge and on-shell conditions have been relaxed. This is necessary in lower dimensions for understanding how off-shell symmetries of certain quantum mechanical models are realized on the gravity side of a holographic duality \cite{Grumiller:2017qao}. We regard this as sufficient motivation for considering a similar set-up in four dimensions. Second, the corner charges described in the previous section are associated with total derivatives in the symplectic potential that might be discarded in an integral over the full conformal boundary of the spacetime. But whether such total derivative terms can be discarded requires an analysis of how they behave as $u \to \infty$. To address this, we consider a ``wedge'' of spacetime with an explicit corner $\partial B$ at the intersection of $B$ with a constant-$u$ (on a neighborhood of $B$) hypersurface in $M$. This corresponds to a hard cut-off at some value of $u$ on $B$, similar to the regulating cutoff on $r$ that defines $B \subset M$. The presence of a corner $\partial B$ is a first step towards studying the $u$-dependence of the fields on $B$, and asking whether novel features like corner charges persist in the limit $u \to \infty$ that recovers the full spacetime. We find that an explicit corner associated with a wedge of asymptotically AdS$_4$ spacetime requires new terms in the action. In order to avoid new boundary conditions that restrict sub-leading terms in the large-$r$ expansion of the metric, we are forced to adjust a $\delta$-exact term in our prescription for the finite corner term in the presymplectic potential.

Before considering the case of Dirichlet boundary conditions in detail, we note that other boundary conditions \cite{Compere:2008us, Odak:2021axr, Anninos:2024xhc} may be studied in the same manner. The construction in that case would begin by expressing $\widetilde{\Theta}^{r}$ as a part linear in variations of appropriate boundary data, plus a $\delta$-exact remainder. Then one would supplement the bulk action with surface terms so that the variation of the full action vanishes on-shell (or, as is the case here, partially on-shell) for all field variations compatible with the boundary conditions.

The calculations in this section can be generalized to $\Lambda > 0$ in a straightforward manner. However, the physical interpretation is very different in that case, so we restrict our attention to asymptotically AdS$_4$ spacetimes with $\Lambda < 0$.

\subsection{The Action}

For the action, we supplement the Einstein-Hilbert bulk term with the standard AdS surface terms on $B$ \cite{Henningson:1998gx, Emparan:1999pm, Balasubramanian:1999re, deHaro:2000xn, Papadimitriou:2005ii}, along with two new terms which must be added on the corner $\partial B$. 
\begin{align}\label{eq:FullAction}
    \Gamma = &\,\, \frac{1}{2}\,\int_{M}\nts \dd^{4}x\,\sqrt{-g}\,\Big(R - 2\Lambda\Big) + \int_{B}\nts \dd^{3}x\,\sqrt{-h}\,\left(K - \frac{2}{\ell} - \frac{\ell}{2}\,{}^{3}R(h)\right) \\ \nonumber
    & \,\, + \int_{\partial B}\bns \dd^{2}x\,\sqrt{\gamma}\,\left(-\frac{1}{2} + \frac{\ell}{2}\,\KK \right) ~.
\end{align}
Here $K$ is the trace of the extrinsic curvature of $B$, ${}^{3}R(h)$ is the Ricci scalar intrinsic to $B$, and $\KK$ is the trace of the extrinsic curvature of $\partial B$. The length scale $\ell$ is related to the (negative) cosmological constant by $\ell = \sqrt{-3/\Lambda}\,$, and the units are such that $\kappa^{2} = 8\pi G = 1$.

Varying the action and imposing the partially on-shell conditions, contributions to $\delta\Gamma$ with support on $B$ are
\begin{align}\label{eq:ActionVariation}
    \delta \Gamma \Big|_{B} = &\,\, \int_{B}\nts \dd^{3}x \,\sqrt{\gamma_0}\,e^{2\beta_0} \left[ \,\frac{1}{2}\,\widehat{\TT}^{AB}\,\delta \gamma^{0}_{AB} +  \TT_{A}\,\delta \UU_{0}^{\,A} + \frac{1}{2}\,\TT\,\delta \Big(\ln\sqrt{\gamma_0}-4\,\beta_0 \Big)\right] \\ \nonumber
    &\,\, + \int_{\partial B}\bns \dd^{2}x \left[\, \sqrt{\gamma_0}\left(\frac{1}{2}\,\hat{\tau}^{AB}\,\delta\gamma^{0}_{AB} + \frac{1}{2}\,\tau\,\delta \ln\sqrt{\gamma_0} + \chi\,\delta\Big(\ln\sqrt{\gamma_0} - 4 \, \beta_0 \Big) \right) 
    + \delta\Big\{- 2\, \sqrt{\gamma_0} \, \chi \Big\}\right] ~.
\end{align}
The $\delta$-exact terms in $\widetilde{\Theta}^{r}$ with support on $B$ have canceled against the variation of the usual AdS$_4$ counterterms. This includes both the divergent $\delta$-exact part of $\widetilde{\Theta}^{r}$ described in \cite{McNees:2024iyu}, as well as the finite term $\delta A_{B}$ in \eqref{eq:ABexact}. There are no divergent corner terms in the shifted presymplectic potential, but varying the terms on $B$ in \eqref{eq:FullAction} generates divergences with support on $\partial B$. These divergences are addressed by the new corner terms. The only $\delta$-exact piece that remains is a finite corner contribution proportional to $\chi$. This term (which happens to vanish for Bondi-Sachs gauge) is related to our prescription for the finite corner term in $\widetilde{\Theta}^{r}$.

For the moment, let us ignore the $\delta$-exact corner term and focus on the other parts of \eqref{eq:ActionVariation}. They are only sensitive to field variations that (in the Hamilton-Jacobi sense) change the boundary data. Field variations which fall off faster than the leading terms in the metric do not contribute, as expected for an action admitting a well-defined variational problem with Dirichlet boundary conditions. But in AdS$_4$ we also expect invariance under field variations that shift the boundary data within the conformal class of $h^{0}_{ij}$, since there is no conformal anomaly on the three dimensional conformal boundary. Indeed, the terms on $B$ in \eqref{eq:ActionVariation} vanish for a Weyl rescaling of $h^{0}_{ij}$, which changes the fields as
\begin{gather}
    \gamma^{0}_{AB} \to e^{2\sigma} \, \gamma^{0}_{AB} \qquad \beta_0 \to \beta_0 + \frac{1}{2}\,\sigma \qquad \UU_{0}^{A} \to \UU_{0}^{A} ~.
\end{gather}
This is not the case on the two-dimensional corner $\partial B$: the $\tau \, \delta\sqrt{\gamma_0}$ term is sensitive to the rescaling of $\gamma^{0}_{AB}$. Adding a finite local functional of the Dirichlet boundary data to the action cannot remove this trace term from $\delta \Gamma$.\,\footnote{This is well-known for a local functional of the metric on $\partial B$. One can, however, construct local functionals of the boundary data on $B$ with first variation that shifts or even cancels the $\tau\,\delta\sqrt{\gamma_0}$ term on $\partial B$. But this also generates terms on $B$ which are not invariant under Weyl rescalings.} Thus, the variational problem in the presence of a corner must be addressed in a manner similar to an odd-dimensional asymptotically AdS spacetime, as discussed in \cite{Papadimitriou:2005ii}.

Returning to the $\delta$-exact corner term in \eqref{eq:ActionVariation}, it is given by
\begin{gather}\label{eq:ExtraCornerVariation}
    \delta\Big(-2\,\sqrt{\gamma_0} \, \chi\Big) = \delta\left( \frac{1}{2}\,\sqrt{\gamma_0}\,\gamma_2 - \frac{1}{4} \, \sqrt{\gamma_0}\, \hat{\gamma}_{1}^{AB} \hat{\gamma}^{1}_{AB} \right) ~.
\end{gather}
This vanishes in Bondi-Sachs gauge, but not, for example, in Newman-Unti gauge. The variation of the part involving the shear does not change the character of the usual Dirichlet boundary value problem for AdS$_4$. The constraint \eqref{eq:Constraintshear} relates it to $\delta(\ln\sqrt{\gamma_0} - 4\,\beta_0)$, $\delta \UU_{0}^{A}$, and the traceless part of $\delta\gamma^{0}_{AB}$, as
\begin{align}
    \delta\Big(\sqrt{\gamma_0} \,\hat{\gamma}_{1}^{AB} \hat{\gamma}^{1}_{AB} \Big) = & \,\, \sqrt{\gamma_0} \, \left( \frac{6}{\Lambda} \,\hat{\gamma}_{1}^{AB} \big(\Dv - \theta\big) \big(\delta\gamma^{0}_{AB} - \gamma^{0}_{AB}\,\delta\ln\sqrt{\gamma_0} \, \big) + \frac{12}{\Lambda} \, \hat{\gamma}^{1}_{AB} \DD^{A}\delta\UU_{0}^{B} \right. \\ \nonumber
    & \qquad \quad \left. + \hat{\gamma}_{1}^{AB} \hat{\gamma}^{1}_{AB} \, \delta\big(\ln\sqrt{\gamma_0} - 4\,\beta_0 \big) \vphantom{\frac{6}{\Lambda} } \right) ~,
\end{align}
which vanishes under the same conditions as the terms on $B$ in $\delta\Gamma$. But the $\delta \gamma_2$ term in \eqref{eq:ExtraCornerVariation} would require an additional condition on a sub-leading term in the large-$r$ expansion of the fields, which is not consistent with Dirichlet boundary conditions. We are not aware of a surface term which could be added to the action to address this $\delta$-exact term, without also generating new contributions to $\delta\Gamma$ at orders $r$ and $r^2$. Instead, we note that \eqref{eq:ExtraCornerVariation} is directly related to our prescription for the finite part of the corner term $\vartheta^{ru}$ in the shifted presymplectic potential $\widetilde{\Theta}^{r} = \Theta^{r} + \partial_{u}\vartheta^{ru}$. The shift used in \cite{McNees:2024iyu} specified the finite part of $\vartheta^{ru}$ as 
\begin{gather}\label{eq:OldPrescription}
    \Big(\vartheta^{ru}\Big)_\text{\tiny FIN} = \Big(\delta \sqrt{\gamma} + \sqrt{\gamma}\,\delta\beta \Big)_\text{\tiny FIN} ~.
\end{gather}
Eliminating the $\delta$-exact part here removes the offending term in \eqref{eq:ActionVariation}. That is, changing the prescription for the finite corner term in the shifted presymplectic potential to 
\begin{gather}\label{eq:NewPrescription}
    \Big(\vartheta^{ru}\Big)_\text{\tiny FIN} = \Big(\sqrt{\gamma}\,\delta\beta \Big)_\text{\tiny FIN} ~,
\end{gather}
the variation of the action \eqref{eq:FullAction} with the partially on-shell conditions enforced is
\begin{align}\label{eq:NewActionVariation}
    \delta \Gamma = &\,\, \int_{B}\nts \dd^{3}x \,\sqrt{\gamma_0}\,e^{2\beta_0} \left[ \,\frac{1}{2}\,\widehat{\TT}^{AB}\,\delta \gamma^{0}_{AB} +  \TT_{A}\,\delta \UU_{0}^{\,A} + \frac{1}{2}\,\TT\,\delta \Big(\ln\sqrt{\gamma_0}-4\,\beta_0 \Big)\right] \\ \nonumber
    &\,\, + \int_{\partial B}\bns \dd^{2}x\, \sqrt{\gamma_0}\left[\frac{1}{2}\,\hat{\tau}^{AB}\,\delta\gamma^{0}_{AB} + \frac{1}{2}\,\tau\,\delta \ln\sqrt{\gamma_0} + \chi\,\delta\Big(\ln\sqrt{\gamma_0} - 4 \, \beta_0 \Big) \right]  ~.
\end{align}
This vanishes for all field variations that preserve the boundary data, without additional conditions on sub-leading terms in the fields. However, the presence of the two-dimensional corner still spoils invariance under Weyl rescalings of the metric $h^{0}_{ij}$ on $B$, similar to the case of the conformal anomaly for asymptotically AdS$_3$ spacetimes.

In the rest of this section we update our prescription for the finite part of the corner term in the shifted presymplectic potential, from \eqref{eq:OldPrescription} to \eqref{eq:NewPrescription}. With this choice, the $\delta$-exact corner term \eqref{eq:AdBexact} in $\widetilde{\Theta}^{r}$ becomes
\begin{gather}\label{eq:NewAdBexact}
        A_{\partial B} = \sqrt{\gamma_0} \left( - \frac{1}{8}\, \hat{\gamma}_{1}^{AB} \, \hat{\gamma}^{1}_{AB} - \frac{3}{4\,\Lambda}\,\RR \right) ~.
\end{gather}
The finite corner contribution \eqref{eq:NewPrescription} precisely cancels $\delta \gamma_2$ and $\delta \gamma_1$ terms coming from other parts of the action, without it we would be forced to impose additional boundary conditions on subleading terms in the fields. The variational problem for different boundary conditions might require a different choice for the finite part of $\vartheta^{ru}$.

\subsection{Ward identities and the Einstein Equations}
\label{sec:WardIdentities}

In this section we consider the action with the partially on-shell conditions applied, as a functional of Dirichlet boundary data on $B$. Its response $\delta_{\xi}\Gamma$ to a diffeomorphism, which follows from \eqref{eq:NewActionVariation} and the transformations of the fields given in section \ref{sec:transfos}, can be expressed as terms on $B$ and $\partial B$ which are in general non-zero. Since we consider the Dirichlet problem with arbitrary boundary data, the terms on $B$ reproduce the components of the Einstein equations that have not been enforced. When the fields are fully on-shell this leads to the usual Ward identities for the dual theory.

First consider the terms on $B$ in $\delta_{\xi}\Gamma$. Starting from \eqref{eq:FullThetarOnB} with $\delta_{\xi} h^{0}_{ij} = \pounds_{\xi} h^{0}_{ij} + 2\,h\,h^{0}_{ij}$, we have
\begin{gather}
    \delta_{\xi} \Gamma \, \Big|_{B} = \int_{B} \nts \dd^{3}x \, \sqrt{-h_0}\,\frac{1}{2}\,T^{ij}\,\Big( {}^{3}\nabla_{i}\xi_j + {}^{3}\nabla_{j}\xi_i + 2\,h\,h^{0}_{ij} \Big) ~. 
\end{gather}
The partially on-shell conditions \eqref{eq:PartiallyOnShellConditions} ensure that $T^{ij}$ is traceless, so there is no dependence on the Weyl rescaling $h$ associated with a PBH diffeomorphism. Likewise, the sub-leading terms $k_1$ and $k_2$ in $\xi^{r}$ do not contribute. Integration by parts gives
\begin{gather}\label{eq:GeneralGammaDiffeoB}
    \delta_{\xi} \Gamma \, \Big|_{B} = \int_{B} \nts \dd^{3}x \, \sqrt{-h_0} \, \Big( \xi^{i} \, W_{i} + \ldots \Big) ~,
\end{gather}
where $\ldots$ indicates total derivatives that will contribute on $\partial B$. The partially on-shell conditions do not force $T_{ij}$ to be conserved, so the coefficients $W_{i} = - {}^{3}\nabla^{j}T_{ij}$ are in general non-zero functions of the fields and their derivatives along $B$.

The coefficients $W_i$ are precisely the evolution equations for the mass and angular momentum, obtained from the components of the Einstein equations conjugate to the gauge-fixed metric components $g_{rr} = g_{rA} = 0$. Let us first project tangent and normal to a constant-$u$ surface $\CC$ as
\begin{gather}
    \xi^{i} \, W_{i} = \Big(\xi^{A} - \UU^{A} \, \xi^{u} \Big) \, W_{A} + \xi^{u} \, \Big( W_{u} + \UU^{A} \, W_{A} \Big)  = \big(\!\perp\!\xi\big)^{A}\,W_{A} + \xi^{v} W_{v} ~,
\end{gather}
with $\xi^{v} = - v_{i}\,\xi^{i} = e^{2\beta_0} f$ and $W_{v} = e^{-2\beta_0}(W_{u} + \UU_{0}^{\,A}W_{A})$. Using the expressions for the components of the $2+1$ decomposition of $T^{ij}$, the coefficient of a diffeomorphism tangent to $\CC$ is
\begin{align}\label{eq:MomentumEvolutionWard}
    W_{A} = &\,\, \frac{1}{2} \,\Big( \Dv + \theta \Big) \!\left(\PP_{A} - \frac{6}{\Lambda} \, \JJ_{A} \right) - e^{-6\beta_0} \DD_{A}\Big(e^{6\beta_0}\,\MM \Big) \\ \nonumber
    & \,\, + \frac{1}{2}\,e^{-2\beta_0} \DD^{B} \left( e^{2\beta_0} \left( \frac{6}{\Lambda} \, \NN_{AB} + \frac{\Lambda}{6} \, \EE_{AB} \right) \right) ~.
\end{align}
This is in fact an evolution equation for $\PP_{A}$, rather than $\PP_{A} - (6/\Lambda) \JJ_{A}$. Recall from \eqref{eq:JDefinition} that $\JJ_{A}$ is given by $-(\Lambda/6) \PP_{A}$ plus non-dynamical pieces involving the shear and the news. The derivative $\Dv$ acting on these other parts of $\JJ_{A}$ cancels against similar terms coming from $\DD^{B}(e^{2\beta_0}\NN_{AB})$. One can write this out explicitly, but it is not especially illuminating compared to the compact result \eqref{eq:MomentumEvolutionWard}. The coefficient of a diffeomorphism $\xi^{v}$ normal to $\CC$ is
\begin{align}\label{eq:MassEvolutionWard}
    W_{v} = & \,\, 2 \left(\Dv + \frac{3}{2}\,\theta \right) \!\MM + \frac{\Lambda}{6} \,e^{-4\beta_0} \DD^{A} \left( e^{4\beta_0}\left(\PP_{A} - \frac{6}{\Lambda} \, \JJ_{A} \right)\right) - \frac{\Lambda}{12} \, \hat{\gamma}_{1}^{AB} \left( \frac{6}{\Lambda} \, \NN_{AB} + \frac{\Lambda}{6} \, \EE_{AB} \right) ~.
\end{align}
The expressions for $W_{v}$ and $W_{A}$ encode the components of the Einstein equations which are not included in the partially on-shell conditions \eqref{eq:PartiallyOnShellConditions}. To check this, we expanded the $\mathcal{G}^{rr}$ and $\mathcal{G}^{rA}$ components of the Einstein equations to fourth subleading order in the $1/r$ expansion, verified that terms at all preceding orders vanish,\footnote{Including the polyhomogeneous terms $r^{-3}\,\ln r$, $r^{-4}\,(\ln r)^{2}$, and $r^{-4}\,\ln r$.} simplified the remaining part using tensor contraction identities specific to two dimensions, and confirmed that the resulting expressions were equal to \eqref{eq:MomentumEvolutionWard} and \eqref{eq:MassEvolutionWard}. Those calculations were carried out using the xAct suite of packages for the Wolfram Language \cite{Martin-Garcia-xAct, Brizuela:2008ra, Nutma:2013zea}, along with custom code for the 2+1+1 decomposition of terms in the large-$r$ expansions of the fields and Einstein equations. 

The general equations \eqref{eq:MomentumEvolutionWard} and \eqref{eq:MassEvolutionWard} agree with previous results in the literature that placed restrictions on the fields. Specifically, we can compare with the evolution equations given in \cite{Compere:2019bua}. In that reference, the authors consider field configurations with $\beta_0=0=\UU^A_0$ and $\partial_r(r^{-4}\,\det(\gamma_{AB}))=0$. The last condition restricts the traces according to $\gamma_1 = \chi = 0$. Imposing those conditions on the fields, the equations $W_v=0$ and $W_A=0$ match equations (2.52) and (2.38) of \cite{Compere:2019bua}, respectively, provided the following dictionary  $M^{(\Lambda)}=\MM $; $N_A^{(\Lambda)}=\frac12(\PP_A-\frac1{\Lambda}\JJ_A)$, $J^{AB}=-\frac6{\Lambda^2} (\NN^{AB}+\frac{\Lambda^2}{36}\EE^{AB})$.

Now let us turn to the corner terms in $\delta_{\xi}\Gamma$. The introduction of the corner $\partial B$ explicitly breaks three-dimensional diffeomorphism invariance through the condition $u=$ constant. This shows up in $\delta_{\xi}\Gamma$ as corner terms which are finite but non-zero, even in the $r \to \infty$ limit and with the fields fully on-shell. With the prescription \eqref{eq:NewPrescription} for the shifted presymplectic potential, the corner terms in $\delta_{\xi}\Gamma$ take the form
\begin{gather}\label{eq:DiffeoCornerTerms}
    \delta_{\xi} \Gamma \, \Big|_{\partial B} = \int_{\partial B} \bns \dd^{2}x \,\sqrt{\gamma_0} \, \left( \xi^{v} \, \WW_{v} + \big(\!\perp\!\xi\big)^{A} \, \WW_{A} -2 \,\chi \, \Dv \xi^{v} + h \, \frac{3}{2\Lambda} \, \RR \right)  ~. 
\end{gather}
In addition to terms linear in $\xi^{i}$, there is a term proportional to $D_v \xi^{v}$, the derivative of $\xi^{v}$ in the direction normal to $\partial B$, and the term associated with the conformal anomaly on $\partial B$. The coefficients $\WW_{v}$ and $\WW_{A}$ of diffeomorphisms normal and tangent to $\partial B$ receive contributions from corner terms in \eqref{eq:ActionVariation} and total derivatives in \eqref{eq:GeneralGammaDiffeoB}. Collecting these, we find
\begin{align}
\label{eq:NormalCornerTerm}
    \WW_{v} = & \,\,-2  \,\MM + \frac{1}{4}\,\lambda_{AB} \, \hat{\gamma}_{1}^{AB} + \frac{3}{4\Lambda}\,\theta \, \RR  + \frac{\Lambda}{48} \, \gamma_{1} \, \hat{\gamma}_{1}^{AB} \hat{\gamma}^{1}_{AB} + \theta \, \chi \\
\label{eq:TangentCornerTerm}
    \WW_{A}  = & \,\, - \PP_{A} + \frac{1}{8} \,\hat{\gamma}_{1}^{BC} \DD_{A} \hat{\gamma}^{1}_{BC} + \frac{1}{4} \, \hat{\gamma}^{1}_{AB} \DD_{C} \hat{\gamma}_{1}^{BC} -\frac{1}{8} \, \DD^{B}\Big( \gamma_1 \, \hat{\gamma}^{1}_{AB} \Big) \\ \nonumber
    & \,\, - \DD_{A} \chi - 4\,\chi \, \DD_{A}\beta_0 ~.
\end{align}
These depend on the dynamical quantities $\MM$ and $\PP_{A}$ and in general are non-zero. The expressions are especially simple for the Dirichlet example described in section \ref{sec:examplesofBC}, which fixes $\gamma^{0}_{AB}$ as the round metric $q_{AB}$ on the two sphere. In that case, accounting for the relationships $2 \,\partial_u f = \DD_{A} Y^{A}$ and $2\,h = - \DD_{A} Y^{A}$ in \eqref{DirichletBCAKV}, and discarding total angular derivatives, the integral \eqref{eq:DiffeoCornerTerms} reduces to
\begin{gather}\label{eq:DirichletExampleCornerTerms}
    \delta_{\xi} \Gamma \, \Big|_{\partial B} = - \int_{\partial B} \bns \dd^{2}x \, \sqrt{q}\,\Big( 2\,\MM \, f + \PP_{A} \, Y^{A} \Big) ~.
\end{gather}
Essentially, the cost of shifting and rotating the corner is set by the mass aspect and angular momentum. A similar expression (including contributions related to the corner anomaly) holds for conformal boundary conditions.

The terms in \eqref{eq:DiffeoCornerTerms} are sensitive to finite (as $r\to\infty$) surface terms on $B$ or $\partial B$ which may be added to the action, as well as our prescription \eqref{eq:NewPrescription} for the finite part of the presymplectic potential. However, they depend on dynamical quantities -- the mass and angular momentum -- and hence cannot be eliminated by adding some functional of boundary data to the action. In general, for suitable $u$-dependence of the fields and compatible diffeomorphisms, the terms in \eqref{eq:DiffeoCornerTerms} are expected to vanish in the limit $u \to \infty$ and recover the usual AdS$_4$ results. As mentioned earlier in this section, the point of considering an action with an explicit corner is setting up a framework for addressing questions about these fall-off conditions for the fields, and whether corner charges may persist as the corner is removed to infinity. This analysis is left for future work.


\section{Conclusion}
In this work, we have investigated locally (A)dS spacetimes in four dimensions with asymptotic boundaries that may be leaky and possess corners. We worked in partial Bondi gauge, which represents a relaxation of the familiar Bondi-Sachs or Newman-Unti gauges, and allowed a completely general boundary metric. Let us first summarize the main results before commenting on some open questions and directions for future work.

Charges associated with residual symmetries were obtained using a generalization of the covariant phase space formalism that allows for ``partially on-shell'' field configurations. In PBG such field configurations need not satisfy the on-shell evolution equations for the mass or angular momentum aspects of the spacetime. This builds on previous work \cite{McNees:2023tus,McNees:2024iyu} proposing a prescription for a manifestly finite symplectic current. The symplectic potential from which the current is obtained was split into boundary and corner contributions, see \eqref{eq:ThetaSplit}. The former leads to Brown-York charges that are constrained by the Einstein equations when the fields are fully on-shell. 
Charges associated with the corner terms are not constrained in this way, unless additional boundary conditions are imposed.  example of such charges is the Weyl charge.

The procedure described in \cite{McNees:2023tus, McNees:2024iyu}, which can be carried out before fixing a specific choice of boundary conditions, yields a manifestly finite symplectic current from a bulk Lagrangian. Equivalently, it yields a symplectic potential which is finite up to a $\delta$-exact part. Once boundary conditions are fixed, the full action for the variational principle can be constructed by including appropriate surface terms that cancel the $\delta$-exact part of the symplectic potential. For the case of Dirichlet conditions and a boundary with an explicit corner, we showed that the usual surface terms required for holographic renormalization of the action \cite{Henningson:1998gx, Emparan:1999pm, Balasubramanian:1999re, deHaro:2000xn, Papadimitriou:2005ii} must be supplemented with new corner terms, as in \eqref{eq:FullAction}. This analysis also led us to revise an earlier proposal for a finite corner ambiguity in the symplectic potential, from \eqref{eq:OldPrescription} to \eqref{eq:NewPrescription}.

Lastly, we computed the Ward identities. Since our partially on-shell conditions lead to a boundary stress tensor that is traceless but not conserved, these are expected to be proportional to the components of the bulk Einstein equations that have not been enforced. Indeed, the Ward identities are precisely the equations governing the evolution of the mass and angular momentum aspects for on-shell field configurations. Since the boundary metric in our setup is completely general, with no conditions that fix its determinant or constrain its dependence on $u$, this provides a novel derivation of the remaining equations of motion, expressed in a form that does not make any simplifying assumptions or otherwise restrict the form of the fields.

This work opens some interesting new avenues to pursue. So far, we have obtained charges in PBG allowing for an arbitrary boundary metric, including potential corners, subject to partial on-shell conditions that relax the evolution equations for the mass and angular momentum. It would be interesting to pin down boundary conditions compatible with the Wald-Zoupas prescription \cite{Wald:1999wa}, including the stationarity condition that the first variation of the action vanish in the absence of radiation. Doing this would require a definition of gravitational radiation for (A)dS spacetimes. However, several different proposals exist in the literature \cite{Ashtekar:2015lla,Szabados:2015wqa,Bishop:2015kay,Chrusciel:2016oux,Saw:2017amv,Mao:2019ahc,He:2017dzb,Saw:2017zks,PremaBalakrishnan:2019jvz,Dobkowski-Rylko:2022dva,Compere:2019bua,Compere:2020lrt, Bonga:2023eml,Ciambelli:2024kre,Fernandez-Alvarez:2021yog,Compere:2023ktn,Compere:2024ekl,Arenas-Henriquez:2025rpt}; we refer to the work \cite{RadGeiller} that discusses and compares them.

For Dirichlet and related boundary conditions, the components of the Brown-York stress tensor on the boundary are given by projections of the electric part of the bulk Weyl tensor. An obvious question is whether one can find boundary conditions such that activate the magnetic components of the Weyl tensor in the charges.

We have identified new charges due to the relaxation of the gauge obtained by going from BS or NU gauges to PBG. This is similar to the three-dimensional examples where relaxing the gauge yields additional charges \cite{Troessaert:2013fma,Grumiller:2016pqb,Geiller:2021vpg,Ciambelli:2023ott,Alessio:2020ioh,Cardenas:2025qqi,Perez:2016vqo,Ruzziconi:2020wrb}; see also \cite{Barnich:2011mi,Campiglia:2014yka,Compere:2018ylh,Henneaux:2018cst,Freidel:2021fxf,Fuentealba:2022xsz,Fiorucci:2024ndw,Geiller:2024amx,Geiller:2025dqe,CovLogSuper} for four-dimensional analyses. In this work, we have made some progress in understanding their origin, showing that they arise purely from a corner term in the symplectic potential. The number of charges depends on both the form of this corner term and the boundary conditions, and fewer charges may be apparent working in certain gauges. In \cite{Ruzziconi:2020wrb}, which focused on three dimensions, a discrepancy was found in the number of such charges accessible from the Fefferman-Graham gauge and Bondi gauge. The authors of \cite{Ciambelli:2024vhy} offered an explanation for this, showing that the diffeomorphism to go between the two gauges was field-dependent and yielded a corner term in the symplectic potential that led to a non-trivial charge. It would be interesting to determine if their proposal applies in the four dimensional case, as well. 

The relation between the corner charges we have found here -- particularly the Weyl charges -- and the results of \cite{Anninos:2024xhc} should be investigated. In that work, the authors uncover a new corner charge associated with Weyl rescaling. But the relevant modes vanish in the limit of an asymptotic boundary, so their charge only seems to exist for a finite boundary. We will leave this investigation for future work.

One advantage of working in PBG is that it is easier to translate exact solutions obtained in other gauges to PBG, than it would be to reach BS or NU gauge. As an example, we present Kerr-AdS$_4$ in PBG in Appendix~\ref{app:Kerr}, based on the ``generalized Bondi coordinates'' description of Kerr-dS$_4$ given in \cite{Hoque:2021nti}. However, this solution does not exhibit any interesting $u$-dependence in the boundary data. In a future publication we will apply our analysis to something like AdS$_4$ Robinson-Trautman \cite{Bakas:2014kfa,Ciambelli:2017wou,Adami:2024mtu}, which allows for boundary data with a non-trivial time evolution.

An important advantage of Bondi coordinates is that the flat limit $\Lambda \to 0$ is well-defined, which is not the case for Fefferman-Graham coordinates. While several results have been obtained via a $\Lambda \to 0$ limit, see for instance the BMS group recovered from $\Lambda-$BMS \cite{Compere:2019bua,Compere:2020lrt} or flux balance laws from the conservation of the AdS stress tensor \cite{Campoleoni:2023fug}, there are some aspects of flat space physics that are difficult to capture this way. For example, subleading terms in the large-$r$ expansion of the fields may include log-branch contributions for $\Lambda = 0$ which are not accessible via the flat limit of the $\Lambda \neq 0$ solution \cite{McNees:2024iyu}. More generally, implementing the flat limit of $\Lambda \neq 0$ quantities may involve assumptions about their dependence on $\Lambda$, and require scaling out factors of $\Lambda$ or $1/\Lambda$ to recover the expected results as $\Lambda \to 0$. The procedure we have followed here suggests a simpler approach which avoids some of these subtleties. The symplectic potential in section \ref{sec:beforeconstrain} is perfectly well defined for $\Lambda = 0$. It is free of $\Lambda \to 0$ divergences and compatible with the $\Lambda = 0$ equations of motion. It is the next step, rewriting the potential as a function of free boundary data after enforcing the constraint \eqref{eq:Constraintshear} between the normal shear and the intrinsic shear of the boundary, that introduces $\Lambda \to 0$ divergences. Rather than taking the $\Lambda \to 0$ limit after enforcing the constraint, the result \eqref{eq:PresymplecticPotentialBefore} -- which retains the normal shear as an independent quantity -- is a better starting point for an analysis of the flat limit. This is supported by the recent works \cite{Hartong:2025jpp,Fiorucci:2025twa}, which establish a stress tensor in asymptotically flat spacetimes based on the variation of an action with respect to the Carrollian geometry of the boundary. A key element in their constructions is that the action is varied with respect to both the tetrad and (part of) the connection. In Carrollian geometry, the normal shear appears as part of the connection but is independent of the boundary tetrad, and therefore contributes as an independent variable.\,\footnote{Treating the connection as an independent variable was already used in the construction of the symplectic potential for teleparallel gravity where the flat connection was used to encode the vacuum structure of asymptotically flat spacetimes \cite{Girelli:2024jhm}.} Likewise, the result \eqref{eq:PresymplecticPotentialBefore} is compatible with the independent r\^oles played by the boundary metric and shear in the flat limit.

Another motivation for considering corners comes from the holography of a boundary CFT (BCFT), where the field theory dual of the gravitational theory is defined on a manifold with boundary; see \cite{Takayanagi:2011zk} and references therein. This could include theories defined on a finite or semi-infinite portion of the timelike boundary of AdS. Since the asymptotic symmetry algebra plays a central role in formulating the holographic dictionary, our careful treatment of asymptotic boundaries with corners may be useful for investigating the bulk counterparts of such BCFTs.

\section*{Acknowledgments}
CZ thanks Marc Geiller for collaboration on related topics and for numerous discussions over the years. CZ thanks the participants of workshop Carrollian Physics and Holography hosted by the Erwin Schroedinger Institute in Vienna for discussions on this upcoming work, in particular Adrien Fiorucci and Romain Ruzziconi. CZ also thanks Dio Anninos, Max Ba\~nados, Luca Ciambelli and Sk Jahanur Hoque for discussions. The authors are grateful to Adrien Fiorucci for his comments on the draft. Research at Perimeter Institute is supported in part by the Government of Canada through the Department of Innovation, Science and Economic Development Canada and by the Province of Ontario through the Ministry of Colleges and Universities.

\appendix

\section{Geometry}  
\label{app:Geometry}

In partial Bondi gauge (PBG) the line element of a four-dimensional spacetime $(M,g)$ takes the form
\begin{gather}\label{eq:appPBG}
    \dd s^{2} = -2\,e^{2\beta}\,\dd u \,\dd r + e^{2\beta}\,\VV\,\dd u^{2} + \gamma_{AB}\,\Big(\dd x^{A} - \UU^{A}\,\dd u\Big)\,\Big(\dd x^{B} - \UU^{B}\,\dd u\Big) ~.
\end{gather}
Our primary interest is the behavior of the fields at large $r$, working in coordinates adapted to the foliation of a timelike constant-$r$ hypersurface $B$ by a family of spacelike surfaces of constant $u$. In this appendix we establish notation and collect several results related to the $4 \to 3 + 1 \to 2 + 1 + 1$ decompositions of the fields, their derivatives, and the equations of motion.

In the following, $x^{\mu} = (r,u,x^{A})$ are spacetime coordinates on $M$, $x^{i} = (u,x^{A})$ are coordinates on the constant-$r$ hypersurface $B \subset M$, and $x^{A}$ are the two remaining coordinates spanning a spacelike constant-$u$ surface $\CC \subset B$. Conventions for curvature tensors and other geometric quantities used here and throughout the text are the same as those given in \cite{McNees-Useful}.

\subsection{Embeddings}

A constant-$r$ hypersurface $B$ embedded in $M$ is described by the outward-pointing spacelike unit vector with components
\begin{gather}\label{eq:appSpacelikeNormal}
    n_{\mu} = \delta_{\mu}{}^{r} \frac{e^{\beta}}{\sqrt{-\VV}} \qquad n^{\mu} = \delta^{\mu}{}_{r} \, e^{-\beta}\,\sqrt{-\VV} - \delta^{\mu}{}_{i} \, \frac{1}{e^{\beta}\,\sqrt{-\VV}}\,\Big(\delta^{i}{}_{u} + \delta^{i}{}_{A}\,\UU^{A} \Big) ~.
\end{gather}
 The extrinsic curvature associated with this embedding is
\begin{gather}
    K_{\mu\nu} = \frac{1}{2}\,P_{\mu}{}^{\lambda} \, P_{\nu}{}^{\sigma} \Big(\nabla_{\lambda} n_{\sigma} + \nabla_{\sigma} n_{\lambda} \Big) ~,
\end{gather}
where $\nabla_{\mu}$ is the four-dimensional covariant derivative compatible with \eqref{eq:appPBG}, and $P^{\mu}{}_{\nu} = \delta^{\mu}{}_{\nu} - n^{\mu} \,n_{\nu}$ projects tangent to $B$. Restricting the PBG line element to $r = $ constant, the induced metric on $B$ is
\begin{gather}\label{eq:3DInducedMetric}
    h_{ij} \,\dd x^{i} \dd x^{j} = e^{2\beta}\,\VV\,\dd u^{2} + \gamma_{AB}\,\Big(\dd x^{A} - \UU^{A}\,\dd u\Big) \Big(\dd x^{B} - \UU^{B}\,\dd u\Big) ~.
\end{gather}
This is in the ADM form with lapse $e^{\beta}\,\sqrt{-\VV}$ and shift vector $-\,\UU^{A}$.  The covariant derivative compatible with $h_{ij}$ is ${}^{3}\nabla_{i}$ and the volume element is $\sqrt{-h} = \sqrt{\gamma}\,e^{\beta}\,\sqrt{-\VV}$. The intrinsic Ricci tensor constructed from $h$ and its derivatives along $B$ is ${}^{3}R_{ij}$, and the Ricci scalar, which appears in section \ref{sec:FiniteAction} as a surface term in the action, is denoted ${}^{3}R$. At large $r$ the leading behavior of this metric is
\begin{gather}\label{eq:Leading3DMetric}
    h_{ij} \,\dd x^i \dd x^j = r^{2}  \left( e^{2\beta_0}\,\VV_{0}\,\dd u^{2} + \gamma^{0}_{AB}\,\Big(\dd x^{A} - \UU_{0}^{A}\,\dd u\Big) \Big(\dd x^{B} - \UU_{0}^{B}\,\dd u\Big) + \ldots \right)
\end{gather}
where $\ldots$ indicates terms which are sub-leading in $1/r$.\footnote{The expansion is polyhomogeneous in $1/r$ and $\ln r$ beginning at order $\ln r / r^3$.} This leading part $h^{0}_{ij}$ is a representative of a conformal class of metrics $[h^{0}]$ associated with the conformal boundary at $r \to \infty$. The equations of motion imply $\VV_0 = e^{2\,\beta_0} \, \Lambda/3 = - e^{2\beta_0}/\ell^2$, so that the leading part of the volume element on $B$ is $\sqrt{-h_0} = \sqrt{\gamma_0}\,e^{2\beta_0}/\ell$. However, in sections \ref{sec:BeforeAndAfterConstraints}-\ref{sec:WardIdentities}, we usually take the volume element to be $\sqrt{\gamma_0}\,e^{2 \beta_0}$. The relative factor of $1/\ell$ is accounted for in various results, below.

The hypersurface $B$ is foliated by a family of spacelike surfaces of constant $u$, which we denote $\CC$. They are defined by a future-pointing, timelike unit vector with components
\begin{gather}\label{eq:appTimelikeNormal3D}
    v_{i} = -e^{\beta}\,\sqrt{-\VV}\,\delta_{i}{}^{u} \qquad v^{i} = \frac{1}{e^{\beta}\sqrt{-\VV}}\,\Big(\delta^{i}{}_{u} + \delta^{i}{}_{A}\,\UU^{A} \Big) ~.
\end{gather}
The induced metric on $\CC$ is $\gamma_{AB}$, with covariant derivative $D_{A}$ and curvature ${\!}^{2}R$. The extrinsic curvature $\KK_{AB}$ of the embedding in $B$ is defined with the same sign convention as above. This curvature and its trace are given by
\begin{gather}\label{eq:CurlyKExtrinsicCurvature}
    \KK_{AB} = \frac{1}{2\,e^{\beta}\sqrt{-\VV}}\,\Big(\partial_{u} +\pounds_{\UU} \Big)\gamma_{AB} \qquad
    \KK = \frac{1}{e^{\beta}\,\sqrt{-\VV}}\,\Big(\partial_{u} \ln\sqrt{\gamma} + D_{A}\UU^{A} \Big)  ~.
\end{gather}
At large $r$, the leading behavior of the metric on $\CC$ is 
\begin{gather}
    \gamma_{AB} = r^{2} \,\gamma^{0}_{AB}  +  \ldots ~.
\end{gather}
As in previous expressions, $\ldots$ indicates terms which are subleading in the large-$r$ expansion. The covariant derivative compatible with the metric $\gamma^{0}_{AB}$ is $\DD_{A}$, and its intrinsic curvature is $\RR$. We assume that $\CC$ is closed, so that total derivatives $\partial_{A}(\ldots)$ may be discarded in integrals, but make no further assumptions about the metric $\gamma^{0}_{AB}$.

\subsection{Projections}

Most results in the main text are expressed in terms of fields and their derivatives on $B$ which have been projected normal or tangent to $\CC$. Tensors are projected normal to $\CC$ by contracting with $v^{i}$, and tangent to $\CC$ by contracting with the projector $\perp^{i}{\!}_{j} = \delta^{i}{\!}_{j} + v^{i} \, v_{j}$. From \eqref{eq:appTimelikeNormal3D} the components of $\perp^{i}{\!}_{j}$ are
\begin{gather}
    \perp^{u}{\!}_{i} = 0 \qquad \perp^{A}{\!}_{B} = \delta^{A}{}_{B} \qquad \perp^{A}{\!}_{u} = - \,\UU^{A} ~.
\end{gather}
For example, a vector $X^{i}$ and tensor $X^{ij}$ on $B$ may be expressed
\begin{align}
    X^{i} = &\,\, -v^{i} \, v_{j} \, X^{j} + \perp^{i}{\!}_{j} X^{j} \\ \nonumber
    X^{ij} = &\,\, v^{i}v^{j} \,v_{k}v_{l} \, X^{kl} - v^{i} \perp^{j}{\!}_{l}\Big(v_{k} X^{kl} \Big) - v^{j} \perp^{i}{\!}_{k}\Big(X^{kl} v_{l} \Big) + \perp^{i}{\!}_{k} \perp^{j}{\!}_{l} \,X^{kl} ~.
\end{align}
An important example is the three-dimensional stress tensor $T^{ij}$ on $B$. 
\begin{gather}
    T^{ij} = v^{i}v^{j} \,v_{k}v_{l} \, T^{kl} - v^{i} \perp^{j}{\!}_{l}\Big(T^{kl} v_{k} \Big) - v^{j} \perp^{i}{\!}_{k}\Big(T^{kl} v_{l} \Big) + \perp^{i}{\!}_{k} \perp^{j}{\!}_{l} \,T^{kl} ~.
\end{gather}
Using the expressions for $v_{i}$ and $\perp^{i}{\!}_{j}$ the components are
\begin{align}
    v_{k} v_{l} \, T^{kl} = &\,\, e^{2\beta}(-\VV) \, T^{uu} \\
    \perp^{A}{\!}_{k}\Big(T^{kl} v_{l} \Big) = & \,\, e^{\beta}\,\sqrt{-\VV} \, \Big( T^{uu} \, \UU^{A} - T_{uA} \Big) \\
    \perp^{A}{\!}_{k} \perp^{B}{\!}_{l} \, T^{kl} = & \,\, T^{AB} - T^{uA} \, \UU^{B} - T^{Au} \, \UU^{B} + \UU^{A} \UU^{B} \, T^{uu} ~.
\end{align}
We are interested in extracting the leading behavior of these projections at large $r$. Recall that the Brown-York stress tensor is defined
\begin{gather}\label{app:LargerTij}
    \frac{2}{\sqrt{-h}} \, \frac{\delta \Gamma}{\delta h_{ij}} = \frac{1}{r^2}\,\left(T_{0}^{ij} + \frac{1}{r}\,T_{1}^{ij} + \ldots + \frac{1}{r^3}\,T_{3}^{ij} + \ldots \right)~.
\end{gather}
The generic leading behavior is $1/r^2$ followed by an expansion in powers of $1/r$, with additional $(\ln r)^{m}/r^{n}$ terms, $ m \leq n-2$, appearing for $n \geq 3$. Appropriate surface terms in the action ensure that the first several terms in this expansion vanish, so that the first non-zero term is $T_{3}^{ij}$. In the main text we refer to $T_{3}^{ij}$ simply as $T^{ij}$ to avoid additional notation, but we retain it here and in appendix \ref{app:ConventionsForExpansions} for clarity. After accounting for the $r$-dependence of all quantities, and imposing the traceless condition $h^{0}_{ij} \, T_{3}^{ij} = 0$, the projections above are related to the quantities $\TT$, $\TT_{A}$, and $\TT^{AB}$ in section \ref{sec:BeforeAndAfterConstraints} by
\begin{align} \label{eq:appTT}
    \TT = & \,\, \frac{1}{\ell}\,e^{2\beta_0}\,(-\VV_0)\,T_{3}^{uu} = \frac{1}{\ell^{3}}\,e^{4\beta_0} \, T_{3}^{uu} \\ \label{eq:appTTA}
    \TT_{A} = & \,\, \frac{1}{\ell}\,\Big(T_{3}^{uu}\,\gamma^{0}_{AB} \UU_{0}^{B} - \gamma^{0}_{AB}\,T_{3}^{uB} \Big) = - \frac{1}{\ell}\,T_{3}^{\,u}{}_{A}\\ \label{eq:appTTAB}
    \TT^{AB} = & \,\, \frac{1}{\ell}\,\Big(T_{3}^{AB} - T_{3}^{uA}\,\UU_{0}^{B} - T_{3}^{uB}\, \UU_{0}^{A} + \UU_{0}^{A} \UU_{0}^{B}\,T_{3}^{uu} \Big) ~.
\end{align}
An overall factor of $1/\ell$ in each expression comes from our use of $\sqrt{\gamma_0}\,e^{2\beta_0}$, rather than $\sqrt{-h_0} = \sqrt{\gamma_0}\,e^{2\beta_0}/\ell$, for the volume element on $B$.

Evolution equations and other results in this paper involve similar projections of three dimensional derivatives. Derivatives tangent to $\CC$ are written in terms of $\partial_{A}$ or $\DD_{A}$. Derivatives normal to $\CC$ act on scalars and covariant spatial (orthogonal to $v^{i}$) tensors as 
\begin{gather}\label{eq:DerivativeNormalToC}
 \frac{1}{e^{\beta}\sqrt{-\VV}} \, \Big(\partial_{u} + \pounds_{\UU} \Big) ~.
\end{gather}
Here $\pounds_{\UU}$ is the Lie derivative along $\UU^{A}$ and $e^{\beta}\sqrt{-\VV}$ is the lapse. Given a covariant spatial tensor $S_{i j \ldots}$ the Lie derivative along a three-dimensional vector $X^k$ can then be written
\begin{gather}
    \pounds_{X} S_{i j \ldots} = \big( -v_{k} \,X^{k} \big) \, \frac{1}{e^{\beta}\sqrt{-\VV}} \Big(\partial_{u} + \pounds_{\UU} \Big)  S_{i j \ldots} + \pounds_{\perp X}  S_{i j \ldots} ~.
\end{gather}
Since we are interested in the behavior of the fields at large $r$, our results are usually expressed in terms of the leading part of the derivative \eqref{eq:DerivativeNormalToC}, which we denote $\Dv$
\begin{gather}\label{eq:DvDefinition}
 \Dv \defeq e^{-2\beta_0}\Big(\partial_{u} + \pounds_{\UU_{0}} \Big) ~.   
\end{gather}
Note that $e^{2\beta_0}$ is, up to a factor of $r/\ell$, the leading part of the lapse $e^{\beta}\sqrt{-\VV}$. Several results in the main text include terms where the derivative $\Dv$ acts on $\ln\sqrt{\gamma_0}$. This quantity is denoted $\theta$ and is given by
\begin{gather}\label{eq:thetaDefinition}
    \theta = \Dv \ln \sqrt{\gamma_0} = e^{-2\beta_0} \Big( \, \partial_{u} \ln\sqrt{\gamma_0} + \DD_{A} \UU_{0}^{\,A} \, \Big) ~.
\end{gather}
From \eqref{eq:CurlyKExtrinsicCurvature} this is just the leading term in the $1/r$ expansion of the trace of the extrinsic curvature. The constraint \eqref{eq:Constraintshear} is simply the statement that the shear $\hat{\gamma}^{1}_{AB}$ is the traceless part of $\KK_{AB}$ at leading order.

\section{Asymptotic Expansions of the Fields}
\label{app:AsymptoticExpansions}

In this appendix we present expressions for subleading terms in the large-$r$ expansions of the fields in PBG (with the partially on-shell conditions \eqref{eq:PartiallyOnShellConditions} imposed) and derive certain quantities in the Newman--Penrose formalism which are used throughout the main text. We also record a few additional results concerning the diffeomorphisms \eqref{eq:AKV} that preserve partial Bondi gauge. They are not needed in the main text but may be helpful when verifying our results.

\subsection{The fields at large $r$}\label{app:FieldsAtLarger}
The partially on-shell conditions \eqref{eq:PartiallyOnShellConditions} enforce the components of the Einstein equations conjugate to non-zero terms in the metric \eqref{eq:appPBG}, but not the components conjugate to the gauge-fixed variables $g_{rr} = g_{rA} = 0$. This is sufficient to fix the large-$r$ behavior of the fields, which admit expansions in powers of $1/r$ up to the second subleading order relative to their leading dependence on $r$. The expansions are polyhomogeneous beyond that point, with terms of the form $(\ln r)^{m} / r^{n}$ appearing for $0 \leq m \leq n-2$ when $n \geq 3$. Additional discussion can be found in \cite{Geiller:2022vto,McNees:2024iyu}. 

Most results in the text rely only on terms up to third subleading order in the large-$r$ expansion of the fields. Contributions to charges and other quantities from higher-order terms will vanish as $r \to \infty$. The relevant terms are
\begin{align}
    \gamma_{AB} = & \,\, r^{2}\,\left( \gamma^{0}_{AB} + \frac{1}{r} \, \gamma^{1}_{AB} + \frac{1}{r^2} \, \gamma^{2}_{AB} + \frac{\ln r}{r^3} \, \AA^{3}_{AB} + \frac{1}{r^3} \, \gamma^{3}_{AB} + \ldots \right) \\ 
    \VV = & \,\, r^{2} \, \left( \VV_{0} + \frac{1}{r} \, \VV_{1} + \frac{1}{r^2} \, \VV_{2} + \frac{\ln r}{r^3} \, \tilde{\VV}_{3} + \frac{1}{r^3} \, \VV_3 + \ldots  \right) \\
    \beta = & \,\, \beta_0 + \frac{1}{r}\,\beta_1 + \frac{1}{r^2} \, \beta_2 + \frac{\ln r}{r^3} \, \tilde{\beta}_3 + \frac{1}{r^3} \, \beta_3 + \ldots \\
    \UU^{A} = & \,\, \UU_{0}^{A} + \frac{1}{r} \, \UU_{1}^{A} + \frac{1}{r^2} \, \UU_{2}^{A} + \frac{\ln r}{r^3} \, \tilde{\UU}_{3}^{A} + \frac{1}{r^3} \, \UU_{3}^{A} + \ldots
\end{align}
The leading terms $\gamma^{0}_{AB}$, $\beta_0$, and $\UU_{0}^{A}$ comprise the boundary data in our treatment of Dirichlet boundary conditions. The traces $\gamma_n = \gamma_{0}^{AB} \gamma^{n}_{AB}$ and $\AA_{3} = \gamma_{0}^{AB} \, \AA^{3}_{AB}$ are not constrained by the partially on-shell conditions. Otherwise, terms at the first few orders in these expansions are determined by the boundary data and traces. The quantities $\gamma^{3}_{AB}$, $\VV_3$, and $\UU_{3}^{A}$ appearing at third subleading order are independent of the boundary data and traces and encode dynamical information about the spacetime. 

Let us summarize how terms in the large-$r$ expansions of the fields depend on the boundary data and traces. For the trace-free tensors in the expansion of $\gamma_{AB}$ we have
\begin{align}
    \frac{\Lambda}{3}\, \hat{\gamma}^{1}_{AB} = & \,\, \Big( \Dv \gamma^{0}_{AB} - \gamma^{0}_{AB} \, \theta \Big) \\
    \hat{\gamma}^{2}_{AB} = & \,\, \frac{1}{4} \, \gamma_{1} \, \hat{\gamma}^{1}_{AB} \\
    \hat{\AA}^{3}_{AB} = & \,\, 0 ~,
\end{align}
where $\Dv$ is the derivative normal to $\CC$ defined in \eqref{eq:DvDefinition}. Not all of these results have a smooth limit as $\Lambda \to 0$. In particular, there is a $\Lambda = 0$ degree of freedom in $\hat{\gamma}^{2}_{AB}$ which is strictly zero for $\Lambda \neq 0$. The terms in the expansion of $\VV$ are 
\begingroup
\allowdisplaybreaks
\begin{align}
    \VV_{0} = & \,\, \frac{\Lambda}{3} \, e^{2\beta_0} \\
    \VV_{1} = & \,\, \frac{\Lambda}{6}\, e^{2\beta_0} \gamma_{1} - e^{2\beta_0} \theta \\
    \VV_{2} = & \,\, -\frac{1}{2}\,e^{2\beta_0} \left(\Dv  + \frac{1}{2}\,\theta \right) \gamma_{1} - \DD_{A} \DD^{A} e^{2\beta_0} - \frac{1}{2}\,e^{2\beta_0}  \RR \\ \nonumber
     & \qquad + \frac{\Lambda}{4}\,e^{2\beta_0} \left(\gamma_{2} - \frac{1}{24}\,(\gamma_1)^{2} - \frac{3}{4} \, \hat{\gamma}_{1}^{AB} \hat{\gamma}^{1}_{AB} \right) \\
    \tilde{\VV}_{3} = & \,\, \frac{\Lambda}{6} \, e^{2\beta_0} \AA_{3} ~.
\end{align}
\endgroup
The subleading terms in the expansion of $\beta$ are
\begingroup
\allowdisplaybreaks
\begin{align}
    \beta_1 = & \,\, 0 \\ \label{eq:Beta2Condition}
    \beta_2 = & - \frac{1}{8}\,\left( \gamma_2 - \frac{1}{8}\,(\gamma_1)^{2} - \frac{1}{4} \, \hat{\gamma}_{1}^{AB} \hat{\gamma}^{1}_{AB}\right) \\
    \tilde{\beta}_{3} = & - \frac{1}{4} \, \AA_{3} \\
    \beta_3 = & \,\, \frac{1}{8} \,\AA_{3} - \frac{1}{4} \, \gamma_3 + \frac{1}{16} \, \gamma_1 \gamma_2 - \frac{1}{128} \, (\gamma_1)^{3} - \frac{1}{64} \, \gamma_1 \, \hat{\gamma}_{1}^{AB} \hat{\gamma}^{1}_{AB} ~.
\end{align}
\endgroup
Finally, subleading terms in the expansion of the vector $\UU^{A}$ are
\begingroup
\allowdisplaybreaks
\begin{align}
    \UU_{1}^{A} = & \,\, \DD^{A}\big(e^{2\beta_0} \big) \\
    \UU_{2}^{A} = & - \frac{1}{2} \, \DD_{B} \left(e^{2\beta_0} \hat{\gamma}_{1}^{AB} - \frac{1}{2} \, e^{2\beta_0} \, \gamma_{1} \right) - \frac{1}{2} \, \gamma_{1} \DD^{A}\big(e^{2\beta_0} \big) \\
    \tilde{\UU}_{3}^{A} = & \,\, 0 ~.
\end{align}
\endgroup
The $\ln r/r^3$ terms in these expansions are pure trace -- both $\AA^{3}_{\langle A B \rangle}$ and $\tilde{\UU}_{3}^{A}$ vanish. 

Terms at fourth subleading order in these expansions \emph{are} needed when verifying the remaining equations of motion in section \ref{sec:WardIdentities}. Those expressions are long and unwieldy, so we do not include them here. 

\subsection{Newman--Penrose formalism in partial Bondi gauge}
\label{app:NPFormalism}
Many expressions in the main text can be simplified by writing them in terms of the Newman-Penrose scalars and related quantities.\,\footnote{See \cite{Mao:2019ahc} for a purely Newman--Penrose treatment of the solution space, albeit with stronger restrictions on the boundary data.} Using the convention of \cite{Geiller:2022vto}, we employ the null tetrad 
\begin{align}
e_1&=\ell=\partial_r,\quad
e_2=n=e^{-2\beta}\left( \partial_u+\frac{\VV}{2}\partial_r+\UU^A\partial_A\right),\\
e_3&=\hat{m}=\sqrt{\frac{\gamma_{\theta\theta}}{2\gamma}}\left(\frac{\sqrt{\gamma}+i\gamma_{\theta\phi}}{\gamma_{\theta\theta}}\,\partial_\theta-i\partial_\phi\right),\quad
e_4=\hat{\bar{m}}.
\end{align}
It is useful to introduce the spin coefficient $\sigma$ and $\lambda$ carrying information about the normal and transverse shear  
\begin{align}
    \sigma& =-\frac1{2r^2}\hat\gamma_{AB}^1 m^Am^B +o(r^{-2})\\
    \lambda&=-\frac{\Lambda}{12} \hat\gamma^1_{AB} \bar m^A\bar m^B-\frac1{2r} \left( \lambda_{AB} +\frac{\Lambda}{12}\gamma_1\, \hat\gamma_{AB}^1\right)\bar m^A\bar m^B+ o(r^{-1})
\end{align}
where $m^A=r\,\hat m^A$, $\theta = \Dv \ln\sqrt{\gamma_0} $ as in \eqref{eq:thetaDefinition}, and
\begin{equation}\label{eq:NPlamAB}
\lambda_{AB}=\left(\Dv - \frac{1}{2}\,\theta \right)\hat{\gamma}^{1}_{AB}  - \frac{\Lambda}{6}\,\gamma^{0}_{AB}\,\hat{\gamma}_{1}^{CD} \, \hat{\gamma}^{1}_{CD} + 2\,e^{-2\beta_0}\,\DD_{\langle A}\DD_{B\rangle} e^{2\beta_0} ~.
\end{equation}
The first few terms in $\lambda_{AB}$ are simply $e^{-2\beta_0}\,N_{AB}$, so we sometimes refer to $N_{AB}$ and $\lambda_{AB}$ collectively as ``news tensors.'' The NP scalars are 
\begin{subequations}
    \begin{align}
        \Psi_4&=\frac1r \,\mathcal N_{AB} \,\bar m^A \bar m^B  +o(r^{-1})\,,\quad
        \Psi_3=\frac1{r^2} \, \mathcal J_{A} \, \bar m^A  +o(r^{-2})\\
        \Psi_2&= \frac1{r^3} \, \big( \MM + i \widetilde{\MM} \big) +o(r^{-3})\,,\quad 
         \Psi_1=\frac1{r^4} \, \PP_A \, m^A + o(r^{-4}) \\ 
        \Psi_0 & = \frac1{r^5} \, \EE_{AB} \, m^Am^B + o(r^{-5})
    \end{align}
\end{subequations}
where 
\begin{subequations}\label{NPscalars1}
    \begin{align}
    \NN_{AB} = &\,\, \frac{1}{2} \, \Dv  \lambda_{AB} - \frac{\Lambda}{12}\,\gamma^{0}_{AB}\,\hat{\gamma}_{1}^{CD} \, \lambda_{CD} - \frac{1}{2}\,e^{-2\beta_0}\,\DD_{\langle A}\DD_{B \rangle} \Big( e^{2\beta_0} \theta \Big)    + \left(\frac{\Lambda}{6}\right)^{2} \EE_{AB}  \\ \nonumber
    &\,\, -\frac{1}{2} \left( \frac{\Lambda}{6} \right)^{2}  \hat{\gamma}^{1}_{AB} \, \hat{\gamma}_{1}^{CD} \, \hat{\gamma}^{1}_{CD}  - \frac{\Lambda}{6}\,\left( \RR \, \hat{\gamma}^{1}_{AB} - \frac{1}{2} \, \DD^{2}\hat{\gamma}^{1}_{AB} \right) - \frac{\Lambda}{12}\,\hat{\gamma}^{1}_{AB}\,e^{-2\beta_0}\,\DD^2 e^{2\beta_0} \\ \nonumber
    & \,\, + \frac{2\Lambda}{3}\,\DD_{\langle A}\beta_0 \, \DD^{C}\hat{\gamma}^{1}_{B \rangle C}   \\   \label{eq:JDefinition}
    \mathcal{J}_{A} = &\,\, \frac{1}{2}\,\DD^{B} \lambda_{AB} + \frac{1}{4}\,\DD_{A}\RR - \frac{\Lambda}{6}\,\PP_{A} + \frac{\Lambda}{24}\,\hat{\gamma}_{1}^{BC} \DD_{A} \hat{\gamma}^{1}_{BC} + \frac{\Lambda}{12}\,\hat{\gamma}^{1}_{AB} \DD_{C}\hat{\gamma}_{1}^{BC}  \\   
    \widetilde\MM = & \,\, \frac{1}{8}\,\lambda^{A}{}_{B} \, \eps^{BC} \, \hat{\gamma}^{1}_{AC} - \frac{1}{4} \, \DD^{A} \DD_{B} \Big( \eps^{BC} \hat{\gamma}^{1}_{AC} \Big)
    \end{align}
\end{subequations}
and
\begin{subequations}\label{NPscalars2}
\begin{align}
     \MM = &\,\, \frac{1}{2}\,e^{-2\beta_0}\,\VV_{3} + \frac{\Lambda}{24} \, \tilde{\AA}_{3} - \frac{\Lambda}{6}\,\gamma_{3} - \DD_{B}\hat{\gamma}_{1}^{AB}\,\DD_{A}\beta_0 + \frac{1}{2}\,\DD_{A}\gamma_{1}\,\DD^{A}\beta_0 \\ \nonumber
    & \,\, + \frac{1}{4} \, \Big(\Dv + \theta \Big) \left( \gamma_{2} - \frac{1}{8}\,(\gamma_1)^{2} - \frac{1}{4}\,\hat{\gamma}_{1}^{AB}\,\hat{\gamma}^{1}_{AB}\right)\\
    \PP_{A} = &  -\frac{3}{2} \, e^{-2\beta_0} \, \gamma^{0}_{AB} \, \UU_{3}^{\,B} + \frac{3}{32} \, e^{2\beta_0} \, \DD_A\left(e^{-2\beta_0}\Big( 4 \, \gamma_2-\gamma_{AB}^1\gamma^{AB}_1\Big) \right) \\\nonumber
    & + \frac{3}{4} \, \gamma^1_{AB} \Big( \DD_C + \DD_{C} \beta_0 \Big) \gamma_{1}^{CB} - \frac{3}{4} \, \gamma^1_{AB} \, \DD^B\gamma_1 \\
 \EE_{AB}  = &\,\, 3 \, \hat\gamma^3_{AB}-\frac3{4}\,\hat\gamma^1_{AB}\left(\gamma_2 - \frac{1}{8}\,(\gamma_1)^{2} - \frac{1}{4}\,\hat{\gamma}_{1}^{CD}\,\hat{\gamma}^{1}_{CD} \right)
\end{align}
\end{subequations}
The expressions \eqref{NPscalars2} were already computed in \cite{Geiller:2022vto}, while the results \eqref{NPscalars1} generalize equations (2.43) and (2.44) of \cite{Geiller:2022vto}.

\subsection{Diffeomorphisms and Transformations}
\label{app:Diffeos}
The general form of the component $\xi^{A}$  is given in \eqref{xiA}. Sub-leading terms in $\xi^{A}$ are easily extracted from the large-$r$ expansions of fields satisfying the partially on-shell conditions. Starting with the $\OO(r^0)$ part $Y^{A}$, the expansion proceeds in powers of $1/r$ before the first $\ln r$ factor appears at $\OO(r^{-4}\ln r)$. The results in this paper only depend on the terms up to $\OO(r^{-3})$, which are
\begin{gather}\label{eq:sublxiA}
    \xi_{1}^{A} = -e^{2\beta_0}\,\DD^{A} f \qquad \xi_{2}^{A} = \frac{1}{2}\,e^{2\beta_0} \left( \hat{\gamma}_{1}^{AB} + \frac{1}{2}\,\gamma_{1} \, \gamma_{0}^{AB} \right)  \DD_{B} f \\ \nonumber
    \xi_{3}^{A} = \frac{1}{4}\,e^{2\beta_0}  \gamma_{0}^{AB}\,\left( \gamma_2 - \frac{3}{8}\,\gamma_{1}^{\,2} - \frac{3}{4}\,\hat{\gamma}_{1}^{CD} \hat{\gamma}^{1}_{CD} \right) - \frac{1}{4}\,e^{2\beta_0} \gamma_{1} \hat{\gamma}_{1}^{AB} 
\end{gather}
These expressions, along with the other terms in \eqref{eq:AKV}, are sufficient for determining the transformations of the fields out to $\OO(r^{-3})$ in the large-$r$ expansion. 

The transformations of $\UU_{0}^{A}$ and $\beta_0$  are given in \eqref{eq:UTransformation} and \eqref{eq:BetaTransformation} with the same general form used for the transformations of other fields. However, their geometric origin as the three-dimensional shift and lapse, respectively, means their transformations can be written in a simpler form as
\begin{align}
    \delta_{\xi} \UU_{0}^{A} = & \,\, - e^{2\beta_0} \Dv (\perp\!\xi)^{A} - \frac{\Lambda}{3}\,X^{A} \\ 
    \delta_{\xi}\beta_0 = & \,\, \pounds_{\perp\xi} \beta_0 + \frac{1}{2}\,\Big(h + \Dv \xi^{v} \Big) ~.
\end{align}
These forms are especially useful in obtaining transformations of quantities involving the field-dependent operator $\Dv$. For example, suppose a quantity $F$ transforms under a diffeomorphism as
\begin{gather}\label{eq:FTransform}
    \delta_{\xi} F = \Big( \xi^{v}\,\Dv + \pounds_{\perp\xi} + w\,h \Big) F + \tilde{\delta}_{\xi} F ~,
\end{gather}
where $w$ is the conformal weight. Then the derivative of $F$ normal to $\CC$ transforms as
\begin{gather}\label{eq:DvFTransform}
    \delta_{\xi} \Dv F = \Big( \xi^{v}\,\Dv + \pounds_{\perp\xi} + (w-1) \,h \Big) \Dv F + \Dv \tilde{\delta}_{\xi} F + w\,F \,\Dv h - \frac{\Lambda}{3} \, e^{-2\beta_0} \pounds_{X} F ~.
\end{gather}
This applies for any $F$ that transforms as in \eqref{eq:FTransform}, whether it is a scalar, tensor, or density. An example is the transformation of $\theta = \Dv \ln\sqrt{\gamma_0}$. Since $\ln\sqrt{\gamma_0}$ has conformal weight zero and $\tilde{\delta}_{\xi}\ln\sqrt{\gamma_0} = 2\,h$, we have
\begin{gather}
    \delta_{\xi} \theta = \Big( \xi^{v}\,\Dv + \pounds_{\perp\xi} - \,h \Big) \theta + 2 \, \Dv h - \frac{\Lambda}{3} \, e^{-2\beta_0} \, \DD_{C} X^{C} ~.
\end{gather}
The result \eqref{eq:DvFTransform} is useful when working out the transformations of quantities such as $\hat{\gamma}^{1}_{AB}$, $\lambda_{AB}$, and $\NN_{AB}$ which involve derivatives normal to $\CC$.

\section{Electric and Magnetic parts of the Weyl Tensor}
\label{sec:WeylParts}

To identify the components of the electric and magnetic part of the Weyl tensor, we start with an orthonormal basis comprising vectors $n^{\lambda}$, $v^{\lambda}$, $\kappa^{\lambda}$, and $\eta^{\lambda}$ that satisfy
\begin{gather}
    n^{\lambda} n_{\lambda} = \kappa^{\lambda} \kappa_{\lambda} = \eta^{\lambda} \eta_{\lambda} = 1 \qquad v^{\lambda} v_{\lambda} = -1 ~,
\end{gather}
with all other contractions equal to zero. The vector $n^{\lambda}$ is the outward-pointing unit vector normal to surfaces of constant $r$, given in \eqref{eq:appSpacelikeNormal}. A convenient choice for the timelike unit vector $v^{\lambda}$ is to take a vector with covariant components spanning just the $r$ and $u$ directions. Then the conditions $P^{\mu}{}_{\lambda} v^{\lambda} = v^{\mu}$ and $v^{\lambda} \, v_{\lambda} = -1$ give 
\begin{gather}\label{eq:appTimelikeNormal4D}
    v_{\lambda} = - \delta_{\lambda}{}^{r} \frac{e^{\beta}}{\sqrt{-\VV}} - \delta_{\lambda}{ }^{u} \,e^{\beta} \sqrt{-\VV} \qquad v^{\lambda} = \frac{1}{e^{\beta}\,\sqrt{-\VV}} \, \Big(\delta^{\lambda}{}_{u} + \delta^{\lambda}{}_{\!A} \,\UU^{A} \Big) ~.
\end{gather}
This is an especially useful choice, since the contravariant components are just the unit normal vector \eqref{eq:appTimelikeNormal3D} associated with the foliation of $B$ by the family of constant-$u$ surfaces $\CC$. Note that $v^{\lambda}$, which is timelike, is \emph{not} normal to constant-$u$ hypersurfaces in $M$, which are null. The combinations $n_{\lambda} \pm v_{\lambda}$ are (up to an overall normalization) the elements $e_1$ and $e_2$ of the null tetrad used to identify the Newman-Penrose scalars and related quantities in appendix \ref{app:NPFormalism}. Next we take $\kappa^{\lambda}$ and $\eta^{\lambda}$ to be any orthogonal pair of unit vectors spanning the directions associated with the two dimensional coordinates $x^{A}$. At large $r$ such vectors fall off as 
\begin{gather}\label{eq:KappaNormalization}
    \kappa^{A} = \frac{1}{r} \, \kappa_{0}^{A} + \ldots  ~,
\end{gather}
with the factor of $r^{-1}$ coming from the normalization condition $\gamma_{AB} \,\kappa^{A} \kappa^{B} = 1$.

The Weyl tensor in four dimensions is related to the Riemann tensor and its contractions by
\begin{gather}
    C_{\lambda \mu \sigma \nu} = R_{\lambda \mu \sigma \nu} + g_{\lambda \nu} S_{\mu\sigma} - g_{\lambda \sigma} S_{\mu\nu} + g_{\mu\sigma} S_{\lambda \nu} - g_{\mu\nu} S_{\lambda \sigma} \\
    S_{\mu\nu} = \frac{1}{2}\,R_{\mu\nu} - \frac{1}{12} \, g_{\mu\nu} R ~,
\end{gather}
where $S_{\mu\nu}$ is the Schouten tensor. We define the electric and magnetic parts by projecting (with $P_{\mu}{}^{\nu} = \delta_{\mu}{}^{\nu} - n_{\mu} n^{\nu}$) the following tensors tangent to a surface of constant $r$ .
\begin{align}
    E_{\mu\nu} = & \,\, C_{\lambda \mu \sigma \nu} n^{\lambda} n^{\sigma} \\ 
        = & \,\, R_{\lambda \mu \sigma \nu} n^{\lambda} n^{\sigma} + n_{\nu} n^{\lambda} S_{\mu\lambda} - S_{\mu\nu} + n_{\mu} n^{\lambda} S_{\lambda \nu} - g_{\mu\nu} S_{\lambda\sigma} n^{\lambda} n^{\sigma} \\
    B_{\mu \sigma \nu} = & \,\, C_{\lambda \mu \sigma \nu} n^{\lambda} \\ 
        = & \, R_{\lambda \mu \sigma \nu} n^{\lambda} + n_{\nu} S_{\mu\sigma} - n_{\sigma} S_{\mu\nu} + g_{\mu\sigma} n^{\lambda} S_{\lambda \nu} - g_{\mu\nu} n^{\lambda} S_{\lambda \sigma} ~.
\end{align}
There is no need to write the projections out explicitly, since the components will be determined by contracting with the other vectors $v$, $\kappa$, and $\eta$ in the basis, which are all orthogonal to $n$.\,\footnote{The tensor $E_{\mu\nu}$ is already tangent to surfaces of constant $r$, since contracting either index with $n^{\lambda}$ gives zero by the symmetries of the Weyl tensor.} The resulting three-dimensional tensors are both traceless with respect to the induced metric on $B$.

At large $r$ we find
\begin{align}\label{eq:Evv}
    E_{\mu\nu} \,v^{\mu} v^{\nu} = &\,\, \frac{1}{r^3}\,(-2\,\MM) + \OO(r^{-4})\\ \label{eq:Eva}
    E_{\mu\nu} \,v^{\mu} \kappa^{\nu} = &\,\, \frac{1}{r^{3}} \, \sqrt{-\frac{3}{\Lambda}}  \left( -\JJ_{A} + \frac{\Lambda}{6}\,\PP_{A} \right) \, \kappa_{0}^{A}  + \OO(r^{-4})   \\ \label{eq:EAB}
    E_{\mu\nu} \,\kappa^{\mu} \eta^{\nu} = &\,\, \frac{1}{r^{3}}\, \frac{1}{2} \left(\frac{6}{\Lambda} \, \NN_{AB} + \frac{\Lambda}{6} \, \EE_{AB} - 2 \, \gamma^{0}_{AB} \, \MM \right) \, \kappa_{0}^{A} \, \eta_{0}^{B} + \OO(r^{-4})
\end{align}
As expected, the electric part of the Weyl tensor consists of the same combinations of Newman-Penrose scalars and related quantities that appear in the traceless three dimensional stress tensor on $B$. For these contractions, terms at orders $r^0$, $r^{-1}$, $r^{-2}$, and $r^{-3} \ln r$ vanish, and the first non-zero terms appear at $\OO(r^{-3})$. Contributions at order $r^{-4}$ (with or without multiple $\ln r$ factors) and higher may or may not vanish. In particular, the $\OO(r^{-4})$ contributions come from the next-order terms in the components of the Riemann tensor, along with the leading terms in the large $r$ expansions of $R_{uA} + \Lambda\,\gamma_{AB}\,\UU^{B}$ and $R_{uu} + R_{uA}\,\UU^{A} - \Lambda\,e^{2 \, \beta}\,\VV$. These components of the Einstein tensor, which are not fixed by the partially on-shell conditions, encode the flux balance equations for the mass and angular momentum. We have not checked whether the $\OO(r^{-4})$ terms in \eqref{eq:Evv}-\eqref{eq:EAB} vanish or are non-zero.

For the magnetic part of the Weyl tensor we find
\begin{align}
    B_{\mu\sigma A} v^{\mu} v^{\sigma} \kappa^{A} = & \,\, \frac{1}{r^3} \, \sqrt{-\frac{3}{\Lambda}} \, \left( \JJ_{A} + \frac{\Lambda}{6} \, \PP_{A} \right) \, \kappa_{0}^{A} + \OO(r^{-4}) \\
    B_{A \sigma B} \kappa^{A} v^{\sigma} \eta^{B} = & \,\, \frac{1}{r^3}\,\frac{1}{2}\,\left( -\frac{6}{\Lambda}\,\NN_{AB} + \frac{\Lambda}{6}\,\EE_{AB} - 2\,\epsilon_{AB} \, \widetilde{\MM} \,\right) \kappa_{0}^{A} \, \eta_{0}^{B}  + \OO(r^{-4}) \\
    B_{\mu A B} \, v^{\mu} \kappa^{A} \eta^{B} = & \,\, \frac{1}{r^3}\,\Big( - 2 \, \epsilon_{AB} \, \widetilde{\MM} \, \Big) \, \kappa_{0}^{A} \, \eta_{0}^{B}  + \OO(r^{-4}) \\
    B_{ABC} \, \kappa^{A} \eta^{B} \lambda^{C} = & \,\, \frac{1}{r^3} \,\sqrt{-\frac{3}{\Lambda}} \left[ \gamma^{0}_{AB} \, \left(\JJ_{C} + \frac{\Lambda}{6} \, \PP_{C} \right) - \gamma^{0}_{AC} \, \left(\JJ_{B} + \frac{\Lambda}{6} \, \PP_{B} \right) \right]   \kappa_{0}^{A} \, \eta_{0}^{B} \, \lambda_{0}^{C} + \OO(r^{-4})  
\end{align}
for arbitrary (but normalized) two-dimensional vectors $\kappa^{A}$, $\eta^{B}$, and $\lambda^{C}$. Notice that the relative sign between the $\JJ_{A}$ and $\PP_{A}$ terms, and the $\NN_{AB}$ and $\EE_{AB}$ terms, is the opposite of the combinations appearing in the stress tensor and the electric part $E_{\mu\nu}$. Again, sub-leading terms may appear at $\OO(r^{-4})$, though in this case we have confirmed that the $r^{-4}\ln r$ terms in the magnetic part all vanish.

It is convenient to relate the results above to terms in the large-$r$ expansion of tensors $E_{ij}$ and $B_{ijk}$ on $B$. The large-$r$ expansions for the fields imply the following expansions for these projections of the Weyl tensor
\begin{align}
    E_{ij} = & \,\, r^{2} \left( E^{0}_{ij} + \frac{1}{r} \, E^{1}_{ij} + \frac{1}{r^2}\,E^{2}_{ij} + \frac{\ln r}{r^3}\,\tilde{E}^{3}_{ij} + \frac{1}{r^3} \, E^{3}_{ij} + \ldots \right) \\
    B_{ijk} = & \,\, r^{3} \left( B^{0}_{ijk} + \frac{1}{r} \, B^{1}_{ijk} + \frac{1}{r^2}\,B^{2}_{ijk} + \frac{\ln r}{r^3}\,\tilde{B}^{3}_{ijk} + \frac{1}{r^3} \, B^{3}_{ijk} + \ldots \right) ~.
\end{align}
Subsequent terms in the expansions are polyhomogeneous in $\ln r$ and $1/r$. With the partially on-shell conditions enforced, $E^{n}_{ij} = B^{n}_{ijk} = 0$ for $n\leq 2$ and $\tilde{E}^{3}_{ij} = \tilde{B}^{3}_{ijk} = 0$. The first non-zero terms are $E^{3}_{ij}$ and $B^{3}_{ijk}$ at the third sub-leading order. Writing the large-$r$ expansion of the timelike unit vector normal to $\CC$ as
\begin{gather}
    v_{i} = r \, v^{0}_{i} + \ldots \qquad v^{i} = \frac{1}{r}\, v_{0}^{i} + \ldots ~,
\end{gather}
where `$\ldots$' indicates subleading terms, we can project $E^{3}_{ij}$ and $B^{3}_{ijk}$ normal and tangent to $\CC$ by contracting with $v_{0}^{i}$ or projecting with 
\begin{gather}
    \perp_{i}{\!}^{j} = \delta_{i}{}^{j} + v^{0}_{i} \, v_{0}^{j} ~.
\end{gather}
Contraction with $-v_{0}^{i}$ or $-v^{0}_{i}$ is denoted with a subscript or superscript $v$, respectively. Then the electric components are
\begin{align} \label{eq:SecondEvv}
    E^{3}_{vv} = & \,\, -2\,\MM \\ \label{eq:SecondEvA}
    E^{3}_{vA} = & \,\, \sqrt{-\frac{3}{\Lambda}}  \left(-\JJ_{A} + \frac{\Lambda}{6}\,\PP_{A}\right) \\ \label{eq:SecondEAM}
    E^{3}_{AB} = & \,\, \frac{1}{2} \left(\frac{6}{\Lambda} \, \NN_{AB} + \frac{\Lambda}{6} \, \EE_{AB} - 2 \, \gamma^{0}_{AB} \, \MM \right) ~,
\end{align}
and the magnetic components are
\begin{align} \label{eq:SecondBvvA}
    B^{3}_{vvA} = &\,\, \sqrt{-\frac{3}{\Lambda}}  \left( \JJ_{A} + \frac{\Lambda}{6} \, \PP_{A} \right) \\ \label{eq:SecondBAvB}
    B^{3}_{AvB} = & \,\, \frac{1}{2} \left(-\frac{6}{\Lambda}\,\NN_{AB} + \frac{\Lambda}{6}\,\EE_{AB} - 2\,\epsilon_{AB} \, \widetilde{\MM} \right) \\ \label{eq:SecondBvAB}
    B^{3}_{vAB} = & \,\, - 2 \, \epsilon_{AB} \, \widetilde{\MM} \\ \label{eq:SecondBABC}
    B^{3}_{ABC} = & \,\, \sqrt{-\frac{3}{\Lambda}} \left[\gamma^{0}_{AB} \, \left(\JJ_{C} + \frac{\Lambda}{6} \, \PP_{C} \right) - \gamma^{0}_{AC} \, \left(\JJ_{B} + \frac{\Lambda}{6} \, \PP_{B} \right) \right] ~.
\end{align}
It is straightforward to check that these tensors are traceless with respect to $h^{0}_{ij}$, and that the components of $B^{3}_{ijk}$ satisfy $B^{3}_{ijk} + B^{3}_{jki} + B^{3}_{kij} = 0$, consistent with the cyclic property of the Riemann tensor.  Our expressions coincide with \cite{RadGeiller} when restricted to Bondi-Sachs gauge.

Using the transformations of various quantities under the diffeomorphisms considered in section \ref{sec:charges}, the components of the electric part of the Weyl tensor transform as
\begin{align}\label{eq:EvvTransform}
    \delta_{\xi} E_{vv} = & \,\, \Big( e^{2\beta_0} f \, \Dv + \pounds_{\perp\xi} -3\,h \Big) \, E_{vv} + 2 \sqrt{-\frac{\Lambda}{3}} \, E_{vA} \, e^{-2\beta_0} X^{A} \\ 
    \delta_{\xi} E_{vA} = & \,\, \Big( e^{2\beta_0} f \, \Dv + \pounds_{\perp\xi} - 2\,h \Big) \, E_{vA} + \sqrt{-\frac{\Lambda}{3}}\,\Big(E_{AB} + \gamma^{0}_{AB} E_{vv} \Big) e^{-2\beta_0} X^{B} \\ 
    \delta_{\xi} E_{AB} = & \,\,  \Big( e^{2\beta_0} f \, \Dv + \pounds_{\perp\xi} - h \Big) \, E_{AB} + \sqrt{-\frac{\Lambda}{3}} \, e^{-2\beta_0} \Big( E_{vA} \,X_{B} + E_{vB} \, X_{A} \Big) ~.
\end{align}
Similar expressions can be worked out for the magnetic components using \eqref{eq:NPDiffeoTransformations}. Equivalent results are given in the next appendix for asymptotic expansions in powers of $\ell/r$ rather than $1/r$.

\section{Conventions for Asymptotic Expansions}
\label{app:ConventionsForExpansions}

In theories with asymptotically flat boundary conditions one typically expands quantities in powers of $1/r$ on some neighborhood of $r \to \infty$. Many of the results used in the main text are obtained with this sort of expansion. When a cosmological constant is present, as is the case here, it is conventional to expand in powers of $\ell/r$ for $r$ much larger than the associated length scale $\ell = \sqrt{-3/\Lambda}$ (or $\ell = \sqrt{3/\Lambda}$ for a positive cosmological constant). In this appendix we present a few of our results as they would appear in an $\ell/r$ expansion, to simplify comparison with literature that follows this convention.

The induced metric on $B$ has the form
\begin{align}
    ds^{2}\Big|_{r} = & \,\, r^{2}\,\left( h^{0}_{ij} + \frac{1}{r}\,h^{1}_{ij} + \frac{1}{r^2}\,h^{2}_{ij} + \ldots \right) \dd x^i \dd x^j \\
    = & \,\, \frac{r^{2}}{\ell^{2}} \, \left( \bar{h}^{0}_{ij} + \frac{\ell}{r}\, \bar{h}^{1}_{ij} + \frac{\ell^{2}}{r^2}\, \bar{h}^{2}_{ij} + \ldots \right) \dd x^i \dd x^j
\end{align}
The dimensionless quantity $\ell/r$ is a (non-unique) choice of defining function such that the conformal boundary at $\ell/r \to 0$ is equipped with the representative $\bar{h}^{0}_{ij}$ of a conformal class of metrics $[\bar{h}^0]$. Terms appearing in the $\ell/r$ expansion are written with a bar, to distinguish them from terms in the $1/r$ expansion. It is straightforward to convert results expressed in terms of the $h^{n}_{ij}$ to equivalent results in terms of $\bar{h}^{n}_{ij}$. Here we record a few results where care must be taken with the various factors of $\ell = \sqrt{-3/\Lambda}$.

First, note that the timelike unit vector that is normal to a constant-$u$ surface $\CC \subset B$  
\begin{gather}
    v_{i} = r \, v^{0}_{i} + \ldots = \frac{r}{\ell} \, \bar{v}^{0}_{i} + \ldots \\
    v^{0}_{i} = - \frac{1}{\ell} \, e^{2\beta_0} \, \delta_{i}{}^{u} \qquad \bar{v}^{0}_{i} = - \, e^{2\beta_0} \, \delta_{i}{}^{u} ~.
\end{gather}
The leading behavior of the metric on $\CC$ is 
\begin{gather}
    \gamma_{AB} = r^{2} \, \gamma^{0}_{AB} + \ldots = \frac{r^2}{\ell^2} \, \bar{\gamma}^{0}_{AB} + \ldots ~,
\end{gather}
and the volume element on $B$ is
\begin{gather}
    \sqrt{-h} = r^{3} \, \sqrt{\gamma_0}\,e^{2\beta_0} \, \frac{1}{\ell} + \ldots = \frac{r^3}{\ell^3} \, \sqrt{\bar{\gamma}_0} \, e^{2\beta_0} + \ldots ~.
\end{gather}
The stress tensor is obtained by varying the (partially) on-shell action as
\begin{gather}
    T^{ij} = \frac{2}{\sqrt{-h}} \, \frac{\delta \Gamma}{\delta h_{ij}} ~,
\end{gather}
so the leading behavior is $T^{ij} \sim r^2$, with an expansion 
\begin{gather}
    T_{ij} = r^{2} \, \left( T^{0}_{ij} + \frac{1}{r} \, T^{1}_{ij} + \ldots + \frac{1}{r^{3}}\,T^{3}_{ij} + \ldots \right) ~.
\end{gather}
Raising and lowering indices on $\bar{T}^{n}_{ij}$ with $\bar{h}^{0}_{ij} = \ell^2 \, h^{0}_{ij}$, we have
\begin{gather}
    \bar{T}^{n}_{ij} = \ell^{2-n} \, T^{n}_{ij} \qquad \bar{T}_{n}^{ij} = \ell^{-(n+2)} \, T_{n}^{ij} ~.
\end{gather}
When the action includes appropriate surface terms, $T^{n}_{ij} = 0$ for $n=0,1,2$, along with a vanishing term at order $(\ln r)/r$, and the first non-zero term is $T^{3}_{ij}$.

Accounting for the various factors of $\ell$, we have
\begin{align}
    \bar{\TT} = & \,\, \frac{1}{\ell} \, e^{2\beta_0} \,(-\bar{\VV}_0) \, \bar{T}_{3}^{uu} = \frac{1}{\ell^{3}}\,\TT \\
    \bar{\TT}_{A} = & \,\, \frac{1}{\ell}\,\Big( \bar{T}_{3}^{uu} \, \bar{\gamma}^{0}_{AB} \,\UU_{0}^{\,B} - \bar{T}_{3}^{uB} \, \bar{\gamma}^{0}_{AB} \Big) = \frac{1}{\ell^{3}}\,\TT_{A} \\
     \bar{\TT}^{AB} = & \,\, \frac{1}{\ell}\,\Big(\bar{T}_{3}^{AB} - \bar{T}_{3}^{uA} \, \UU_{0}^{B} - \bar{T}_{3}^{uB} \, \UU_{0}^{A} + \UU_{0}^{A} \UU_{0}^{B}\, \bar{T}_{3}^{uu} \Big) = \frac{1}{\ell^{5}} \, \TT^{AB} ~.
\end{align}
In terms of these quantities, the transformations from the previous appendix can be written
\begin{align}\label{eq:TTbartransform}
    \delta_{\xi}\bar{\TT} = & \,\, \Big(f \, \Dv + \pounds_{\perp\xi} -3\,h \Big) \, \bar{\TT} +2\,\bar{\TT}_{A} \, e^{4\beta_0} \bar{\DD}^{A} f \\ \label{eq:TTAbartransform}
    \delta_{\xi}\bar{\TT}_{A} = & \,\, \Big(f \, \Dv + \pounds_{\perp\xi} -3\,h \Big) \, \bar{\TT}_{A} - \bar{\TT}_{A} \, \Dv f +  \left( \bar{\TT}_{AB} + \bar{\gamma}^{0}_{AB} \, \bar{\TT} \right) \bar{\DD}^{B} f \\ \label{eq:TTABbartransform}
    \delta_{\xi}\bar{\TT}_{AB} = & \,\, \Big(f \, \Dv + \pounds_{\perp\xi} - h \Big) \, \bar{\TT}_{AB} + \bar{\TT}_{A} \, e^{4\beta_0} \DD_{B} f + \bar{\TT}_{B} \, e^{4\beta_0} \DD_{A} f
\end{align}
Likewise, the components of the electric part of the Weyl tensor are rescaled when switching to the $\ell/r$ expansion. Including the factor of $\ell$ relating $\bar{v}^{0}_{i}$ and $v^{0}_{i}$ when defining the components, we have
\begin{align}
    \bar{E}_{vv} = \frac{1}{\ell^{3}}\,E_{vv} \qquad \bar{E}_{vA} = \frac{1}{\ell^2} \, E_{vA} \qquad \bar{E}_{AB} = \frac{1}{\ell} \, E_{AB} ~.
\end{align}
The transformations from the previous appendix become
\begin{align}\label{eq:EvvbarTransform}
    \delta_{\xi} \bar{E}_{vv} = & \,\, \Big(f \, \Dv + \pounds_{\perp\xi} -3\,h \Big) \, \bar{E}_{vv} + 2 \, \bar{E}_{vA} e^{2\beta_0} \bar{\DD}^{A} f \\
    \delta_{\xi} \bar{E}_{vA} = & \,\, \Big(f \, \Dv + \pounds_{\perp\xi} -3\,h \Big) \, \bar{E}_{vA} + \Big(\bar{E}_{AB} + \bar{\gamma}^{0}_{AB} \bar{E}_{vv} \Big) e^{2\beta_0} \bar{\DD}^{B} f \\
    \delta_{\xi} \bar{E}_{AB} = & \,\,  \Big(f \, \Dv + \pounds_{\perp\xi} - h \Big) \, \bar{E}_{AB} + e^{2\beta_0} \Big( \bar{E}_{vA} \, \DD_{b} f + \bar{E}_{vB} \, \DD_{A} f \Big) ~.
\end{align}
In both cases, the inhomogeneous parts of the transformations no longer involve factors of $\Lambda$.

\section{A Conserved Current from Diffeomorphism Invariance}
\label{app:Conserved}

Covariant phase space methods begin with the derivation of a presymplectic potential $\Theta$ and associated current $\omega$ which is conserved for field variations that satisfy the linearized equations of motion. Here, we obtain a related current which is conserved for all field configurations and reduces on-shell to the usual symplectic current $\omega$ contracted with an asymptotic symmetry $\xi$. This result gives the codimension-2 form $k_{\xi}$ used in section \ref{sec:charges}. The focus is on Einstein gravity but the result may be generalized to gravity coupled to arbitrary fields, see appendix A of \cite{McNees:2024iyu}. 

Consider the bulk Einstein-Hilbert action for gravity with a cosmological constant. Its linearized response to a small change in the metric $g \to g + \delta_{1} g$ is
\begin{gather}
    \delta_{1} I_{M} = \int_{M} \nts d^{4}x \, \Big(\partial_{\mu}\Theta^{u}(g, \delta_1 g) + \sqrt{-g} \, \mathcal{G}^{\mu\nu} \, (\delta_{1} g)_{\mu\nu} \Big) ~.
\end{gather}
where $(\delta_1 g)_{\mu\nu}$ is an infinitesimal but otherwise arbitrary deformation of the metric, and $\mathcal{G}^{\mu\nu}$ was defined in \eqref{eq:EinsteinEquations}. Linearizing a second time and anti-symmetrizing gives
\begin{gather}
    \big[ \delta_{2},\delta_{1} \big] I_{M} = \int_{M} \nts d^{4}x \, \Big[\partial_{\mu}\omega^{\mu}(g,\delta_{1}g,\delta_{2}g) + \delta_{2}\Big(\sqrt{-g} \, \mathcal{G}^{\mu\nu} \Big) \, (\delta_{1} g)_{\mu\nu} - \delta_{1}\Big(\sqrt{-g} \, \mathcal{G}^{\mu\nu} \Big) \, (\delta_{2} g)_{\mu\nu} \Big] ~.
\end{gather}
We assume that the linearization procedure targets only the dependence on the metric $g$ so that $\big[ \delta_{2},\delta_{1} \big] I_{M} = 0$. Therefore, we have
\begin{gather}\label{eq:AntiSymmSecondVariationResult}
    \partial_{\mu}\omega^{\mu}(g,\delta_{1}g,\delta_{2}g) + \delta_{2}\Big(\sqrt{-g} \, \mathcal{G}^{\mu\nu} \Big) \, (\delta_{1} g)_{\mu\nu} - \delta_{1}\Big(\sqrt{-g} \, \mathcal{G}^{\mu\nu} \Big) \, (\delta_{2} g)_{\mu\nu} = 0 ~.
\end{gather}
When the deformations satisfy the linearized equation of motion, the current $\omega^{\mu}$ is conserved. This is the usual starting point for the covariant phase space formalism \cite{Crnkovic:1986ex, Lee:1990nz,Iyer:1994ys,Wald:1993nt,Wald:1999wa}. 

Instead of imposing the linearized equations of motion, suppose the first deformation has the form of an infinitesimal vector $\xi$ acting on the metric, $(\delta_1 g)_{\mu\nu} = \pounds_{\xi} g_{\mu\nu}$, and the second is given by a variational operator $\delta$ acting on $g_{\mu\nu}$. Then \eqref{eq:AntiSymmSecondVariationResult} becomes
\begin{gather}\label{eq:InitialCurrent}
    0 = \partial_{\mu} \omega_{\xi}^{\mu} + \delta \Big(\sqrt{-g} \, \mathcal{G}^{\mu\nu} \Big) \, \pounds_{\xi} g_{\mu\nu} - \pounds_{\xi} \Big(\sqrt{-g} \, \mathcal{G}^{\mu\nu} \Big) \, \delta g_{\mu\nu} ~,
\end{gather}
where $\omega_{\xi}^{\mu}$ is shorthand for $\omega^{\mu}(g, \pounds_{\xi} g, \delta g)$. Integration by parts of both the Lie derivative and $\delta$ (using its Leibniz property) leads to
\begin{gather}\label{eq:ConservedDiffeoCurrent}
    0 = \partial_{\mu} \Big[ \,\omega_{\xi}^{\mu} + 2\,\delta\big( \sqrt{-g}\,\mathcal{G}^{\mu\lambda} g_{\lambda\nu} \big) \, \xi^{\nu} - \xi^{\mu}\,\sqrt{-g}\,\mathcal{G}^{\nu\lambda} \, \delta g_{\nu\lambda} \,\Big] ~.
\end{gather}
This defines a conserved current for all field configurations as a result of diffeomorphism invariance, since the non total derivative terms generated by integration by parts of \eqref{eq:InitialCurrent} vanish by the contracted Bianchi identity $\nabla_{\nu} \mathcal{G}^{\mu\nu} = 0$. Defining the weakly vanishing Noether current
\begin{gather}
    \SS_{\xi}^{\mu} = 2\,\sqrt{-g} \, \mathcal{G}^{\mu\lambda}\,g_{\lambda\nu}\,\xi^{\nu} ~,
\end{gather}
the second term in the conserved current is $\delta(\SS_{\xi}) - \SS_{\delta\xi}$. The third term in the conserved current vanishes by the partially on-shell conditions. It follows from \eqref{eq:ConservedDiffeoCurrent} that there exists a family of codimension-2 forms $k_{\nu}^{\mu\nu}$ such that 
\begin{gather}\label{eq:OriginOfk}
    \partial_{\nu} k_{\xi}^{\mu\nu} = \omega_{\xi}^{\mu} + 2\,\delta\big( \sqrt{-g}\,\mathcal{G}^{\mu\lambda} g_{\lambda\nu} \big) \, \xi^{\nu} ~.
\end{gather}
When the fields are fully on-shell, $\mathcal{G}^{\mu\nu} = 0$, this reduces to the usual symplectic current contracted with an asymptotic symmetry $\xi$.

To arrive at the result \eqref{eq:boundarycharge} for $k_{B}^{ri}[\xi]$ we begin with the expression \eqref{eq:FullThetarOnB} for $\widetilde{\Theta}_{B}^{r}$. Then the relevant part of $\omega_{\xi}^{r}$ is
\begin{gather}
    \omega_{B,\xi}^{r} = \frac{1}{2}\,\delta\Big(\sqrt{-h_0}\,T^{ij} \Big) \delta_{\xi} h^{0}_{ij} - \frac{1}{2}\,\delta_{\xi}\Big( \sqrt{-h_0}\,T^{ij} \Big) \delta h^{0}_{ij} ~.
\end{gather}
Using the transformations \eqref{eq:3DVolumeTransform}, \eqref{transfoh0}, and \eqref{eq:transfTij}, then integrating by parts with respect to both $\partial_i$ and $\delta$, leads to
\begin{align}\label{eq:omegaBxi}
    \omega_{B,\xi}^{r} = &\,\,\partial_{i}\left[ \delta \Big(\sqrt{-h_0} \,T^{ik} h^{0}_{kj} \xi^{j} \Big) - \sqrt{-h_0} \,T^{ik} h^{0}_{kj} \delta \xi^{j} - \frac{1}{2}\,\xi^{i} \sqrt{-h_0} \,T^{jk}\,\delta h^{0}_{jk} \right] \\ \nonumber
    & \,\, - \xi^{j}\,\delta\Big(\sqrt{-h_0}\,{}^{3}\nabla^{i} T_{ij} \Big) ~.
\end{align}
Terms involving the PBH component of the diffeomorphism have canceled by the traceless condition $T^{ij}\,h^{0}_{ij} = 0$. Since the Brown-York tensor is not conserved for partially on-shell fields, the $\omega_{B, \xi}^{r}$ contribution to the conserved current is not a total derivative on its own. But the second line of \eqref{eq:omegaBxi} is precisely canceled in \eqref{eq:OriginOfk} by the contribution originating with the weakly vanishing Noether current. The remainder is the total derivative in the first line of \eqref{eq:omegaBxi}, which results in \eqref{eq:boundarycharge}.

\section{Kerr-AdS$_4$}
\label{app:Kerr}

The Kerr-de Sitter metric was presented in ``generalized Bondi coordinates'' in \cite{Hoque:2021nti}. Adapting these results to a negative cosmological constant, the leading part of the metric on $\CC$ is
\begin{align}
    \gamma^{0}_{AB}\,\dd x^{A} \dd x^{B} = & \,\, \frac{1}{\Delta_{\theta}}  \left(1 - \frac{\alpha^{2} \, \Xi\,\cos^{2}\theta}{\Delta_{\theta}}\right) \, \dd\theta^{2} + 2\, \frac{\alpha^{2} \cos\theta \, \sin^{2}\theta}{\Xi^{\frac{1}{2}} \, \Delta_{\theta}}\, \dd\theta \dd\phi + \frac{\sin^{2}\theta}{\Xi} \, \dd \phi^{2}
\end{align}
with
\begin{gather}
    \Xi = 1 - \alpha^{2} \qquad \Delta_{\theta} = 1 - \alpha^{2} \cos^{2}\theta \qquad \sqrt{\det \gamma_0} = \frac{\sin\theta}{\Xi^{\frac{1}{2}}}~.
\end{gather}
Here $\alpha = a/\ell$ is the usual Kerr rotation parameter made dimensionless with a factor of $\ell = \sqrt{-3/\Lambda}$. The leading components of the vector $\UU_{0}^{A}$ and scalar $\beta_0$ are
\begin{gather}
    \UU_{0}^{\theta} = \frac{\alpha \cos\theta}{\ell \, \Xi^{\frac{3}{2}}} \qquad \UU_{0}^{\phi} = - \frac{\alpha}{\ell \, \Xi \, \Delta_{\theta}} \qquad     e^{2\beta_0} = \Xi^{-\frac{3}{2}} ~.
\end{gather}
The Cotton tensor vanishes for the three-dimensional metric $h^{0}_{ij}$, so the metric on $B$ is conformal to $\mathbb{R} \times S^{2}$. The next term in the $1/r$ expansion of the metric on $\CC$ gives the shear $\hat{\gamma}^{1}_{AB}$ and trace $\gamma_1$, which are
\begin{align}
    \hat{\gamma}^{1}_{\theta\theta} = & \,\, \frac{\alpha\,\ell}{\sin\theta}\,\left( 1 - \frac{\alpha^{2} \cos^{2}\theta \, \sin^{2}\theta}{ \Delta_{\theta}^{\,2}}\right) \\
    \hat{\gamma}^{1}_{\theta\phi} = & \,\, -\alpha^{3} \ell \, \frac{\cos\theta \, \sin\theta}{\Xi^{\frac{1}{2}} \Delta_{\theta}} \\
    \hat{\gamma}^{1}_{\phi\phi} = & \,\, - \alpha\,\ell \,\frac{\sin\theta}{\Xi} \\
    \gamma_{1} = & \,\, \frac{2\,\ell\,\alpha}{\sin\theta}\,\Big(1-2\,\cos^{2}\theta\Big) ~.
\end{align}
To compute the charges we also need the subleading terms $\UU_{3}^{\,A}$ and $\VV_{3}$ from $h_{uk}$:
\begin{gather}
    \UU_{3}^{\,\theta} = \,\, -\frac{2\,\ell\,\alpha\,M}{\Xi^{\frac{5}{2}}}\,\cos\theta \qquad
    \UU_{3}^{\,\phi} = \,\, \frac{2\,\ell\,\alpha\,M}{\Xi^{2} \, \Delta_{\theta}} \qquad
    \VV_{3} = \,\, \frac{(2-\alpha^{2})\,M}{\Xi^{\frac{5}{2}}} ~.
\end{gather}
Using the expressions for $\MM$, $\JJ_{A}$, and $\PP_{A}$, we obtain
\begin{align}
    \MM = & \,\, \frac{M\,(2+\alpha^{2})}{2 \, \Xi} \\
    \PP_{\theta} - \frac{6}{\Lambda}\,\JJ_{\theta} = & \,\, \frac{6 \, M \, \alpha \,\ell \, \cos\theta}{ \Delta_{\theta}} \\
    \PP_{\phi} - \frac{6}{\Lambda}\,\JJ_{\phi} = & \,\, -\frac{6 \, M \, \alpha \,\ell \, \sin^{2}\theta}{ \Xi^{\frac{3}{2}}} ~.
\end{align}
These quantities determine the integrand of the stress tensor charges as written in \eqref{eq:StressTensorCharges}. The Kerr-AdS$_4$ solution in generalized Bondi coordinates has Killing vectors $\partial_{\phi}$ and $\partial_{u}$. The angular momentum associated with the former is 
\begin{align}
    Q\big[\!-\partial_{\phi}\big] = & \,\, \int_{0}^{2\pi}\bns \dd\phi \int_{0}^{\pi} \nts\!\! \dd\theta \, \frac{\sin\theta}{\Xi^{\frac{1}{2}}} \, \frac{1}{2\,\kappa^{2}} \left(\frac{6 \, M \, \alpha \,\ell \, \sin^{2}\theta}{ \Xi^{\frac{3}{2}}} \right) \\
    & \,\, = \frac{8 \pi M a}{\kappa^{2}\,\Xi^{2}} ~.
\end{align}
For the mass associated with $\partial_u$ we have $\xi^{v} = e^{2\beta_0}$ and $(\!\perp\!\xi)^{A} = - \UU_{0}^{\,A}$, giving
\begin{align}
    Q\big[\partial_{u} \big] = &\,\, \int_{0}^{2\pi}\bns \dd\phi \int_{0}^{\pi} \nts\!\! \dd\theta \, \frac{\sin\theta}{\Xi^{\frac{1}{2}}} \, \left[ \frac{2}{\kappa^{2}} \, \frac{M\,(2+\alpha^{2})}{2 \, \Xi } \, e^{2\beta_0}  + \frac{1}{2\kappa^{2}} \, \left( \PP_{A} - \frac{6}{\Lambda}\,\JJ_{A} \right) \, \big(-\UU_{0}^{\,A}\big) \right] \\ 
    = & \,\, \frac{8 \pi M}{\kappa^{2} \, \Xi^{2}} ~.
\end{align}
These are precisely the angular momentum and mass for the Kerr-AdS$_4$ black hole as discussed, for instance, in \cite{Gibbons:2004ai, Papadimitriou:2005ii}.
In these coordinates, the timelike Killing vector associated with the mass is $\partial_u$, rather than a linear combination of $\partial_u$ and $\partial_{\phi}$.  

In addition to the charges constructed from $\TT$ and $\TT_{A}$, we can also extract the components of $\widehat{\TT}_{AB}$, the corner stress tensor $\hat{\tau}^{AB}$, and the scalar $\chi$.
\begingroup
\allowdisplaybreaks
\begin{align}
    \widehat{\TT}_{\theta\theta} = &\,\,\frac{3\,M\,\alpha^{2}}{2\, \Xi\,\Delta_{\theta}^{\,2}}\,\Big(\,\Xi^{2}\,\cos^{2}\theta - \sin^{2}\theta \Big) \\
    \widehat{\TT}_{\theta\phi} = &\,\, - \frac{3\,M\,\alpha^{2} \big(2-\alpha^{2}\big) \cos\theta \, \sin\theta}{ 2 \, \Xi^{\frac{3}{2}} \Delta_{\theta}} \\
    \widehat{\TT}_{\phi\phi} = &\,\, \frac{3\,M\,\alpha^{2} \big(1-2\,\cos^{2}\theta\big) \sin^{2}\theta}{ 2 \, \Xi^{2}} \\
    \hat{\tau}^{\theta\theta} = & \,\, \frac{1}{2} \, a^{2} \\
    \hat{\tau}^{\theta\phi} = & \,\, -a^{2} \, \frac{\alpha^{2} \, \Xi^{\frac{1}{2}} \,  \cos\theta}{ 2 \, \Delta_{\theta}} \\
    \hat{\tau}^{\phi\phi} = & \,\, -a^{2} \, \frac{\Xi}{ 2 \, \sin^{2}\theta \, \Delta_{\theta}^{\,2}}\,\Big(\Delta_{\theta}^{\,2} - \alpha^{4} \,\cos^{2}\theta \, \sin^{2}\theta \Big) \\
    \chi = & \,\, a^{2} \, \frac{1}{2}\,\Big(3\,\cos^{2}\theta - 1 \Big) ~.
\end{align}
\endgroup
Finally, only the vector component of the magnetic part of the Weyl tensor is non-zero for Kerr-AdS$_4$. 
\begin{align}
    \JJ_{\theta} + \frac{\Lambda}{6}\,\PP_{\theta} = & \,\, - \frac{3\,M\,\alpha^{3}\,(3-3 \, \alpha^2 + \alpha^4)\,\cos\theta}{\ell\,\Xi^{3}\,\Delta_{\theta}} \\
    \JJ_{\phi} + \frac{\Lambda}{6}\,\PP_{\phi} = & \,\, \frac{3\,M\,\alpha^{3}\,(3-3 \, \alpha^2 + \alpha^4)\,\sin^{2}\theta}{\ell\,\Xi^{\frac{9}{2}}} 
\end{align}
Both $\widetilde{\MM}$ and the combination of $\NN_{AB}$ and $\EE_{AB}$ appearing in the magnetic components vanish.

\providecommand{\href}[2]{#2}\begingroup\raggedright\endgroup

\end{document}